\documentclass[10pt,amsmath,amssymb,prx,superscriptaddress, twocolumn]{revtex4}
\usepackage{multirow}

\usepackage{mathrsfs}

\usepackage{ascmac}
\usepackage[dvipdfmx]{graphicx}
\usepackage{dcolumn}
\usepackage{bm}
\usepackage{times}
\usepackage{booktabs}
\usepackage{color}
\newcounter{one}
\usepackage{algorithm}
\usepackage{algorithmic}

\newcommand{\bbE}{\mathbb{E}}

\newcommand{\half}[1]{{ \rm  h}}

\newcommand{\T}{{\cal T}}

\newcommand{\Tr}[0]{ {\rm{Tr}}}
\newtheorem{theorem}{Theorem}
\newtheorem{lemma}{Lemma}
\newtheorem{proposition}{Proposition}

\newcommand{\cH}{{\cal H}}

\newtheorem{propositiondash}{Proposition}

\def\QED{\mbox{\rule[0pt]{1.5ex}{1.5ex}}}

\def\endproof{\hspace*{\fill}~\QED\par\endtrivlist\unskip}

\newenvironment{proofof}[1]{\vspace*{5mm} \par \noindent
         {\bf Proof of #1:\hspace{2mm}}}{\endproof
}

\def\Label{\label}

\newcommand{\affA}{Center for Emergent Matter Science (CEMS), RIKEN, Wako, Saitama 351-0198 Japan}
\newcommand{\affB}{Graduate School of Mathematics, Nagoya University, Furocho, Chikusa-ku, Nagoya 464-8602, Japan}
\newcommand{\affC}{Centre for Quantum Technology, National University of Singapore,
Singapore 117543}

\begin{document}
\title{Optimal Efficiency of Heat Engines with Finite-Size Heat Baths}
\author{Hiroyasu Tajima}
\affiliation{\affA}
\author{Masahito Hayashi}
\affiliation{\affB}
\affiliation{\affC}

\begin{abstract}
The optimal efficiency of quantum (or classical) heat engines whose heat baths are $n$-particle systems is given by the information geometry and the strong large deviation.
We give the optimal work extraction process as a concrete energy-preserving unitary time evolution among the heat baths and the work storage.
We show that our optimal work extraction turns the disordered energy of the heat baths to the ordered energy of the work storage, by evaluating the ratio of the entropy difference to the energy difference in the heat baths and the work storage, respectively.
By comparing the statistical mechanical optimal efficiency with the macroscopic thermodynamic bound, we evaluate the accuracy of the macroscopic thermodynamics with finite-size heat baths from the statistical mechanical viewpoint.
We also evaluate the quantum coherence effect on the optimal efficiency of the cycle processes without restricting their cycle time, by comparing the classical and quantum optimal efficiencies.
\end{abstract}
 
\maketitle

\section{Introduction}\Label{introduction}

Thermodynamics started as a study to clarify the upper limit of the efficiency of heat engines \cite{Carnot} and has become a huge realm of science which covers from electric batteries \cite{Fermi} to black holes \cite{Bardeen}.   
Statistical mechanics, which is another theory for macroscopic systems, was born as an effort to explain thermodynamic behavior of large systems from microscopic mechanical laws \cite{Maxwell, Boltzmann}. 
It also grew to one of the groundworks of physics, having many applications to other domains of science \cite{ouyoutoukei111}.

Meanwhile, as pointed by Lieb and Yngvason \cite{Lieb1996}, statistical mechanics has not completely succeeded in founding thermodynamics, which is one of the original aims of statistical mechanics.
One of such unfulfilled goals of statistical mechanics is to derive the optimal performance of heat engines from a microscopic mechanical laws \cite[page3]{Shimizu}. All of existing results deriving  the optimal performance assume either of the following two assumptions;$\\$
\textit{Free-resource assumption \cite{Sekimoto, Horodecki, oneshot1, oneshot3, Egloff, Brandao, Car2, Popescu2014, oneshot2};} we can treat the heat baths as a kind of resource which we can use freely. $\\$
\textit{Quasi-static assumption\cite{Ehrenfest, Jarzynski, Croocks, Sagawa2010, Ponmurugan2010, Horowitz2010, Horowitz2011, Sagawa2012, Ito2013, tasaki, Kurchan, Car1, Xiao, Lenard, TLH, BBM1, oldresult5, sagawa1, jacobs, sagawa2, Funo, Morikuni, Parrondo, Izumida, IE1, IE2, IE1.5};} we can perform quasi-static processes, which transforms Gibbs state into another Gibbs state.

The former assumption has been used in the stochastic differential approach \cite{Sekimoto}, which treats the heat baths as the free resource of noise.
Recently, resource theory approach  has been actively studied \cite{Horodecki,oneshot1,oneshot3,Egloff,Brandao,Car2,Popescu2014,Popescu2015,oneshot2}, however, it 
also assumes that the heat baths are the free resource of Gibbs states.
However, in real, there is no free resource, in particular, when 
the work is extracted from two baths with different temperatures.
This is because we need to pay some cost to realize such two bathes in a real heat engine.
(For example, in real engines, we have to burn some fuel to make the temperature difference.)
To reflect such realistic situations, we have to take into account the finiteness of the heat baths.

For this purpose, we need to treat the working body and the heat baths as a closed system, which is called 
the closed dynamics approach \cite{Ehrenfest}. 
Most of the existing studies \cite{Ehrenfest, Jarzynski, Croocks, Sagawa2010, Ponmurugan2010, Horowitz2010, Horowitz2011, Sagawa2012, Ito2013, tasaki, Kurchan, Car1, Xiao, Lenard, TLH, BBM1, oldresult5, sagawa1, jacobs, sagawa2, Funo, Morikuni, Parrondo, Izumida, IE1, IE2, IE1.5} of this approach assume the quasi-static assumption to derive the optimal performance of heat engines.
However, any quasi-static process has not been described 
as a time evolution of a closed system.
Hence, we can't say that the existing studies predicts the right value of the optimal performance that is achieved by a microscopic mechanical dynamics.
For the same reason, this assumption makes the closed dynamics approach be not self-contained.
Therefore, there has been no self-contained theory to recover thermodynamics based on microscopic viewpoint under the realistic situation.
This problem has been considered for the information erasing process in a pioneering research by Reeb and Wolf \cite{Reeb}, but there is no general method to evaluate the finite-size effect affect on the optimal performance of heat engines.

In the present article, we establish the limited-resource theory to treat the thermodynamic features of the finite-size heat baths from microscopic mechanical laws. 
To be concrete, we analyze how the finite-size heat baths affect the optimal efficiency of heat engines by statistical mechanics, by giving an asymptotic expansion for the optimal efficiency without the quasi-static assumption. 
The asymptotic expansion shows how the optimal efficiency goes down from Carnot efficiency when the particle numbers of the heat baths are $n$.

Our theory can be applied to both of the classical and quantum heat engines.
With using this feature, we secondly tackle an interesting problem of the difference between the performances of the classical heat engines and the quantum heat engines.
Today, in various fields including the computer science, it is known that the quantum devices using the quantum coherence can attain the task which is impossible for the classical devices.
Investigating the comparison, we can evaluate the effect by the quantum coherence with respect to work extraction.
Clearly, when we don't restrict the cycle time and the size of the heat baths, the optimal efficiencies should be the same, i.e., Carnot efficiency.
However, it is known that when we restrict the cycle time but don't restrict the size of the heat engine, the quantum coherence makes the optimal efficiency of the quantum heat engine better than that of the classical heat engine \cite{X1,X2}.
Our result give an answer to the opposite case, i.e., the case that we restrict the size of the heat baths, but don't restrict the cycle time.
To clarify the difference,
we give the asymptotic expansions for both of the quantum setup and the classical setup.
By comparing them, we show that the effect of the quantum coherence on the optimal efficiency is sublinear to $Q_{n}/n$, where $Q_{n}$ is the extracted heat from the hot heat bath.

Indeed, we have another merit for discussing classical and quantum heat engines as follows.
Fortunately, both heat engines have common structures.
In this paper, we clarify this similarity, and
study the efficiency of classical heat engine.
Combining them, we investigate the efficiency of quantum heat engine.
That is, in order to derive the efficiency in the quantum statistical mechanical setup, 
we need to discuss in the classical statistical mechanical setup as an essential step.

Our asymptotic expansions of the optimal efficiencies are the predictions purely given by statistical mechanics.
Therefore, by comparing these predictions with the prediction given by thermodynamics, we can evaluate the accuracy of thermodynamics for finite-many body systems from statistical mechanical viewpoint.
For this purpose, we also introduce thermodynamic setup, other than the above two statistical mechanical setups.
And, we compare our asymptotic expansions given by statistical mechanics with the prediction given by thermodynamics. 
As the result, we show that the prediction given by thermodynamics for the optimal efficiency harmonizes with the statistical mechanical prediction up to the order $Q^{2}_{n}/n^2$.

In order to accomplish our purposes, we have to discuss the statistical mechanical formulations and the thermodynamic formulation parallelly.
In thermodynamics, the external work storage is usually treated implicitly, and the work extracted by an adiabatic operation is defined as the difference of the internal energy of the composite system of the heat baths and the working body.
This definition is justified by the first law of thermodynamics.
Therefore, we mainly treat the implicit setup of the external work storage in statistical mechanics.

On the other hand, recently, the study treating the external work storage explicitly is studied hard\cite{Horodecki, oneshot1, oneshot3, Egloff, Brandao, Car2, Popescu2014,Popescu2015, oneshot2}.
In such studies, the work extraction is describe as an energy preserving unitary on the whole system which consists of the internal system and the external work storage.
However, in such an explicit formulation, we face a subtle issue, i.e., the classification between heat and work.
Today, there is no consensus about the statistical definitions of the work, which is the ordered energy,  and the heat, which is the disordered energy.
Some of the above studies \cite{Horodecki, oneshot1, oneshot3, Egloff, Brandao, Car2, oneshot2} employ the notion of the single-shot work extraction, in which the work extraction is defined as a deterministic translation from the ground pure state to the excited pure state in a two-level system.
It defines a work-like energy transfer in the quantum scale well.
However, the correspondence between the shingle-shot work extraction and the work extraction in thermodynamics is not straight forward.
The shingle-shot work extraction defines the work extraction as a deterministic energy transfer in microscopic scale, although the work extraction in a macroscopic device should be fluctuated in the microscopic scale.
The notion of the average work extraction \cite{Popescu2014, Popescu2015}, which is another notion for the work-like energy transfer in the microscopic scale, has a clear correspondence with the work extraction in thermodynamics.
In the average work extraction, the extracted work is defined as the gain of the internal energy of the external work storage.
In the definition, the amount of the extracted work is equal to the loss of the internal energy of the internal system, as with thermodynamics.

However, the average work extraction has a serious drawback of implicit description of the work storage, i.e., there is no clear criterion describing work and heat for the average work extraction in the explicit formulation.
As the last theme of our paper, we tackle this problem.
We firstly show that we can translate the explicit formulation with a whole unitary and the implicit formulation of an average work extraction
each others, by using the translation between the direct measurement and the indirect measurement.
Secondly, we give a novel criterion describing work and heat for the average work extraction in the explicit formulation.
We give the criterion by the ratio of the energy gain to the entropy gain in the external work storage;
when the entropy-energy ratio of the external work storage goes to 0 in the macroscopic limit, we consider the energy gain in the work storage to be ``work.''
Next, we give a concrete protocol which attains the optimal efficiency, in the explicit formulation.
Finally, we show that our optimal protocol translate ``heat'' in the baths to  ``work'' in the work storage.

This paper is organized as follows.
In Section \ref{s1}, we discuss the set ups that we use in this paper. 
In order to accomplish our purposes, we have to compare the predictions in the classical statistical mechanics, the quantum statistical mechanics and thermodynamics.
Therefore, we introduce three setups, i.e.,  the setup of the classical statistical mechanics in Section \ref{s1AA}, the setup of the quantum statistical mechanics  in Section \ref{s1AB}, and a setup in thermodynamics in Section \ref{s1B}.
In Section \ref{s2}, we give our mains results.
For each setup, we derive the asymptotic expansion of the optimal efficiency.
Then, we compare the asymptotic expansions each others.
In Section \ref{s4},  we explain the relationship between the implicit formulation and the explicit formulation.
We also give a concrete protocol which attains the optimal efficiency, and show that our optimal work extraction is a good example of the translation from ``heat'' to ``work''.

\section{Set Up: Cyclic Heat Engines with two baths}\Label{s1}

The purpose of this article is to analyze how the finite-size heat baths affect the optimal efficiencies of heat engine in both of the classical case and the quantum case, and to compare them each others. 
We also compare these two predictions of optimal efficiencies given by statistical mechanics with the prediction given by thermodynamics, in order to verify how close the statistical mechanical predictions are to the thermodynamic prediction with finite-size resources. 
Therefore, we introduce two setups in Subsection \ref{s1A}, i.e.,  the setup of the classical statistical mechanics and the setup of the quantum statistical mechanics.
For the comparison, Subsection \ref{s1B} gives the detailed thermodynamic setup of a heat engine with finite-size heat baths.

\subsection{Formulations in statistical mechanics}\Label{s1A}

\subsubsection{Requirements for the statistical mechanical formulations}
In order to treat the thermodynamic setup and the statistical mechanical setups parallelly, 
we organize the statistical mechanical formulations so that they are as similar to the thermodynamic formulation as possible.
Therefore, we firstly overview the features of the thermodynamic formulations, and translate them into the requirements for the statistical mechanical formulations.

In thermodynamics, the set up the heat engine with finite-size heat baths satisfies the following conditions (We will introduce the detailed setup in the subsection \ref{s1B});
\begin{description}
\item[A: Implicit treatment of the work storage]{We don't treat work storage explicitely. We only treat the composite system of the working body and the finite-size heat baths.}
\item[B: Entropy non-decreasing rule]{During the process, we don't use any extra thermodynamic system. Namely, we perform the adiabatic operation on the composite system. The adiabatic operation does not decrease the thermodynamic entropy regardless of the initial state of the composite system.}
\item[C: Cyclic process]{We assume that the adiabatic process is cyclic for the working body. Namely, the initial and the finial states of the working body are the same. Also, we do not restrict about the process time.}
\item[D: Energy conservation law]{During the process, we assume that the heat from the hot bath and the extracted work are equal to the loss of the internal energy of the hot bath and the loss of the internal energy of the composite system, respectively.}
\end{description}

Therefore, we request  both classical and quantum statistical mechanical setups to satisfy the following conditions.
\begin{description}
\item[A': Implicit treatment of the external work storage]{We don't treat the external work storage explicitly. We only treat the composite system of the working body and the finite-size heat baths.}
\item[B': Entropy non-decreasing rule]{Similar to thermodynamics, we impose the entropy non-decreasing rule, i.e., 
we assume that our operation does not decrease the Shannon (or the von Neumann) entropy.}
\item[C': Cyclic rule]{We assume that our operation is cyclic for the working body. Namely, the initial and the finial states of the working body are the same. Also, we do not restrict about the process time.}
\item[D': Energy conservation law]{During the process, we assume that the heat from the hot bath and the extracted work as to be the loss of the internal energy of the hot bath and the loss of the internal energy of the composite system, respectively.}
\end{description}
In Subsections \ref{s1AA} and \ref{s1AB}, we introduce such statistical mechanical setups.
One might consider that we need to explicitly describe the external work storage, at least in the quantum setup.
However, as explained later, such an implicit treatment of the external work storage is possible.
The relationship between the implicit treatment and the explicit treatment will be explained in the section \ref{s4}.

\subsubsection{Classical heat engine}\Label{s1AA}
We firstly introduce the setup for the classical heat engine with two finite-size heat baths.
We introduce conditions A'--D' in order.

\textit{A': Implicit treatment of the external work storage:}$\\$
Since we discuss only the composite system of the working body and the heat baths in the thermodynamic set up, 
it is natural to formulate a heat engine in the classical statistical mechanics so that 
only the composite system of the working body and the heat baths is discussed.
Such a setup has been used to study classical heat engines \cite{Jarzynski,Sagawa2010,Ponmurugan2010,Horowitz2010,Horowitz2011,Sagawa2012,Ito2013} since Gibbs and Einstein \cite{Ehrenfest}, and the setup is called the classical-standard model \cite[Section I]{1ponme}.
In the classical-standard model, the work extraction process is formulated as an invertible and deterministic dynamics in the classical system.
In this paper, we employ a stochastic extension of the classical-standard model.

As the composite system, we consider three classical systems; an $n$-particle hot bath ${\cal X}$, an $n$-particle cold bath ${\cal Y}$ and a working body (catalyst) ${\cal Z}$.
We refer to the composed system ${\cal X}\times{\cal Y}\times{\cal Z}$ as the internal system ${\cal I}$. 
We assume that each classical bath has $d^{n}<\infty$ microscopic states, and that the working body has $D<\infty$ microscopic states.
We give the Hamiltonian of the baths as a real valued function of the microscopic states; $h^{(n)}_{I}(x,y,z):=h^{(n)}_{X}(x)+h^{(n)}_{Y}(y)+h_{Z}(z)$, where $x$, $y$ and $z$ are the microscopic states of ${\cal X}$, ${\cal Y}$ and ${\cal Z}$, respectively.
We assume that the initial distribution $P^{(n)}_{I}(x,y,z)$ of ${\cal I}$ is the composite distribution of an arbitrary distribution of ${\cal Z}$, and the Gibbs states of ${\cal X}$ and ${\cal Y}$; $P^{(n)}_{I}(x,y,z)=P_{\beta_{H}|h^{(n)}_{X}}(x)P_{\beta_{L}|h^{(n)}_{Y}}(y)P_{Z}(z)$, where $P_{\beta|h}(m):=e^{-\beta h(m)}/(\sum_{m}e^{-\beta h(m)})$.
We also use the abbreviation $P^{(n)}_{\beta_{H}\beta_{L}}(x,y):=P_{\beta_{H}|h^{(n)}_{X}}(x)P_{\beta_{L}|h^{(n)}_{Y}}(y)$.
Then, we describe the stochastic dynamics of the above internal system ${\cal I}$
as a probability transition matrix $\T(x,y,z|x',y',z')$
while a work extraction process 
is formulated as an invertible and deterministic 
dynamics in the classical system originally in the classical-standard mode \cite{Ehrenfest}.

\textit{B': Entropy non-decreasing rule:}$\\$
In the thermodynamics, it is conventional to impose entropy decreasing rule for our operation.
Hence, to discuss the statistical mechanical counterpart,
we assume that the probability transition matrix $\T(x,y,z|x',y',z')$ corresponding to our operation does not decrease the Shannon entropy of $I$;
\begin{align}
S(P^{(n)}_{I})\le S(P^{(n)'}_{I})\Label{0}
\end{align}
for an arbitrary initial distribution $P^{(n)}_{I}$, where $S(P):=\sum_{j}-P(j)\log P(j)$ and $P^{(n)'}_{I}(x,y,z):=\sum_{x',y',z'}\T(x,y,z|x',y',z')P^{(n)}_{I}(x',y',z')$.
The condition \eqref{0} is equivalent to the condition that $\T(x,y,z|x',y',z')$ is a bi-stochastic probability transition matrix;
\begin{equation}
\sum_{x,y,z}\T(x,y,z|x',y',z')=\sum_{x',y',z'}\T(x,y,z|x',y',z')=1.\Label{1}
\end{equation}
Also, a matrix $\T(x,y,z|x',y',z')$ is bi-stochastic if and only if $\T(x,y,z|x',y',z')$ is given as a probabilistic mixture of invertible and deterministic dynamics.
So, to satisfy the similarity with the thermodynamics,
we adopt the above formulation as a natural extension of the classical standard model \cite{Ehrenfest}.

\textit{C': Cyclic rule:}$\\$
In response to the condition of the cyclic process in the thermodynamic setup, 
we also assume that the dynamics $\T$ is cyclic for the working body ${\cal Z}$;
\begin{align}
\sum_{x,y}P^{(n)'}_{I}(x,y,z)=P_{Z}(z).\Label{2}
\end{align}

\textit{D': Energy conservation law:}$\\$
In response to the condition of the energy conservation law in the thermodynamic setup, we define the work and the heat as the loss of the internal energy of the composite system $I$, and the hot bath $X$, which is the standard definitions in the classical standard model.
Namely, the amount of extracted work $W^{(n)}(\beta_{H},\beta_{L},\T)$ 
and the endothermic energy amount $Q^{(n)}_H(\beta_{H},\beta_{L},\T)$
are given as
\begin{align}
&W^{(n)}(\beta_{H},\beta_{L},P_{Z},\T):=\left<h^{(n)}_{I}\right>_{P^{(n)}_{I}}-\left<h^{(n)}_{I}\right>_{P^{(n)'}_{I}},\\
&Q^{(n)}_H(\beta_{H},\beta_{L},P_{Z},\T):=\left<h^{(n)}_{X}\right>_{P^{(n)}_{I}}-\left<h^{(n)}_{X}\right>_{P^{(n)'}_{I}},
\Label{4}
\end{align}
where for an arbitrary function $A(z)$ and a probability $P(z)$, $\left<A\right>_{P}:=\sum_{z}A(z)P(z)$.
So, we define the efficiency $\eta^{(n)}_C(\beta_{H},\beta_{L},P_{Z},\T)$ in the following way;
\begin{align}
\eta^{(n)}_C (\beta_{H},\beta_{L},P_{Z},\T) :=
\frac{W^{(n)}(\beta_{H},\beta_{L},P_{Z},\T)}{Q^{(n)}_H(\beta_{H},\beta_{L},P_{Z},\T)}.
\end{align}
Therefore, when we fix the endothermic energy amount $Q^{(n)}_H(\beta_{H},\beta_{L},P_{Z},\T)$ at an arbitrary real number $Q$,
the optimal efficiency $\eta^{(n)}_C[\beta_{H},\beta_{L},Q]$ is given as
\begin{align}
\eta^{(n)}_C[\beta_{H},\beta_{L},Q]:= \sup_{\T,P_{Z}: \substack{\eqref{1},\enskip\eqref{2}\\ Q^{(n)}_{H}=Q}}
\eta^{(n)}_C (\beta_{H},\beta_{L},P_{Z},\T).
\end{align}

As a special case of above, we can consider the work extraction without the working body ${\cal Z}$.
In this case, we perform a bi-stochastic dynamics $T(x,y|x',y')$ on ${\cal X}\times{\cal Y}$.
In order to compare the case with the working body and the case without the working body, 
we refer to the extracted work, the endothermic amount, the efficiency and the optimal efficiency of the case without the working body as $W^{(n)}_{\lnot}(\beta_{H},\beta_{L},\T)$, $Q^{(n)}_{H\lnot} (\beta_{H},\beta_{L},\T)$, $\eta^{(n)}_{C\lnot} (\beta_{H},\beta_{L},\T)$, and $\eta^{(n)}_{C\lnot}[\beta_{H},\beta_{L},Q]$.
Trivially, we have the relation
$\eta^{(n)}_{C\lnot}[\beta_{H},\beta_{L},Q]
\ge \eta^{(n)}[\beta_{H},\beta_{L},Q]$.
Comparing them, we can evaluate the effect by a working body (catalyst).

\subsubsection{Quantum heat engine}\Label{s1AB}
We secondly introduce our setup for the quantum heat engine with two finite-size heat baths.
As in the classical formulation, we introduce the conditions A'--D' in order.

\textit{A': Implicit treatment of the external work storage:}$\\$
As in the thermodynamic set up, we only treat the composite system $I$ of the working body and the heat baths in the quantum setup. 
As the components of the composite system, we introduce three quantum systems; 
an $n$-particle hot bath $H$, an $n$-particle cold bath $L$, and a catalyst (working body) $C$. 
We refer to the Hilbert spaces of $H$, $L$ and $C$ as $\cH_{H}$, $\cH_{L}$ and $\cH_{C}$, and refer to the Hilbert space of the internal system $I:=HLC$ as $\cH_{I}:=\cH_{H}\otimes\cH_{L}\otimes\cH_{C}$.
We assume that each bath is composed of $n$-particles of $d$-level, i.e., the dimension of $\cH_{L}$ is $d^{n}<\infty$.
We also assume that the dimension of $\cH_{C}$ is $D<\infty$. 
We give the Hamiltonian of $I$ as $\hat{H}^{(n)}_{I}:=\hat{H}^{(n)}_{H}+\hat{H}^{(n)}_{L}+\hat{H}_{C}$.
In order to treat the quantum setup in parallel with the classical setup,
we assume that 
\begin{align}
&\hat{H}^{(n)}_{H}=\sum_{x}h^{(n)}_{X}(x)|x\rangle_{H}~_{H}\langle x|,\enskip
\hat{H}^{(n)}_{L}=\sum_{y}h^{(n)}_{Y}(y)|y\rangle_{L}~_{L}\langle y|,\nonumber\\
&\hat{H}_{C}=\sum_{z}h_{Z}(z)|z\rangle_{C}~_{C}\langle z|,\Label{(8)}
\end{align}
where $|x\rangle_{H}$, $|y\rangle_{L}$ and $|z\rangle_{C}$ are the energy pure eigenstates whose eigenvalues are $h^{(n)}_{X}(x)$, $h^{(n)}_{Y}(y)$, and $h_{Z}(z)$ respectively.
We assume that the initial state of $I$ is the product state of an arbitrary state $\rho_{C}$ of $C$ and the Gibbs states of $H$ and $L$; $\rho^{(n)}_{I}:=\rho_{\beta_{H}|\hat{H}^{(n)}_{H}}\otimes\rho_{\beta_{L}|\hat{H}^{(n)}_{L}}\otimes\rho_{C}$, where $\rho_{\beta|\hat{H}}:=\exp[-\beta\hat{H}]/\Tr[\exp[-\beta\hat{H}]]$.
We also use the abbreviation $\rho^{(n)}_{\beta_{H}\beta_{L}}:=\rho_{\beta_{H}|\hat{H}^{(n)}_{H}}\otimes\rho_{\beta_{L}|\hat{H}^{(n)}_{L}}$.

As a quantum counterpart of stochastic dynamics, 
we consider an operator ${\cal W}$ mapping from a density operator to another density operator.
Due to the nature of quantum mechanics,
such a map has to satisfy completely positivity and trace preserving property.
So, such a map is called a completely positivity and trace preserving (CPTP) map.

\textit{B': Entropy non-decreasing rule:}$\\$
As a quantum counterpart of the entropy decreasing rule,
we request that ${\cal W}$ does not decrease the von Neumann entropy of $I$, regardless of the initial state of $I$;
\begin{align}
S(\rho^{(n)}_{I})\le S(\rho^{(n)'}_{I}),\enskip\forall \rho^{(n)}_{I}\Label{(B')}.
\end{align}
where $\rho^{(n)'}_{I}={\cal W}(\rho^{(n)}_{I})$.
It is equivalent to the condition that the CPTP map ${\cal W}$ is unital \cite{openquantum};
\begin{align}
{\cal W}(\hat{1}_{I})=\hat{1}_{I},\Label{unital}
\end{align}
where $\hat{1}_{I}$ is the identity operator on $I$.
The condition \eqref{unital} can be regarded as a quantum extension of the bi-stochastic condition \eqref{1} for $\T$ in the classical case.
Therefore, our quantum setup is a natural quantum extension of the classical setup.

\textit{C': Cyclic rule:}$\\$
In response to the condition of the cyclic process in the thermodynamic setup, we assume that our work extraction process is cyclic for $C$;
\begin{align}
\Tr_{H L}\rho^{(n)'}_{I}=\rho_C
\Label{3-14-3eq}.
\end{align}

\textit{D': Energy conservation law:}$\\$
In response to the energy conservation law in the thermodynamic setup, we define the work and the heat as the loss of the internal energy of the composite system $I$, and the hot bath $H$. 
For an arbitrary unital CPTP map ${\cal W}$ and an arbitrary initial state $\rho_{C}$ of $C$, the amount of extracted work $W^{(n)}(\beta_{H},\beta_{L},\rho_{C},{\cal W})$ 
is given as
\begin{align}
W^{(n)}(\beta_{H},\beta_{L},\rho_{C},{\cal W})
&:=\Tr (\rho^{(n)}_{I}- \rho^{(n)'}_{I} ) \hat{H}^{(n)}_{I} \Label{E.C.1}.
\end{align}
Also, the endothermic energy amount 
$Q^{(n)}_{H}(\beta_{H},\beta_{L},\rho_{C},{\cal W})$
is defined as
\begin{align}
Q^{(n)}_{H}(\beta_{H},\beta_{L},\rho_{C},{\cal W})
&:= \Tr (\rho^{(n)}_{I}- \rho^{(n)'}_{I} ) \hat{H}^{(n)}_{H}.\Label{12}
\end{align}
So, the efficiency $\eta^{(n)}_{Q}(\beta_{H},\beta_{L},\rho_{C},{\cal W})$ is given as
\begin{align}
\eta^{(n)}_{Q}(\beta_{H},\beta_{L},\rho_{C},{\cal W}):=
\frac{W^{(n)}(\beta_{H},\beta_{L},\rho_{C},{\cal W})}{Q^{(n)}_{H}(\beta_{H},\beta_{L},\rho_{C},{\cal W})}.\Label{7-6-er}
\end{align}
Therefore, when we fix the endothermic energy amount $Q^{(n)}_{H}(\beta_{H},\beta_{L},\rho_{C},{\cal W})$ at an arbitrary real number $Q$,
the optimal efficiency $\eta^{(n)}_{Q}[\beta_{H},\beta_{L},Q]$ is given as
\begin{align}
\eta^{(n)}_{Q}[\beta_{H},\beta_{L},Q]
:= \sup_{\rho_C,{\cal W}:\substack{
Q^{(n)}_H=Q\\
\eqref{E.C.1},\eqref{unital},\eqref{3-14-3eq}}}
\eta^{(n)}_{Q}(\beta_{H},\beta_{L},\rho_C,{\cal W}).
\end{align}
In summary, 
when a TPCP map ${\cal W}$ satisfies the above conditions,
it is called an {\it average work extraction}
because this description gives only the average of the extracted work.

\textit{A narrow formulation -- formulation for achievability of our bound:}$\\$
The class of average work extractions is too general.
This feature is good in order to guarantee the generality of our asymptotic expansion of the optimal efficiency.
However, the class might be too large to guarantee that any operation corresponds to ``realistic'' work extraction.
To resolve this problem, we need compatible operations with the unitary dynamics on the
whole system including the external battery.
Interestingly, such an operation is given as a measurement process.
Then, as a compatible operation, we introduce 
a {\it CP-instrument work extraction} $\mathscr{W}:=\{{\cal E}_{j},w_{j}\}_{j\in J}$ as a pair of a CP-instrument $\{{\cal E}_{j}\}_{j\in J}$ and the set $\{w_{j}\}_{j\in J}$ of measured outcomes, i.e., amounts of extracted work \cite[Section 3]{1ponme}.
Here, a CP-instrument $\{{\cal E}_{j}\}_{j\in J}$ 
expresses a measurement process, and satisfies the following conditions;
(1) each ${\cal E}_{j}$ is a completely positive (CP) map, 
(2) $\sum_{j}{\cal E}_{j}$ is a completely positive and trace preserving (CPTP) map, and 
(3) $J$ is the set of the outcome, which is a discrete set.

Since we regarded $w_j$ as the amount of extracted work,
the amount of extracted work $W^{(n)}(\beta_{H},\beta_{L},\rho_{C},\mathscr{W})$ 
is given as
\begin{align}
W^{(n)}(\beta_{H},\beta_{L},\rho_{C},\mathscr{W})
:=\sum_j w_j  \Tr {\cal E}_{j}(\rho^{(n)}_{I}).
\end{align}
Similarly, the endothermic energy amount 
$Q^{(n)}_{H}(\beta_{H},\beta_{L},\rho_{C},\mathscr{W})$
and the efficiency $\eta^{(n)}_{Q}(\beta_{H},\beta_{L},\rho_{C},\mathscr{W})$ 
are defined in \eqref{12} and \eqref{7-6-er} by regarding this operation 
$\{{\cal E}_{j}\}_{j\in J}$ as a TPCP map $\sum_{j\in J} {\cal E}_{j}$ so that
$ \rho^{(n)'}_{I} := \sum_{j\in J} {\cal E}_{j} (\rho^{(n)}_{I} )$.
Then, we impose the entropy non-decreasing rule \eqref{(B')} and the cyclic rule \eqref{3-14-3eq} by using this correspondence.
Further, we additionally impose a stronger energy conservative condition as
\begin{align}
{\cal E}_{j}(\Pi_{x}) 
=P_{h_{x}-w_{j}}{\cal E}_{j}(\Pi_{x})P_{h_{x}-w_{j}}
\Label{CP2}
\end{align}
for any initial eigenstate $\Pi_{x}:=|x\rangle\langle x|$ of $\hat{H}_I^{(n)}$,
where $P_{h}$ is the projection to the eigenspace of the Hamiltonian 
$\hat{H}_I^{(n)}$ with eigenvalue $h$.
This condition is equivalent to 
\begin{align}
\hat{H}_I^{(n)} {\cal E}_{j}(\Pi_{x}) 
= {\cal E}_{j}(\Pi_{x}) \hat{H}_I^{(n)}
=(h_{x}-w_{j}){\cal E}_{j}(\Pi_{x}).
\end{align}
Then, the relation \eqref{E.C.1} holds.
In order that the measured value $w_{j}$ reflects the amount of loss of energy of internal system when the measurement outcome is $j$, 
the CP-instrument work extraction $\mathscr{W}$ needs to satisfy this condition \cite[Lemma 16]{1ponme}.

As a special case, we consider CP-instrument-work extractions without the catalyst $C$.
In this case, we refer to the extracted work, the endothermic amount, the efficiency and the optimal efficiency of the case without the working body as $W^{(n)}_{\lnot}(\beta_{H},\beta_{L},\mathscr{W})$, $Q^{(n)}_{H\lnot} (\beta_{H},\beta_{L},\mathscr{W})$, $\eta^{(n)}_{Q\lnot} (\beta_{H},\beta_{L},\mathscr{W})$, and $\eta^{(n)}_{Q\lnot}[\beta_{H},\beta_{L},Q]$.
Trivially, we have the relation $\eta^{(n)}_{Q\lnot}[\beta_{H},\beta_{L},Q]\le \eta^{(n)}_{Q}[\beta_{H},\beta_{L},Q]$.
Comparing them, we can investigate the effect of catalyst and the difference between the average energy conservation law \eqref{E.C.1} and 
the stronger energy conservation law \eqref{CP2}.
In the next section, we investigate the difference.
We will construct a CP-instrument-work extractions without the catalyst $C$ to achieve the efficiency $\eta^{(n)}_{Q}[\beta_{H},\beta_{L},Q]$ in the asymptotic sense.

At the end of this subsection, 
we remark another motivation to introduce a CP-instrument work extraction.
As is mentioned in \cite{1ponme}, 
it is important to discuss the energy transfer from a collection of quantum systems
to a classical system. 
In this situation, it is crucial to determine the amount of extracted work. 
As shown in \cite[(7)]{1ponme}, when a CP-instrument work extraction satisfies \eqref{CP2},
its measurement outcome precisely determines the amount of extracted work.

\subsection{Formulation and prediction in thermodynamics}\Label{s1B}
Next, we introduce the thermodynamic setup for the macroscopic heat engines with finite-size heat baths, and review the prediction of the macroscopic thermodynamics about the optimal efficiency of the heat engines with finite-size heat baths 
when the thermodynamic equilibrium states is given by using
classical and quantum Gibbs states \cite{Tasakitext,Shimizu}.
Unlike the previous two subsections, the setup which is introduced in the present subsection is not the statistical mechanical one.

\subsubsection{Formulation in thermodynamics}
We introduce our setup, and show that the setup satisfies the conditions A--D.

\textit{A. Implicit treatment of the external work storage:}$\\$
We treat three thermodynamic systems; the reservoirs $R_{H}$ and $R_{L}$, and the working body $W$.
The reservoirs $R_{H}$ and $R_{L}$ correspond to the heat baths $H$ and $L$ respectively, and the working body $W$ corresponds to the catalyst $C$.
Unlike	$X$, $Y$, $Z$, $H$, $L$ and $C$, we assume that the thermodynamic systems 
$R_{H}$, $R_L$ and $W$ are the macroscopic thermodynamic systems, i.e., we assume that $R_{H}$, $R_L$ and $W$ obey the laws of macroscopic thermodynamics.
Then, because of the thermalization law, the state of the isolated macroscopic system (the macroscopic system surrounded by the adiabatic wall) becomes the thermodynamic equilibrium after enough long time.
We describe the thermodynamic equilibrium states of $R_{H}$, $R_{L}$ and $W$ as $(T_{R_{H}},\Omega_{R_{H}})$, $(T_{R_{L}},\Omega_{R_{L}})$ and $(T_{W},\Omega_{W})$, where $T$ is the temperature,  and $\Omega$ is the set of other thermodynamic variables.

\textit{B. Entropy non-decreasing rule:}$\\$
Under the above setup, we fix the initial temperatures of the baths $R_{H}$ and $R_{L}$ as 
$\beta_{H}=1/T_{R_{H}}$ and $\beta_{L}=1/T_{R_{L}}$, and perform the following adiabatic operation on $R_{H}R_{L}W$.
Then, we assume the second law of thermodynamics.
Namely, an adiabatic operation from $(T_{R_{H}},\Omega_{R_{H}})\times(T_{R_{L}},\Omega_{R_{L}})\times(T_{W},\Omega_{W})$ to $(T'_{R_{H}},\Omega'_{R_{H}})\times(T'_{R_{L}},\Omega'_{R_{L}})\times(T'_{W},\Omega'_{W})$  is possible only if the following inequality holds;
\begin{align}
&S_{R_{H}}(T_{R_{H}},\Omega_{R_{H}})+S_{R_{L}}(T_{R_{L}},\Omega_{R_{L}})+S_{W}(T_{W},\Omega_{W})\nonumber\\
&\le S_{R_{H}}(T'_{R_{H}},\Omega'_{R_{H}})+S_{R_{L}}(T'_{R_{L}},\Omega'_{R_{L}})+S_{W}(T'_{W},\Omega'_{W}),\Label{loei}
\end{align}
where $S(T,\Omega)$ is the thermodynamic entropy.
We also assume the existence of the adiabatic operation that transforms $(T,\Omega)$ to $(T',\Omega)$ for arbitrary $T<T'$ and $\Omega$.

\textit{C. Cyclic process:}$\\$
We also assume that the adiabatic operation is cyclic for $W$, and does not change $\Omega$ of the baths.
Combining the conditions B and C, we perform the adiabatic operation as follows;
$(T_{R_{H}},\Omega_{R_{H}})\times(T_{R_{L}},\Omega_{R_{L}})\times(T_{W},\Omega_{W})\rightarrow(T'_{R_{H}},\Omega_{R_{H}})\times(T'_{R_{L}},\Omega_{R_{L}})\times(T_{W},\Omega_{W})$.

\textit{D. Energy conservation law:}$\\$
We also assume the first law of thermodynamics.
The extracted work $W_{\mathrm{ad}}$ and the endothermic amount $Q_{\mathrm{ad}}$ of the above adiabatic operation are defined as the differences of the internal energies;
$W_{\mathrm{ad}}=E_{R_{H}}(T_{R_{H}},\Omega_{R_{H}})+E_{R_{L}}(T_{R_{L}},\Omega_{R_{L}})+E_{W}(T_{W},\Omega_{W})-E_{R_{H}}(T'_{R_{H}},\Omega_{R_{H}})-E_{R_{L}}(T'_{R_{L}},\Omega_{R_{L}})-E_{W}(T_{W},\Omega_{W})$ and $Q_{\mathrm{ad}}:=E_{R_{H}}(T_{R_{H}},\Omega_{R_{H}})-E_{R_{H}}(T'_{R_{H}},\Omega_{R_{H}})$, where $E(T,\Omega)$ is the thermodynamic internal energy.
We also assume that the thermodynamic internal energies $E_{R_{H}}(T,\Omega)$ and $E_{R_{H}}(T,\Omega)$ and the thermodynamic entropies $S_{R_{H}}(T,\Omega)$ and $S_{R_{L}}(T,\Omega)$ are continuous functions for the temperature $T$.
This assumption means that the first-order phase transition does not occurs by the change of the temperatures of the baths $R_{H}$ and $R_{L}$.
This assumption also holds for the statistical internal energies and the von-Neumann entropy of $X$, $Y$, $H$ and $L$, because of $d^{n}<\infty$; for example, $\Tr[\hat{H}^{(n)}_{H}\rho_{\beta_{H}|\hat{H}^{(n)}_{H}}]$ and $S(\rho_{\beta_{H}|\hat{H}^{(n)}_{H}})$ are continuous for the inverse temperature $\beta_{H}$.

\subsubsection{Optimal efficiency predicted by thermodynamics}
We define the thermodynamic optimal efficiency $\eta_{T}[\beta_{H},\beta_{L},\Omega_{R_{H}},\Omega_{R_{L}},Q]$ as follows;
\begin{align}
\eta_{T}[\beta_{H},\beta_{L},\Omega_{R_{H}},\Omega_{R_{L}},Q]:=\sup_{\substack{T'_{R_{H}},T'_{R_{L}},T_{W},\Omega_{W}\\:Q_{\mathrm{ad}}=Q,\eqref{loei}}}\frac{W_{\mathrm{ad}}}{Q_{\mathrm{ad}}}\Label{thermover}.
\end{align}
Then, the macroscopic thermodynamics tells the following proposition.
\begin{proposition}\Label{8-18-0}
The efficiency $\eta_{T}[\beta_{H},\beta_{L},\Omega_{R_{H}},\Omega_{R_{L}},Q]$ satisfies
\begin{align}
&\eta_{T}[\beta_{H},\beta_{L},\Omega_{R_{H}},\Omega_{R_{L}},Q]\nonumber\\
&=1-\frac{E_{R_{L}}(T'^{\mathrm{quasi}}_{R_{L}},\Omega_{R_{L}})-E_{R_{L}}(T_{R_{L}},\Omega_{R_{L}})}{Q}\Label{thermover2},
\end{align}
where $T'^{\mathrm{quasi}}_{R_{L}}$ is defined by the following two equalities;
\begin{align}
&E_{R_{H}}(T_{R_{H}},\Omega_{R_{H}})-E_{R_{H}}(T'^{\mathrm{quasi}}_{R_{H}},\Omega_{R_{H}})=Q,\Label{seigenthermo0}\\
&S_{R_{H}}(T_{R_{H}},\Omega_{R_{H}})+S_{R_{L}}(T_{R_{L}},\Omega_{R_{L}})\nonumber\\
&=S_{R_{H}}(T'^{\mathrm{quasi}}_{R_{H}},\Omega_{R_{H}})+S_{R_{L}}(T'^{\mathrm{quasi}}_{R_{L}},\Omega_{L}).\Label{seigenthermo}
\end{align}
\end{proposition}
This proposition will be shown in Section A of Supplementary.

\subsubsection{Preliminary for the comparison between the prediction given by thermodynamics and the prediction given by statistical mechanics}

One of purposes in this article is to compare the above prediction given by thermodynamics with the prediction given by statistical mechanics.
Because the thermodynamic setup and the statistical mechanical setup are separate things, we cannot compare $\eta^{(n)}_{T}[\beta_{H},\beta_{L},Q]$ with the values in the statistical mechanical formulations as it is.
Therefore, we request that the thermodynamic baths $R_{H}$ and $R_{L}$ satisfy the following assumption, and rewrite $\eta^{(n)}_{T}[\beta_{H},\beta_{L},Q]$ in terms of statistical mechanics;$\\$
\textit{Assumption}: the thermodynamic internal energy and the thermodynamic entropy of $R_{H}$ and $R_{L}$ are the same functions of the temperature as the statistical-mechanical internal energy and von-Neumann entropy of $X$, $Y$, $H$ and $L$. 
Namely, we assume that there exist $\Omega^{(n)}_{R_{H}}$ and $\Omega^{(n)}_{R_{L}}$ such that 
\begin{align}
&E_{R_{H}}(T,\Omega^{(n)}_{R_{H}})=\Tr[\rho_{1/T|\hat{H}^{(n)}_{H}}\hat{H}^{(n)}_{H}]=\left<h^{(n)}_{X}\right>_{P_{1/T|h^{(n)}_{X}}},\nonumber\\
&E_{R_{L}}(T,\Omega^{(n)}_{R_{L}})=\Tr[\rho_{1/T|\hat{H}^{(n)}_{L}}\hat{H}^{(n)}_{L}]=\left<h^{(n)}_{Y}\right>_{P_{1/T|h^{(n)}_{Y}}},\nonumber\\
&S_{R_{H}}(T,\Omega^{(n)}_{R_{H}})=S(\rho_{1/T|\hat{H}^{(n)}_{H}})=S(P_{1/T|h^{(n)}_{X}}),\nonumber\\
&S_{R_{L}}(T,\Omega^{(n)}_{R_{L}})=S(\rho_{1/T|\hat{H}^{(n)}_{L}})=S(P_{1/T|h^{(n)}_{Y}}),
\Label{4-3-1eq}
\end{align}
for arbitrary $T$.

Under the condition \eqref{4-3-1eq}, 
similar to the efficiencies in the statistical mechanical setups, 
we can rewrite the efficiency 
$\eta^{(n)}_{T}[\beta_{H},\beta_{L},Q]:=\eta_{T}[\beta_{H},\beta_{L},\Omega^{(n)}_{R_{H}},\Omega^{(n)}_{R_{L}},Q]$
only by substituting \eqref{4-3-1eq} for \eqref{thermover2};
\begin{align}
\eta^{(n)}_{T}[\beta_{H},\beta_{L},Q]
&=1-\frac{\Tr[\rho_{\beta'_{L}|\hat{H}^{(n)}_{L}}\hat{H}^{(n)}_{L}]-\Tr[\rho_{\beta_{L}|\hat{H}^{(n)}_{L}}\hat{H}^{(n)}_{L}]}{Q}\nonumber\\
&=1-\frac{\left<h^{(n)}_{Y}\right>_{P_{\beta_{L}|h^{(n)}_{Y}}}-\left<h^{(n)}_{Y}\right>_{P_{\beta'_{L}|h^{(n)}_{Y}}}}{Q},\Label{(27)}
\end{align}
where $\beta'_{L}:=1/T'^{\mathrm{quasi}}_{R_{L}}$.

With straight forward algebra, we can express \eqref{(27)} in Carnot efficiency and the relative entropy. For example, in quantum case, \eqref{(27)} is equivalent to
\begin{align}
\eta^{(n)}_{T}[\beta_{H},\beta_{L},Q]=
1-\frac{\beta_{H}}{\beta_{L}}-\frac{D(\rho^{(n)}_{\beta'_{H}\beta'_{L}}\|\rho^{(n)}_{\beta_{H}\beta_{L}})}{\beta_{L}Q},\Label{(29)}
\end{align}
where $\beta'_{H}:=1/T'^{\mathrm{quasi}}_{R_{H}}$. We prove \eqref{(29)} in Section A of Supplementary.

In the thermodynamic set up, the efficiency $\eta^{(n)}_{T}[\beta_{H},\beta_{L},Q]$ can be attained; the macroscopic thermodynamics tells that the efficiency $\eta^{(n)}_{T}[\beta_{H},\beta_{L},Q]$ is attained by a quasi-static adiabatic operation, whose existence is guaranteed by the thermalization law.
After the operation, $R_{H}$ and $R_{L}$ becomes  $(T'^{\mathrm{quasi}}_{R_{H}},\Omega^{(n)}_{R_{H}})$ and $(T'^{\mathrm{quasi}}_{R_{L}},\Omega^{(n)}_{R_{L}})$.
On the other hand, in both of the classical and quantum statistical mechanical setups, 
the thermalization law is not assumed, and therefore we cannot guarantee the existence of the quasi-static operation.
Thus, in general, the efficiency $\eta^{(n)}_{T}$ might be different from $\eta^{(n)}_{C}$ or $\eta^{(n)}_{Q}$.
In the next section, we will derive the asymptotic equalities of $\eta^{(n)}_{C}$ and $\eta^{(n)}_{Q}$ without assuming the existence of the quasi-static operation, and compare them with $\eta^{(n)}_{T}$, which can be regarded as a kind of thermodynamic limit.
(Technically, when we assume the existence of the quasi-static operation
in the classical and quantum statistical mechanical setups, 
$\eta^{(n)}_{C}$ and $\eta^{(n)}_{Q}$ equal the thermodynamic efficiency $\eta^{(n)}_{T}$.)

\section{Results for implicit formulation}\Label{s2}
In the present section, we give four results about the heat engines with finite-size heat baths.
(1) We derive general upper bounds for the efficiencies of the classical and quantum work extractions.
(2) We give a total order structure among the five optimal efficiencies $\eta_{C\lnot}$, $\eta_{C}$, $\eta_{Q\lnot}$, $\eta_{Q}$ and $\eta_{T}$.
(3) We asymptotically expand the above five optimal efficiencies, 
and clarify that the statistical mechanical optimal efficiencies are close to the thermodynamic optimal efficiency in the asymptotic regime.
We also evaluate the effect of quantum coherence on the optimal efficiency, by comparing the asymptotic expansion of $\eta_{C}$ and $\eta_{Q}$.
(4) Finally, we give the concrete form of the quantum optimal work extraction, which turns the disordered energy of the baths to the ordered energy of the external work storage.

\subsection{General upper bounds}\Label{3A}
We firstly give two general upper bounds for the efficiencies of the classical and quantum work extractions;
\begin{theorem}\Label{3-14-0T}
Let $Q$ be an arbitrary positive real number, and
$P_{Z}$ be an arbitrary initial distribution of $Z$.
When a classical dynamics ${\cal T}$ satisfies \eqref{1}, \eqref{2} and $Q^{(n)}(\beta_{H},\beta_{L},\T)=Q$,
and the thermodynamic baths $R_{H}$ and $R_{L}$ satisfy \eqref{4-3-1eq},
the following inequalities hold;
\begin{align}
\eta^{(n)}_{C}(\beta_{H},&\beta_{L},P_{Z},\T)
\le 
1-\frac{\beta_{H}}{\beta_{L}}-\frac{D(P^{(n)'}_{XY}\|P^{(n)}_{\beta_{H}\beta_{L}})}{\beta_{L}Q} \Label{CT0}\\
&\le\eta^{(n)}_{T}[\beta_{H},\beta_{L},Q]-\frac{D(P^{(n)'}_{XY}\|P^{(n)}_{\beta'_{H},\beta'_{L}})}{\beta'_{L}Q},\Label{CT}
\end{align}
where $P^{(n)'}_{XY}:=\sum_{z}P^{(n)'}_{I}(x,y,z)$ is the final state of the classical heat baths, and $\beta'_{H}:=1/T'^{\mathrm{quasi}}_{H}$ is the inverse temperature of $T'^{\mathrm{quasi}}_{H}$ in \eqref{seigenthermo0}. 

Similarly, let $Q$ be an arbitrary positive real number, and
$\rho_{C}$ be an arbitrary initial state $\rho_{C}$ of $C$.
When a quantum dynamics ${\cal W}$ satisfies \eqref{E.C.1},  \eqref{unital} and $Q^{(n)}_{H}(\beta_{H},\beta_{L},{\cal W})=Q$,
and the thermodynamic baths $R_{H}$ and $R_{L}$ satisfy \eqref{4-3-1eq},
the following inequalities hold;
\begin{align}
\eta^{(n)}_{Q}(\beta_{H},\beta_{L},&\rho_{C},{\cal W})
\le1-\frac{\beta_{H}}{\beta_{L}}-\frac{D(\rho^{(n)'}_{HL}\|\rho^{(n)}_{\beta_{H}\beta_{L}})}{\beta_{L}Q} \Label{QT0}\\
&
\le \eta^{(n)}_{T}[\beta_{H},\beta_{L},Q]-\frac{D(\rho^{(n)'}_{HL}\|\rho_{\beta'_{H},\beta'_{L}})}{\beta'_{L}Q},\Label{QT}
\end{align}
where $\rho^{(n)'}_{HL}:=\Tr_{C}[\rho^{(n)'}_{I}]$ is the final state of the quantum heat baths.
\end{theorem}
We prove Theorem 1 in Section B of Supplementary.
Because the case without working body is the special case of the classical case,
$\eta_{C\lnot}(\beta_{H},\beta_{L},\T)$ also satisfies \eqref{CT0} and \eqref{CT}.
Similarly, $\eta_{Q\lnot}(\beta_{H},\beta_{L},{\cal W})$ also satisfies \eqref{QT0} and \eqref{QT}.

Let us see the physical meanings of Theorem \ref{3-14-0T}.
We explain only \eqref{CT0} and \eqref{CT}, because \eqref{QT0} and \eqref{QT} have the same meanings as \eqref{CT0} and \eqref{CT}.
The upper bound \eqref{CT0} shows that the classical efficiency $\eta_{C}(\beta_{H},\beta_{L},P_{Z},\T)$ is lower than Carnot's efficiency by $D(P^{(n)'}_{XY}\|P^{(n)}_{\beta_{H}\beta_{L}})$.
The relative entropy $D(P^{(n)'}_{XY}\|P^{(n)}_{\beta_{H}\beta_{L}})$ expresses the disturbance of the heat baths caused by $\T$, because $P^{(n)}_{\beta_{H}\beta_{L}}$ and $P^{(n)'}_{XY}$ are the initial and final states of the classical heat baths. 
Similarly, the upper bound \eqref{CT} shows that the classical optimal efficiency $\eta_{C}(\beta_{H},\beta_{L},P_{Z},\T)$ is lower than the thermodynamic optimal efficiency by $D(P^{(n)'}_{XY}\|P^{(n)}_{\beta'_{H}\beta'_{L}})$.
The relative entropy $D(P^{(n)'}_{XY}\|P^{(n)}_{\beta'_{H}\beta'_{L}})$ expresses the degree of non-equilibrium of the final states of baths, because the state $P^{(n)}_{\beta'_{H}\beta'_{L}}$ corresponds to the thermodynamic final state $(T'^{\mathrm{quasi}}_{R_{H}},\Omega^{(n)}_{R_{H}})\times(T'^{\mathrm{quasi}}_{R_{L}},\Omega^{(n)}_{R_{L}})$ that are predicted by the macroscopic thermodynamics.
Namely, the further the final states of baths are from the equilibrium states predicted by the macroscopic thermodynamics, the lower the efficiency is.

\subsection{A total order structure among the optimal efficiencies}\Label{3B}
Because the relative entropy is non-negative, Theorem \ref{3-14-0T} implies that the statistical mechanical optimal efficiencies $\eta^{(n)}_{Q}$, $\eta^{(n)}_{C}$, $\eta^{(n)}_{Q\lnot}$ and $\eta^{(n)}_{C\lnot}$ are clearly smaller than or equal to the thermodynamic optimal efficiency $\eta^{(n)}_{T}$.
Moreover, with the following proposition, we give a total order structure among the optimal efficiencies $\eta^{(n)}_{Q}$, $\eta^{(n)}_{C}$, $\eta^{(n)}_{Q\lnot}$, $\eta^{(n)}_{C\lnot}$ and $\eta^{(n)}_{T}$; 
\begin{proposition}\Label{C=Q}
\begin{align}
\eta^{(n)}_{C\lnot}[\beta_{H},\beta_{L},Q]&=\eta^{(n)}_{Q\lnot}[\beta_{H},\beta_{L},Q],\Label{C=Q1}\\
\eta^{(n)}_{C}[\beta_{H},\beta_{L},Q]&\le\eta^{(n)}_{Q}[\beta_{H},\beta_{L},Q]\Label{C=Q2}
\end{align}
\end{proposition}
This proposition will be shown in Section C of Supplementary.

With Proposition \ref{C=Q}, we can easily obtain the total order structure as follows.
The inequality $\eta^{(n)}_{C\lnot}\le\eta^{(n)}_{C}$ follows from the fact that
the case without working body is the special case of the case with working body.
The inequality $\eta^{(n)}_{Q}\le\eta^{(n)}_{T}$ follows from the inequality \eqref{QT} in Theorem \ref{3-14-0T}.
Hence, we obtain the following theorem;
\begin{theorem}\Label{totalorder}
\begin{align}
\eta^{(n)}_{C\lnot}[\beta_{H},\beta_{L},Q]=\eta^{(n)}_{Q\lnot}&[\beta_{H},\beta_{L},Q]\le\eta^{(n)}_{C}[\beta_{H},\beta_{L},C]\nonumber\\
&\le\eta^{(n)}_{Q}[\beta_{H},\beta_{L},Q]\le\eta^{(n)}_{T}[\beta_{H},\beta_{L},Q].\Label{order}
\end{align} 
\end{theorem}

\subsection{Asymptotic expansion of optimal efficiencies}\Label{3C}
In the present subsection, we show that the five optimal efficiencies in \eqref{order} have the same asymptotic expansion, and evaluate how close thermodynamic optimal efficiency is to the statistical mechanical optimal efficiencies.
Unlike in Section 3-A and 3-B, now we request another assumption for the heat baths.
We assume that the baths ${\cal X}$, ${\cal Y}$, $H$ and $L$ are the composed of $n$ uncorrelated identical subsystem, i.e., the following equalities hold;
\begin{eqnarray}
&h^{(n)}_{X}(x)=\sum^{n}_{k=1}h_{1}(x_{k}),\enskip h^{(n)}_{Y}(y)=\sum^{n}_{k=1}h_{1}(y_{k}),\Label{id1}\\
&\hat{H}^{(n)}_{H}=\sum^{n}_{k=1}\hat{H}^{[k]}_{H},\enskip
\hat{H}^{(n)}_{L}=\sum^{n}_{k=1}\hat{H}^{[k]}_{L},\Label{id2}
\end{eqnarray}
where $\hat{H}^{[k]}_{H}=\hat{H}_{1}:=\sum^{d}_{s=1}h_{1}(s)|s\rangle\langle s|$ and $\hat{H}^{[k]}_{L}=\hat{H}_{2}:=\sum^{d}_{s=1}h_{2}(s)|s\rangle\langle s|$ are the Hamiltonians of the $k$th particle of $H$ and $L$, respectively.
We assume this assumption in order to make the von Neumann entropy of the heat baths satisfy additivity, which is satisfied by the thermodynamic entropy.
If there exists correlation between some subsystems, then the von Neumann entropy of the heat baths does not satisfy additivity.

In order to describe the asymptotic behavior of the optimal efficiency,
we introduce the energy variance $\sigma^{2}_{\hat{H}}(\beta)$
and the energy skewness $\gamma_{\hat{H}}(\beta)$ of the Gibbs state $\rho_{\beta|\hat{H}}$; 
\begin{align}
\sigma^{2}_{\hat{H}}(\beta)&:=\Tr[\rho_{\beta|\hat{H}}\hat{H}^{2}_{\hat{H}}]-\Tr[\rho_{\beta|\hat{H}}\hat{H}]^2,\\
\gamma_{\hat{H}}(\beta)&:=\Tr[\rho_{\beta|\hat{H}}(\hat{H}-\Tr[\rho_{\beta|\hat{H}}\hat{H}])^3]/\sigma^{3}_{\hat{H}}(\beta).
\end{align} 

Let us give the asymptotic expansions of the optimal efficiencies.
We firstly expand $\eta^{(n)}_{T}$ in the form of \eqref{(29)}, which is the largest one in the total order structure \eqref{order};
\begin{proposition}\Label{8-18-2}
Let $\{Q_{n}\}^{\infty}_{n=1}$ be arbitrary positive real numbers which satisfy
$\lim_{n\rightarrow\infty}Q_{n}/n=0$.
When the thermodynamic baths $R_{H}$ and $R_{L}$ satisfy \eqref{4-3-1eq},
the thermodynamic optimal efficiency $\eta^{(n)}_{T}$ is asymptotically calculated as
\begin{align}
\eta^{(n)}_{T}[\beta_{H},\beta_{L},Q_{n}]
=1-\frac{\beta_{H}}{\beta_{L}}-\sum^{2}_{k=1}c^{(k)}_{\beta_{H},\beta_{L}}\frac{Q^{k}_{n}}{n^{k}}+O\left(\frac{Q^{3}_{n}}{n^3}\right),\Label{B1}
\end{align}
where $c^{(1)}_{\beta_{H},\beta_{L}}$ 
and $c^{(2)}_{\beta_{H},\beta_{L}}$ are given by
\begin{align}
c^{(1)}_{\beta_H,\beta_L}
:=&\left(\frac{1}{2\beta^2_{H}\sigma^{2}_{\hat{H}_{1}}(\beta_{H})}+\frac{1}{2\beta^2_{L}\sigma^{2}_{\hat{H}_{2}}(\beta_{L})}\right)\frac{\beta^{2}_{H}}{\beta_{L}},\Label{c1} \\
c^{(2)}_{\beta_H,\beta_L}
:=&
\biggl(-\frac{\gamma_{\hat{H}_{1}}(\beta_H)}{6 \beta_H^3 \sigma^{3}_{\hat{H}_{1}}(\beta_H)}
+\frac{\gamma_{\hat{H}_{2}}(\beta_L)}{6 \beta_L^3 \sigma^{3}_{\hat{H}_{2}}(\beta_L)}
+\frac{1}{2 \beta_L^4 \sigma^{4}_{\hat{H}_{2}}(\beta_L)}\nonumber\\
&\quad +\frac{1}{2 \beta_H^2 \beta_L^2 \sigma^{2}_{\hat{H}_{1}}(\beta_H)\sigma^{2}_{\hat{H}_{2}}(\beta_L)}\biggl)\frac{\beta^{3}_{H}}{\beta_{L}}.\Label{c2}
\end{align}
\end{proposition}
This proposition will be shown in Section D of Supplementary.
The inequality \eqref{B1} is an extension of Carnot's inequality.
The first term is the Carnot efficiency, and the second and third terms show how the efficiency decreases when the heat baths are finite-size.
However, Proposition \ref{8-18-2} just gives an asymptotic expansion of thermodynamic efficiency $\eta^{(n)}_{T}$.
It does not tell us whether \eqref{B1} is a good approximation of the true optimal efficiencies  $\eta^{(n)}_{Q}$, $\eta^{(n)}_{C}$, $\eta^{(n)}_{Q\lnot}$ and $\eta^{(n)}_{C\lnot}$ or not.
In order to solve this problem, we give a concrete classical work extraction $\T^{\mathrm{opt}}_{Q_{n}}$ as Protocol \ref{protocol1} and asymptotically expand its efficiency $\eta^{(n)}_{C\lnot}(\beta_{H},\beta_{L},\T^{\mathrm{opt}}_{Q_{n}})$, which is a lower bound for the lefthand-side of the total order structure \eqref{order};

\begin{Protocol}                  
\caption{Classical optimal work extraction}         
\label{protocol1}      
\begin{algorithmic}
\LECTURE  We consider the combination of the following three steps $g_{1n}$, $g_{2n}$, and $g_{1n}^{-1}$, i.e., 
$f_n := g_{1n}^{-1} \circ g_{2n}\circ g_{1n}$.
Hence, the classical optimal work extraction 
${\cal T}^{opt}_{Q_n}$
is given as the following deterministic and invertible dynamics;
\begin{align}
\T^{\mathrm{opt}}_{Q_n}(x,y|x',y'):=\delta_{(x,y),f_{n}(x',y')}.\Label{Topt} 
\end{align}             
\STEPONE  First, we convert the elements $x$ and $y$ in ${\cal X}$ and ${\cal Y}$, which are described as the pair of integers $(i,j)$ in the set $\mathbb{Z}_{d^n}^2$,
by using the following two functions $g_{1nX}$ and $g_{1nY}$ 
from $\mathbb{Z}_{d}^n$ to $\mathbb{Z}_{d^n}$
as 
\begin{align}
P^{n}_{\beta_{H}|h^{(n)}_{X}}(x)
&=P^{n\downarrow}_{\beta_{H}|h^{(n)}_{X}}(g_{1nX}(x)),
\nonumber\\
P^{n}_{\beta_{L}|h^{(n)}_{Y}}(y)
&=P^{n\downarrow}_{\beta_{L}|h^{(n)}_{Y}}(g_{1nY}(y))
\Label{g1ndef},
\end{align}
where $P^{n\downarrow}_{\beta_{H}|h^{(n)}_{X}}$ and $P^{n\downarrow}_{\beta_{L}|h^{(n)}_{Y}}$ are the descending reordered distributions of the distributions
$P^{n}_{\beta_{H}|h^{(n)}_{X}}$ and $P^{n}_{\beta_{L}|h^{(n)}_{Y}}$,
respectively. 
That is, we apply the function $g_{1n}(x,y):=(g_{1nX}(x),g_{1nY}(y))$
to the classical system ${\cal X}\times{\cal Y}$.
\STEPTWO Next, we apply the function $g_{2n}$ defined as
\begin{align}
g_{2n}:&
(i_A d^{m_n} + j_B ,i_B d^{n-m_n}+ j_A ) \nonumber\\
&\mapsto 
(i_A +i_B d^{m_n},j_A +j_B d^{n-m_n}) 
\Label{g2ndef}
\end{align}
for $i_A , j_B\in \mathbb{Z}_{d^{m_n}} $ and $i_B , j_A\in \mathbb{Z}_{d^{n-m_n}} $,
where
\begin{align}
m_{n}&:=\left\lfloor 
\frac{\beta_{H} Q_{n}
+ \frac{Q_{n}^2}{2 n \sigma_X^2(\beta_{H})}}{\log d}
\right\rfloor.
\end{align}
\STEPTHREE Finally, we apply the inverse function $g_{1n}^{-1}$.

The physical meaning of the classical optimal work extraction is the same as that of the quantum optimal work extraction.
We will explain them at once, in the section \ref{3B1}.
\end{algorithmic}
\end{Protocol}

\begin{proposition}\Label{3-14-2T}
Assume that the baths ${\cal X}$ and ${\cal Y}$ are composed of $n$ uncorrelated identical particles, i.e., the equality \eqref{id1} holds. 
Then, the classical work extraction $\T^{\mathrm{opt}}_{Q_n}$ given in Protocol \ref{protocol1} satisfies
\begin{align} 
\eta^{(n)}_{C\lnot}(\beta_{H},\beta_{L},\T^{\mathrm{opt}}_{Q_n})
=
1-\frac{\beta_{H}}{\beta_{L}}
-
c^{(1)}_{\beta_{H},\beta_{L}} \frac{Q_{n}}{n}
+O(\frac{1}{n})
+O(\frac{Q_{n}^2}{n^2})\Label{arxsecondorder2x} ,
\end{align}
and
\begin{align} 
Q^{(n)}_{H\lnot}(\beta_{H},\beta_{L},\T^{\mathrm{opt}}_{Q_n})
=Q_{n}+ o(\frac{Q^{2}_{n}}{n^3}).\Label{arxsecondorder2.2x}
\end{align}

Further,
when $h_X$ and $h_Y$ are non-lattice,
i.e., when there does not exist a positive number $t$
such that $\{h(z)-h(z')\}_{z,z'} \subset t\mathbb{Z}$ for $h^{(n)}_{X}(x)$ nor $h^{(n)}_{Y}(y)$, 
the classical work extraction $\T^{\mathrm{opt}}_{Q_n}$ has more detailed asymptotic expansions as
\begin{align} 
\eta^{(n)}_{C\lnot}(\beta_{H},\beta_{L},\T^{\mathrm{opt}}_{Q_n})
=&
1-\frac{\beta_{H}}{\beta_{L}}
-\sum^{2}_{k=1}
c^{(k)}_{\beta_{H},\beta_{L}} \frac{Q^{k}_{n}}{n^{k}}
-
d^{(1)}_{\beta_{H},\beta_{L}} \frac{Q_{n}}{n^2}
\nonumber\\
&+O\left(\frac{Q^2_{n}}{n^{5/2}}\right)
+O\left(\frac{Q^{3}_{n}}{n^{3}}\right)\Label{arxsecondorder2} ,
\end{align}
and
\begin{align} 
Q^{(n)}_{H\lnot}(\beta_{H},\beta_{L},\T^{\mathrm{opt}}_{Q_n})
=Q_{n}+ O(\frac{Q^{3}_{n}}{n^4}),\Label{arxsecondorder2.2}
\end{align}
where  
\begin{align}
d^{(1)}_{\beta_{H},\beta_{L}}
:=
&
\Biggl((\frac{\gamma_1(\beta_{H})}{2 \beta_{H} \sigma(\beta_{H})}
+\frac{1}{\beta_{H}^2 \sigma^2(\beta_{H})})^2
\nonumber \\
\quad &+
(\frac{\gamma_1(\beta_{L})}{2 \beta_{L} \sigma(\beta_{L})}
+\frac{1}{\beta_{L}^2 \sigma^2(\beta_{L})})^2
\Biggr)\frac{\beta^2_{X}}{\beta_{L}}.\Label{dbeta}
\end{align}
\end{proposition}
Proposition \ref{3-14-2T} will be shown in Sections E and F of Supplementary.

As in \eqref{arxsecondorder2.2}, the endothermic amount $Q^{(n)}_{H\lnot}(\beta_{H},\beta_{L},\T^{\mathrm{opt}}_{Q_n})$ is not strictly equal to $Q_{n}$.
However, we can easily take $\tilde{\T}^{\mathrm{opt}}_{Q_n}$ which satisfies \eqref{arxsecondorder2x}, \eqref{arxsecondorder2} and $Q^{(n)}_{H\lnot}(\beta_{H},\beta_{L},\tilde{\T}^{\mathrm{opt}}_{n})=Q_{n}$, by taking the probabilistic mixture of $\T^{\mathrm{opt}}_{Q'_n}$ and $\T^{\mathrm{opt}}_{Q''_n}$ that satisfy $Q^{(n)}_{H\lnot}(\beta_{H},\beta_{L},\T^{\mathrm{opt}}_{Q'_n})<Q_{n}<Q^{(n)}_{H\lnot}(\beta_{H},\beta_{L},\T^{\mathrm{opt}}_{Q''_n})$.
Therefore, we obtain 
\begin{align}
\eta^{(n)}_{C\lnot}[
\beta_{H},\beta_{L},Q_n]
\ge 
1-\frac{\beta_{H}}{\beta_{L}}
-c^{(1)}_{\beta_{H},\beta_{L}}\frac{Q_{n}}{n}
+O(\frac{1}{n})+O(\frac{Q^{2}_{n}}{n^2})\Label{8-17-1}
\end{align}
for the general case,
and
\begin{align}
\eta^{(n)}_{C\lnot}[
\beta_{H},\beta_{L},Q_n]
\ge
1-\frac{\beta_{H}}{\beta_{L}}
-\sum^{2}_{k=1}c^{(k)}_{\beta_{H},\beta_{L}}\frac{Q^{k}_{n}}{n^{k}}\nonumber\\
+O(\frac{Q_{n}}{n^2})+O(\frac{Q^{3}_{n}}{n^3})
\Label{8-17-2}
\end{align}
for the non-lattice case.
Therefore, by combining Theorem \ref{totalorder}, Propositions \ref{8-18-2} and the inequalities \eqref{8-17-1} and \eqref{8-17-2}, we obtain the following theorem, which tells us how accurate the prediction of the macroscopic thermodynamics is;
\begin{theorem}\Label{3-14-3T} 
Assume that the baths $H$, $L$ ${\cal X}$ and ${\cal Y}$ are composed of $n$ uncorrelated identical particles, i.e., the Hamiltonians satisfy \eqref{id1} and \eqref{id2}.
Then, for arbitrary positive real numbers $\{Q_{n}\}^{\infty}_{n=1}$ satisfying
$\lim_{n\rightarrow\infty}Q_{n}/n=0$, the following equality holds:
\begin{align}
&\eta^{(n)}_{Q\lnot}[\beta_{H},\beta_{L},Q_n]
=
\eta^{(n)}_{C\lnot}[
\beta_{H},\beta_{L},Q_n]\nonumber\\
&=\eta^{(n)}_{C}[\beta_{H},\beta_{L},Q_n]+o(\frac{Q_{n}}{n})
\nonumber\\
&=
\eta^{(n)}_{Q}[\beta_{H},\beta_{L},Q_n]
+o(\frac{Q_{n}}{n})\nonumber\\
&=\eta^{(n)}_{T}[\beta_{H},\beta_{L},Q_n]+o(\frac{Q_{n}}{n}). \Label{8-1x}
\end{align}

Further, when the baths' Hamiltonians are non-lattice,
the following equality holds:
\begin{align}
&\eta^{(n)}_{Q\lnot}[\beta_{H},\beta_{L},Q_n]=
\eta^{(n)}_{C\lnot}[
\beta_{H},\beta_{L},Q_n]\nonumber\\
&=\eta^{(n)}_{C}[
\beta_{H},\beta_{L},Q_n]+O(\frac{Q_{n}}{n^2})+O(\frac{Q^{3}_{n}}{n^3})\nonumber\\
&=\eta^{(n)}_{Q}[\beta_{H},\beta_{L},Q_n]
+O(\frac{Q_{n}}{n^2})+O(\frac{Q^{3}_{n}}{n^3}) \nonumber\\
&=\eta^{(n)}_{T}[\beta_{H},\beta_{L},Q_n]
+O(\frac{Q_{n}}{n^2})+O(\frac{Q^{3}_{n}}{n^3}). \Label{8-1}
\end{align}
\end{theorem}


\subsection{Significance of asymptotic expansion}
Now, we can accomplish the main purposes of this article. 
Firstly, we evaluate how close the optimal efficiencies given by statistical mechanics are to the optimal efficiency given by thermodynamics.
Due to Theorem \ref{3-14-3T} ,  in the non-lattice case,  the optimal efficiency predicted by thermodynamics is the same as the optimal efficiency given by statistical mechanics up to the order $o(Q^{2}_{n}/n^2)$, because of $O(Q^{2}_{n}/n^2)<O(Q_{n}/n^2)+O(Q^{3}_{n})/$.
In general case,  the optimal efficiency predicted by thermodynamics is the same as the optimal efficiency given by statistical mechanics up to the order $o(Q_{n}/n)$.
Therefore, in general, the differences between the thermodynamic prediction and the statistical mechanical predictions are at most sublinear order in terms of $Q_{n}/n$.

Second, we evaluate the effect of the quantum coherence on the optimal efficiency.
For this purpose, we compare the optimal efficiencies of the classical and quantum cases.
In this comparison, the classical optimal efficiencies $\eta_{C}$ and $\eta_{C\lnot}$ 
express the case when our operations are 
restricted to the diagonal operations with respect to the eigenbasis of the Gibbs state.
This is because such operations can be written as probability transition matrices.
The quantum optimal efficiencies $\eta_{Q}$ and $\eta_{Q\lnot}$ 
express the case when we are allowed to use all quantum operations on the system, which employ 
the effect of an arbitrary coherence. 
In addition to such an coherent effect, we can discuss the effect of use of catalyst
because the optimal efficiencies $\eta_{C}$ and $\eta_{Q}$ describes the optimal efficiency of the quantum heat engines with the quantum catalyst $C$ whose initial state is arbitrary,
and the optimal efficiencies $\eta_{C\lnot}$ and $\eta_{Q\lnot}$ express the case without catalyst.
In this comparison, the optimal efficiency $\eta_{C\lnot}$ is the minimum and the optimal efficiency $\eta_{Q}$ is the maximum.
Theorem \ref{3-14-3T} show how different $\eta_{C}$ is from $\eta_{Q\lnot}$. 
In the non-lattice case,  the effect of the quantum coherence is up to the order $o(Q^{2}_{n}/n^2)$.
In general case,  the effect of the quantum coherence is up to the order $o(Q_{n}/n)$.
Therefore, in general,  the effect of the quantum coherence is at most sublinear order in terms of $Q_{n}/n$.

\begin{figure}
\includegraphics[scale=0.65,clip]{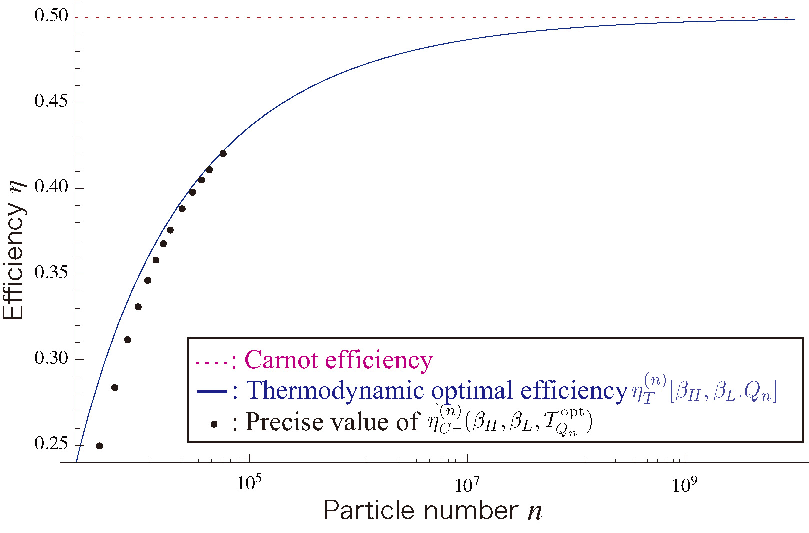}
\caption{We plot Carnot's bound, the thermodynamic optimal efficiency $\eta^{(n)}_{T}$, and the results of numerical calculation for $\eta^{(n)}_{C\lnot}(\beta_{H},\beta_{L},\T^{\mathrm{opt}}_{Q_{n}})$, as the red-dotted-curved line, the blue-curved line and the black dots, respectively.}\Label{gra1}
\end{figure}
\begin{figure}
\includegraphics[scale=0.65,clip]{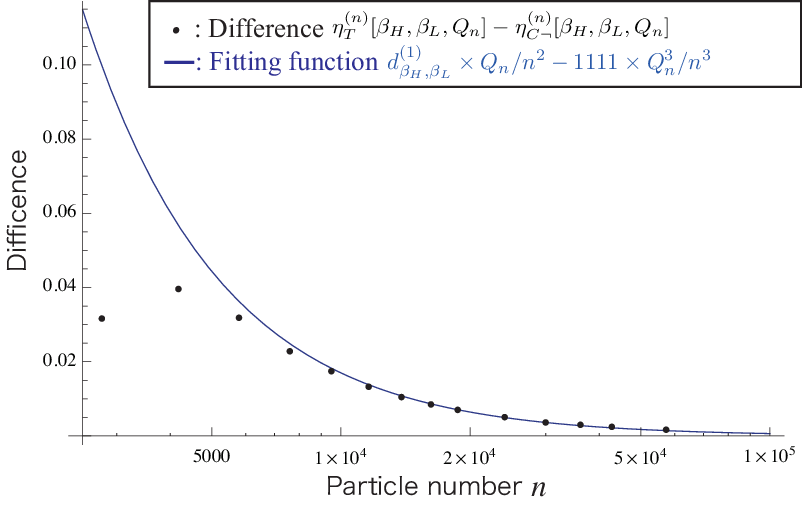}
\caption{We plot the difference $\eta^{(n)}_{T}-\eta^{(n)}_{C\lnot}(\beta_{H},\beta_{L},\T^{\mathrm{opt}}_{Q_{n}})$ as the black dots.
We also plot the function $d^{(1)}_{\beta_{H},\beta_{L}}Q_{n}/n^2-1111Q_{n}/n^{2}$ as the blue line.
The parameter $d^{(1)}_{\beta_{H},\beta_{L}}$ is defined in \eqref{dbeta}, and $1111$ is the fitting parameter.
In the present condition, the parameter $d^{(1)}_{\beta_{H},\beta_{L}}\approx14343$.}\Label{gra2}
\end{figure}

Our results also enable to evaluate quantitatively how the optimal efficiencies get closer to each others.
In order to illustrate this, let us see Figures \ref{gra1} and \ref{gra2}.
In these figures, we treat the $n$-qubit heat baths whose temperatures are $30K$ and $15K$.
We also assume that the energy levels of each qubit of heat baths are $+k_{B}J$ and $-k_{B}J$, and that the extracted heat satisfies $Q_{H,n}=0.3n^{2/3}k_{B}J$. 
In Figure \ref{gra1}, the blue curved line, which expresses the thermodynamic optimal efficiency $\eta^{(n)}_{T}$, is  an upper bound for the four statistical mechanical optimal efficiencies, $\eta^{(n)}_{C}$, $\eta^{(n)}_{C\lnot}$, $\eta^{(n)}_{Q}$ and $\eta^{(n)}_{Q\lnot}$.
The black dots, which are the results of the numerical calculation for $\eta^{(n)}_{C\lnot}(\beta_{H},\beta_{L},\T^{\mathrm{opt}}_{Q_{n}})$, show a lower bound for the four statistical mechanical optimal efficiencies.
Figure \ref{gra1} shows that the thermodynamic optimal efficiency and the four statistical mechanical optimal efficiencies are very close to each others when $n$ is more than $10^4$.
Figure \ref{gra1} also shows that all of the five optimal efficiencies convert to Carnot's bound at the macroscopic limit $n\rightarrow\infty$.
How fast does the difference between the upper bound and the lower bound decrease?
Figure \ref{gra2} answers this question.
In Figure \ref{gra2}, we plot the difference between $\eta^{(n)}_{T}$ and the results of numerical calculation for $\eta^{(n)}_{C\lnot}(\beta_{H},\beta_{L},\T^{\mathrm{opt}}_{Q_{n}})$ as the black dots.
We also plot the function $d^{(1)}_{\beta_{H},\beta_{L}}Q_{n}/n^2-1100Q^{3}_{n}/n^{3}$ as the blue line, where $d^{(1)}_{\beta_{H},\beta_{L}}$ is defined in \eqref{dbeta}, and where $1111$ is the fitting parameter.
Thus, Figure \ref{gra2} shows that the speed of decreasing of the difference between the upper bound and the lower bound is approximately in the form of $O(Q_{n}/n^2)+O(Q^{3}_{n}/n^3)$.

\section{Explicit treatment of external work storage}\Label{s4}
Until now, we have not explicitly discussed the external work storage.
This treatment is caused by our parallel discussion of the statistical mechanical formulations with the thermodynamic formulation because thermodynamics does not describe the external work storage explicitly so that 
the work extracted by an adiabatic operation is defined as the difference of the internal energy of the composite system $R_{H}R_{L}W$ according to the first law of thermodynamics.
However, since the existence of such an adiabatic operation is equivalent to the axiom of thermodynamics,
there is no guarantee for the existence of such an adiabatic operation in the law of quantum theory.
In order to discuss the ultimate performance of a quantum mechanical heat engine,
we need to describe our operations by a unitary on the whole system including the external work storage \cite{Horodecki, oneshot1, oneshot3, Egloff, Brandao, Car2, Popescu2014,Popescu2015, oneshot2}, which has been studied hard recently.
This treatment is called the explicit treatment of the external work storage 
in contrast with the implicit treatment.

In the present section, we tackle this problem.
In the subsection \ref{s4-1}, we firstly review the relationship between our implicit treatment and the explicit treatment.
We show that we can convert the explicit formulation and the CP-instrument model
to each others.
Then, in the subsection \ref{3B1}, 
we give a concrete protocol that attains the optimal efficiency, in the explicit formulation.
Finally, in the subsection \ref{s4-4}, 
we evaluate the quality of extracted energy in our protocol
by discussing the energy and entropy ratio in the work storage.

\subsection{Formulation of explicit treatment of external work storage}\Label{s4-1}
In order to treat external work storage explicitly, 
we need to describe the whole dynamics as a unitary on the composite system of  
the internal system ${\cal H}_I$ and the external work storage system ${\cal H}_{E}$
We refer to the composite system $IE$ as the total system.
As the energy conservation law, we impose the following condition to the unitary $U_{IE}$ describing 
the dynamics of the total system $IE$ as
\begin{align}
[U_{IE},\hat{H}_{I}+\hat{H}_{E}]=0\Label{(A')}.
\end{align}
where  $\hat{H}_{E}$ is the Hamiltonian of $E$.
In order to prevent the work storage $E$ playing a role of ``memory,'' we also request that the dynamics of $I$ is independent of the initial state of $E$, i.e.,
\begin{align}
{\cal W}^{\cal F}(\rho_{I}):=\Tr_{E} [U_{IE} (\rho_I \otimes \rho_{E} )U^{\dagger}_{IE}].\Label{memoryless}
\end{align}
is independent of $\rho_{E}$ 
whenever the initial state of $E$ is an eigenstate of $\hat{H}_E$.

So, under the above conditions,
the tuple ${\cal F}=({\cal H}_E,\hat{H}_E,U_{IE},\rho_E)$ gives an explicit treatment of quantum heat engine,
which is often called a {\it full quantum (FQ) work extraction}.
Then, we can show that ${\cal W}^{\cal F}(\rho_{I})$ satisfies \eqref{E.C.1} and \eqref{unital}
when the difference between the expectations of the Hamiltonian on the initial and final states on the the external work storage $E$ is regarded as the amount of extracted work \cite[Theorem 12]{1ponme}.
That is, an FQ work extraction ${\cal F}$ gives an average work extraction in the sense of Subsection \ref{s1AB} via \eqref{memoryless}.

We also assume the following three conditions to the work storage $E$;$\\$ 
(A): The Hamiltonian $\hat{H}_{E}:=\sum_{e\in \Lambda} e|e\rangle_{E}~_{E}\langle e|$ is non-degenerate, and $\Lambda$ is countable.$\\$
(B): For arbitrary $e\in \Lambda$, $(x,y)\in {\cal X}\times{\cal Y}$ and $(x',y')\in {\cal X}\times{\cal Y}$, there exist $e'\in \Lambda$ such that $e'-e=h_{XY}(x,y)-h_{XY}(x',y')$.$\\$ 
(C): The initial state of the work storage $E$ is a pure energy eigenstate. We refer to the initial state as $|e_{0}\rangle_{E}~_{E}\langle e_{0}|$.$\\$
It is possible to take such an external system $E$ \cite[Section 4]{1ponme}.

As shown in \cite[Lemma 10]{1ponme},
when an FQ work extraction ${\cal F}$ satisfies the above three conditions,
we obtain a CP-instrument work extraction $\mathscr{W}^{\cal F}:=\{{\cal E}^{\cal F}_{e},w_{e}\}$ as
\begin{align}
{\cal E}^{\cal F}_{j}(\rho_{I})&:=\Tr_{E} [|e\rangle_{E}~_{E}\langle e|U_{IE} (\rho_I \otimes \rho_{E} )U^{\dagger}_{IE}|e\rangle_{E}~_{E}\langle e|]\\
w_{j}&:=e-e_{0}.
\end{align}
Conversely, as shown in \cite[Lemma 11]{1ponme},
when a CP-instrument work extraction $\mathscr{W}$ satisfies the condition \eqref{CP2},
it can be realized by a combination of a suitable FQ work extraction satisfying \eqref{(A')}, a suitable initial state on the external work storage $E$,
and a suitable measurement on the external work storage $E$.
That is, the efficiency realized by a CP-instrument work extraction can be attained by a FQ work extraction.
In other word, any CP-instrument work extraction satisfying \eqref{CP2} can be regarded as a time evolution of a closed system including the external work storage $E$.
Hence, it is natural to optimize our efficiency among CP-instrument work extractions satisfying \eqref{CP2}.

\subsection{Quantum optimal work extraction}\Label{3B1}
Next, we give the optimal work extraction for the quantum case.
That is, we give an FQ work extraction ${\cal F}^{\mathrm{opt}}_{Q_n}$ whose CP-instrument work extraction $\mathscr{W}^{\mathrm{opt}}_{Q_n}$ attains \eqref{arxsecondorder2x} and \eqref{arxsecondorder2.2x} in general case, and attains \eqref{arxsecondorder2} and  \eqref{arxsecondorder2.2x} in the non-lattice case.
We show the relations \eqref{arxsecondorder2x}, \eqref{arxsecondorder2.2x}, \eqref{arxsecondorder2} and  \eqref{arxsecondorder2.2}
for  $\mathscr{W}^{\mathrm{opt}}_{Q_{n}}$ 
by showing the relations
\begin{align} 
\eta^{(n)}_{C\lnot}(\beta_{H},\beta_{L},\T^{\mathrm{opt}}_{Q_n})
=
\eta^{(n)}_{Q\lnot}(\beta_{H},\beta_{L},\mathscr{W}^{\mathrm{opt}}_{Q_n})\Label{W=T}
\end{align}
and
\begin{align} 
Q^{(n)}_{H\lnot}(\beta_{H},\beta_{L},\T^{\mathrm{opt}}_{Q_n})
=
Q^{(n)}_{H\lnot}(\beta_{H},\beta_{L},\mathscr{W}^{\mathrm{opt}}_{Q_n})
\Label{W=T2}.
\end{align}
For this purpose, we give an FQ work extraction ${\cal F}^{\mathrm{opt}}_{Q_n}$ 
by using a canonical conversion from a deterministic classical work extraction to a quantum work extraction that is introduced in Ref.\cite[Section 4]{1ponme}.

For the invertible function $f_n$ on $({\cal X} \times {\cal Y})^n$ given as the classical optimal work extraction,
we introduce an external system ${\cal H}_E$ with $\hat{H}_E$ satisfying  the condition (A)--(C) in Section \ref{s4-1},
and an energy preserving unitary $U_{f_n}$ on $IE$ as follows;
\begin{align}
U_{f_n}:=\sum_{x,y}\sum_{e\in \Lambda}|&f_{n}(x,y)\rangle_{HL}
|e'(x,y)\rangle_{E}~_{HL}\langle x,y|~ _{E}\langle e|,\Label{Ufn}
\end{align}
where $|e'(x,y)\rangle_{E}:=|e+h^{(n)}_{XY}(x,y)-h^{(n)}_{XY}(f_{n}(x,y))\rangle_{E}$.
The existence of such an external system ${\cal H}_E$ with $\hat{H}_E$ is shown in \cite[]{1ponme}.
Clearly, the unitary $U_{f_n}$ satisfies \eqref{(A')}, which implies that
the tuple ${\cal F}^{\mathrm{opt}}_{Q_n}=({\cal H}_E,\hat{H}_E,U_{f_n},\rho_E)$ is an FQ work extraction.
Then, from the FQ work extraction ${\cal F}^{\mathrm{opt}}_{Q_n}$,
we obtain the CP-instrument work extraction $\mathscr{W}^{\mathrm{opt}}_{Q_n}=\{{\cal E}_{e},w_{e}\}_{e\in \Lambda}$ as follows;
\begin{align}
{\cal E}_{e}(\rho_{HL})&:=~ _{E}\langle e| U_{f_{n}} (\rho_{HL}\otimes|e_{0}\rangle~ _{E}\langle e_{0}|) U_{f_{n}}^{\dagger}|e\rangle_{E},\\
w_{e}&=e-e_{0}.
\end{align}

To consider the physical meaning of our FQ work extraction ${\cal F}^{\mathrm{opt}}_{Q_n}$, 
we divide the unitary $U_{f_{n}}$ into three steps, which correspond to 
the functions $g_{1n}$, $g_{2n}$, and $g_{1n}^{-1}$ in Protocol \ref{protocol1}.
For this purpose, we need to describe the resultant systems of the unitary corresponding to $g_{1n}$ as $n$-fold tensor product systems of the qudit system.
This is because the ranges of the functions $g_{1nX}$ and $g_{1nY}$
are $\mathbb{Z}_{d^n}$ not the $n$-dimensional vector space of $\mathbb{Z}_d$.
To resolve this problem, 
we make the following correspondence between
the set $\mathbb{Z}_{d^n}$ and the $n$-dimensional vector space of $\mathbb{Z}_d$.
For a vector $x:=(x_1, \ldots,x_n) \in \mathbb{Z}_{d}^n$,
we define the integer $g_{3n}({x}) \in \mathbb{Z}_{d^n}$ by
we make a correspondence
\begin{align}
g_{3n}(x)= \sum_{k=1}^n x_k d^{k-1}. \Label{4-19-1eq}
\end{align}
Hence, we can decompose $U_{f_{n}}$ into the following three steps;
\begin{description}
\item[Step 1 \protect{($U_{g_{3n}^{-1} \circ g_{1n}}$)}:]
We perform $U_{g_{3n}^{-1} \circ g_{1n}}$, which is given by substituting $g_{3n}^{-1} \circ g_{1n}$ for $f$ in \eqref{Ufn}. 
By the unitary $U_{g_{3n}^{-1} \circ g_{1n}}$, 
we reorder the diagonal elements of $\rho^{\otimes n}_{\beta_{H}|\hat{H}_{H}}$ and $\rho^{\otimes n}_{\beta_{L}|\hat{H}_{L}}$ 
so that the eigenvalues are arrayed in the descending order.
Then, we set the resultant state in the $n$-fold qudit systems with the correspondence \eqref{4-19-1eq}.

\item[Step 2 \protect{(
$U_{g_{3n}^{-1} \circ g_{2n} \circ g_{3n} }$)}:]
Second, we perform $U_{g_{3n}^{-1} \circ g_{2n} \circ g_{3n} }$, 
which is given by substituting ${g_{3n}^{-1} \circ g_{2n} \circ g_{3n} }
$ for $f$ in \eqref{Ufn}. 
By the unitary $U_{g_{3n}^{-1} \circ g_{2n} \circ g_{3n} }$, 
we reorder the particles as the red arrows in Fig.\ref{optope}.

\item[Step 3 \protect{($U_{g_{3n}^{-1}\circ g_{1n}}^{\dagger}$)}:]
Finally, we perform the inverse process of the first step, i.e., we perform $U_{g_{3n}^{-1}\circ g_{1n}}^{\dagger}$.
\end{description}

\begin{figure}
\includegraphics[scale=0.99,clip]{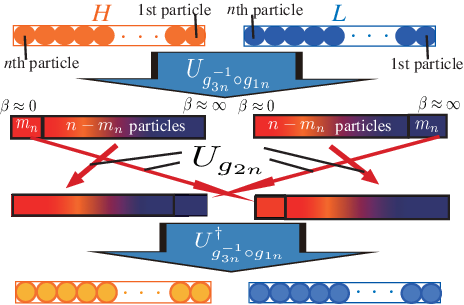}
\caption{structure of $U_{f_{n}}$. 
We omit $E$ in this figure just for simplicity.}\Label{optope}
\end{figure}

The physical meanings of the above steps are given as follows.
To discuss the first step, we notice that $P^{n\downarrow}_{\beta_{H}|h^{(n)}_{X}}(i)\approx0$ when $i$ is large,
and that $\sum^{d^{n-m}}_{i=1}P^{n\downarrow}_{\beta_{H}|h^{(n)}_{X}}(i)\approx d^{-m}$ when $m$ is small.
Under the correspondence \eqref{4-19-1eq}, 
this fact can be interpreted in the following way.
In the hot bath $H$, 
when the particle of our interest is close to the $n$-th particle,
the state is almost the ground state, which is the Gibbs state of the zero temperature, i.e., 
the particle is extremely cold.
Similarly, 
when the particle of our interest is close to the first particle,
the state is almost $\hat{1}/d$, which is the Gibbs state of the infinite temperature, i.e., the particle is extremely hot. 
Therefore, 
by the operation in the first step, 
the energy is compressed into particles with earlier numbers.
The operation in the first step does not change the entropy in the both baths $H$ and $L$
because it just sorts the eigenstates of $H$ and $L$, respectively.
Hence, 
the resultant states have higher energy than the initial states in both baths
because the initial states in both baths have the minimum energy under the same entropy.
So, to realize the first step, we need to insert additional energy from the external system to both baths.

The meaning of the second step is easy.
It just swaps the extremely hot particles in the hot bath and the extremely cold particles in the cold bath.
When both baths have the same Hamiltonian, we need no additional energy in the second step.
Finally, the third step diffuses the energy in both baths
by the inverse operation of the first step.
Then, the resultant states in both baths $H$ and $L$ become very close to the Gibbs states whose temperatures are $\beta'_{H}$ and $\beta'_{L}$.
In the third step, the external system recovers a larger amount of energy 
than the amount of energy inserted in the first step.
The amount of extracted work is the difference between the recovered and inserted energies.

We see intuitively why the operation is optimal in the physical viewpoint as follows.
In order to maximize the efficiency of the heat engine, 
we have to transfer the entropy from the hot bath to the cold bath as efficiently as possible.
Our operation exchanges the maximum entropy state and the zero entropy state between $H$ and $L$.

Now, let us consider how to realize the unitary $U_{f_{n}}$.
Thanks to Lemma 3 in Ref. \cite{1ponme},
any unitary $U$ satisfying \eqref{(A')} can be realized with the sufficiently long time $t$ 
by adding the small interaction Hamiltonian term $B$, which does not change in $0<t<t_0$.
Thus, in order to realize the unitary $U_{f}$, we only have to turn on the interaction $B$ at $t=0$ and to turn off it at $t=t_0$.
From $t=0$ to $t=t_0$, we do not have to control the total system $HLE$ time dependently.
Namely, we can realize a ``clockwork heat engine," which is programmed to perform the unitary transformation $U_{f_{n}}$ automatically.
Similarly, we can also realize respective steps of $U_{f_{n}}$, that are $U_{g_{3n}^{-1} \circ g_{1n}}$,
$U_{g_{3n}^{-1} \circ g_{2n} \circ g_{3n} }$,
and 
$U_{g_{3n}^{-1} \circ g_{1n}}^\dagger
=U_{g_{1n}^{-1}\circ g_{3n}}$, as clockwork processes.

\subsection{Conversion from ``heat'' to ``work'' in the optimal protocol}\Label{s4-4}
The three requirements (A)--(C) in Section \ref{s4-1} for an FQ work extraction ${\cal F}$ are not enough for quantum heat engine
because we need to care the quality of the energy gain in the external work storage.
To assess the quality, we evaluate the order of the ratios of the entropy difference to the energy difference in the external storage $E$.
So, we define the entropy-energy ratio 
$A({\cal F})$ as 
\begin{align}
A({\cal F})&:=\frac{S(\sigma_{E})-S(|e_{0}\rangle_{E}~ _{E}\langle e_{0}|)}{(\sigma_{E}-|e_{0}\rangle_{E}~ _{E}\langle e_{0}|)\hat{H}_{E}}\Label{AEX}
\end{align}
where $\sigma_{E}$ is the resultant state of the unitary $U_{IE}$,
which is defined as
$\sigma_{E}:=\Tr_{HL}[U_{IE}(\rho_{HL}\otimes|e_{0}\rangle~ _{E}\langle e_{0}|) U_{IE}^{\dagger}]$.
Therefore, under the three requirements (A)--(C), 
the order of the ratio $A({\cal F})$ gives a novel criterion 
to decide whether the energy gain of the work storage can be regarded as work or not.

Now, we check this condition for our FQ work extraction ${\cal F}^{\mathrm{opt}}_{Q_n}$.
In order to evaluate the order of the entropy-energy ratios, 
we estimate the difference of the entropy of $E$.
Because the unitary $U_{f_{n}}$ satisfies $[U_{f_{n}},\hat{H}_{HL}+\hat{H}_{E}]$,
and because the initial state of $E$ is energy pure eigenstate, the following inequality holds;\cite[Lemma 6]{1ponme}
\begin{align}
S(\sigma_{E})\le \log N,
\end{align}
where $N$ is the number of eigenvalues of the Hamiltonian $\hat{H}_{HL}$.
The set of eigenvalues of the Hamiltonian $\hat{H}_{HL}$ is written as $\{ h_X^{(n)} (x)+h_Y^{(n)}(y) \}_{(x,y)\in ({\cal X}\times {\cal Y})^n}$, and the number of its elements is  
 less than $ ((n+1)^{(d-1)})^2=(n+1)^{2(d-1)}$.
Therefore, the entropy gain of $E$ is bounded as follows;
\begin{align}
S(\sigma_{E})-S(|e_{0}\rangle_{E}~ _{E}\langle e_{0}|)\le 4(d-1)\log n.\Label{entEx}
\end{align}
Therefore, we obtain the following theorem.
\begin{theorem}
The entropy-energy ratios $A({\cal F}^{\mathrm{opt}}_{Q_n})$ satisfies 
\begin{align}
A({\cal F}^{\mathrm{opt}}_{Q_n})\le O\left(\frac{\log n}{Q_{n}}\right).
\end{align}
\end{theorem}
Thus, when $Q_{n}=an^{b}$ holds with the real numbers $0<a$ and $0<b<1$, 
the criterion is sufficiently small, i.e., the energy gained in $E$ has high quality. 
That is, the entropy-energy ratio of the external work storage goes to 0 in the macroscopic limit.
So, we can conclude that
the extracted energy in $E$ can be regarded as work.
That is, in this meaning, we can interpret the optimal work extraction as the entropy filter;
it filters out the entropy from the energy flow into $E$.

To see the total entropy flow in the detail, 
let us evaluate the entropy differences of $H$ and $L$ during $U_{f_{n}}$.
In the unitary  $U_{f_{n}}$, the first step $U_{g^{-1}_{3n}\circ U_{g_{1n}}}$ and the third step $U^{\dagger}_{g^{-1}_{3n}\circ U_{g_{1n}}}$ just reorder the diagonal basis of $H$ and $L$, respectively.
Therefore, these two steps do not change the entropies of $H$ and $L$, respectively.
In the second step $U_{g_{2n}}$, we swap the maximal mixed state of $m_{n}$ particles in $H$ and the ground state of $m_{n}$ particles in $L$, approximately.  
Therefore, we can guess that the entropy differences in $H$ and $L$ are very close to $-m_{n}\log d$ and $m_{n}\log d$, respectively.
In fact, the following proposition holds;
\begin{proposition}\Label{EntP}
The following equalities hold with a proper positive number $\gamma$;
\begin{align}
S(\sigma_{H})-S(\rho_{\beta_{H}|\hat{H}^{(n)}_{H}})&=-m_{n}\log d +O(e^{-n\gamma})\Label{entH}\\
S(\sigma_{L})-S(\rho_{\beta_{L}|\hat{H}^{(n)}_{L}})&=m_{n}\log d +O(e^{-n\gamma})\Label{entL}
\end{align}
\end{proposition}
This proposition will be shown in Section G of Supplementary.
This proposition shows that the entropy mainly flows from the hot bath $H$ to the cold bath $L$.


\section{Remark: Relation to other formulations}
Since there are several statistical definitions of the work,
we discuss the relation of our formulation with them, here.
One might want to formulate quantum heat engine by using a quantum analogue of an invertible and deterministic dynamics of classical internal system.
In this scenario, it is natural to formulate a quantum heat engine
as a unitary process on the internal system $I$, which might be considered as a natural quantum extension of the classical standard formulation \cite{Ehrenfest}.
This type formulation has been studied for a long time, 
\cite{Lenard,tasaki,Croocks,Car1,sagawa1,jacobs,sagawa2,Funo,Morikuni,Parrondo,IE1,IE2,IE1.5,BBM1,TLH}
and is called as the semi-classical model \cite[Setion V-A]{1ponme}. 
In this model, we consider that to extract work, an external agent performs the external operation as the unitary time evolution $U_{I}:={\cal T}[\exp(\int-i\hat{H}_{I}(t)dt)]$ 
(${\cal T}[...]$ means time-ordering product) on $I$ by time-dependently controlling the control parameter of the Hamiltonian $\hat{H}_{I}(t)$ of the internal system $I$.
Similar to the classical standard model, 
the semi-classical model considers the extracted work to be the average energy loss of the internal system during the time evolution.

For explicit description of external storage,
\r{A}berg \cite[Section II-D in Supplement]{catalyst} showed that 
the internal unitary in the semi-classical scenario can be realized as the approximation of the unitary of the whole system.
However, a serious problem was founded in the semi-classical model 
even with this approximation
\cite[Section 5]{1ponme}.
When we request that the amount of work is detectable by measuring on the external agent, the time evolution of $I$ cannot be close to a unitary in the internal system $I$.
That is, there is a serious trade-off relation between 
the approximation for an internal unitary and
the detectability of the amount of extracted work \cite[Section 5]{1ponme}.
This detectability was resolved under a CP-instrument model with the condition \eqref{CP2}.

As another notion for the work-like energy transfer in the microscopic scale,
several researchers \cite{Horodecki, oneshot1, oneshot3, Egloff, Brandao, Car2, oneshot2}
study single-shot work extraction, in which the work extraction is defined as a deterministic translation from the ground pure state to the excited pure state in a two-level system.
It defines a work-like energy transfer in the quantum scale well.
That is, 
this model well works as an energy transfer in microscopic scale because the deterministic condition is crucial in this scale.
However, the correspondence between the shingle-shot work extraction and the work extraction in thermodynamics is not straight forward as follows.
Since the shingle-shot work extraction defines the work extraction as a deterministic energy transfer in microscopic scale, 
their formulation depends on the property of the microscopic scale so that
they did not care about the measurability of the amount of extracted work from a macroscopic viewpoint.
To discuss work extraction as a macro-micro energy transfer,
we need the measurability of the amount of extracted work from a macroscopic viewpoint \cite[Section I]{1ponme}.
For the measurability, we surpass the three conditions for our FQ model, which generates CP-instrument model with the condition \eqref{CP2}.

Further, we should remark that our result is meaningful even for 
an energy transfer in microscopic scale as follows.
While our optimal unitary on the whole system is compatible with CP-instrument work extraction,
the unitary itself can be used for an energy transfer in microscopic scale.
In this application, the quality of the energy gain is important.
As a measure of the quality, they proposed the deterministic condition 
for energy transfer in microscopic scale as single-shot work extraction \cite{Horodecki, oneshot1, oneshot3, Egloff, Brandao, Car2, oneshot2}.
However, their condition is too restrictive and is not necessary.
To resolve this issue, as another measure, we proposed our criterion $A ({\cal F})$
to measure the quality of the energy gain in Section \ref{s4-1}. 
That is, it is enough to show that the criterion $A ({\cal F})$ is sufficiently small
to guarantee the quality of the energy gain even in microscopic scale.
Since we have shown that the quantity $A ({\cal F})$ of our optimal unitary goes to zero, 
our optimal unitary works as a good energy transfer in microscopic scale.

\section{Conclusion}
In the present article, we have extended Carnot's theorem to finite-size heat baths, and evaluate the accuracy of the prediction of the macroscopic thermodynamics in the finite-particle systems from the statistical mechanical viewpoint.
We have given an asymptotic expansion of the optimal efficiency of quantum (or classical) heat engines whose heat baths are $n$-particle systems, up to the order of $Q^{2}_{n}/n^2$.
We have described a concrete work extraction process that attains the optimal efficiency as an energy-preserving unitary time evolution among the heat baths and the work storage.
During the unitary, The entropy-energy ratio of the work storage is so negligibly small as compared with that of the heat baths, i.e., our optimal work extraction turns the disordered energy of the heat baths to the ordered energy of the work storage.

We emphasize that our results don't use the widely used assumptions, i.e., the quasi-static assumption and the free-resource assumption.
These assumptions were necessary because of the shortage of the analytical ability of the previous approaches.
In the stochastic differential approach, it is very difficult to treat the noise with memory effect in baths.
Therefore, to calculate the behavior of the system, we need to assume that the heat bath is fixed, which is equivalent to assume the infinite-size heat bath.
In the resource theory, we can obtain computable result only when the particle number of total system is very small, or infinite limit.
On the other hand, the previous closed dynamics approach \cite{Jarzynski,tasaki,Kurchan,Car1,Xiao,oldresult5,Croocks,Sagawa2010,Ponmurugan2010,Horowitz2010,Horowitz2011,Sagawa2012,Ito2013,sagawa1,jacobs,sagawa2,Funo,Morikuni,Parrondo,Izumida,IE1,IE2,IE1.5} could not construct the microscopic dynamics based on the rule of quantum theory that achieves the optimal efficiency.
Hence, they employed the quasi-static assumption.
The weak point of the resource theory is the same as that of the previous closed dynamics approach.
These approaches have the mathematical difficulty of the optimization problem on Hilbert space with extremely high dimension, which is a serious defect of traditional methods of physics.
In this paper,  instead of traditional methods of physics, we focus on the methods of information sciences. 
That is, we have solved this problem by establishing the limited-resource theory based on the information geometry \cite{AN} and the strong large deviation \cite{LD1,BH1}.
In our theory, we have translated the optimization problem to the minimization problem of the relative entropy, and have solved it asymptotically by comparing the tail probabilities of the initial and final distributions.
Therefore, our theory enables us to treat the finite-many particle systems without taking the thermodynamic limit.
As a by-product, our theory gives the accuracy of the prediction of the macroscopic thermodynamics in the finite-particle systems from the statistical mechanical viewpoint for the first time.
We have found that the macroscopic thermodynamics predicts the optimal efficiency within the error of the order $O(Q_{n}/n^2)+O(Q^{3}_{n}/n^3)$.

Finally, we discuss the future work of our results.
We can expect our formulation and method to have many applications.
For example, the following four themes can be considered;
\begin{description}
\item[(1) Refinement of the principle of maximum work: ]~~~~~~~~~
{It is natural to expect that there exists the refined version not only for Carnot's Theorem, but also other
expressions of the second law.}
\item[(2) Variance of the extracted work: ]~\\
{In the present article, we use the entropy gain of the work storage to evaluate how the extracted work is ordered.
However, there is another way to evaluate it; we can use the variance of the extracted work.
We expect that we can evaluate the variance by applying our calculation method.  
}
\item[(3) Higher order expansion of the optimal efficiency: ]
{The thermodynamic prediction for the optimal efficiency is accurate up to the order of $Q_{n}^2/n^2$.
However, there is no guarantee that it is accurate in the orders higher than $Q_{n}^2/n^2$.
If we derive the higher order expansion of the optimal efficiency, we can evaluate the limit of the size of the isolated system in that thermodynamics is accurate.
}
\item[(4) Effect of catalyst under the restricted unitary: ]~\\
{In the present article, we request only three conditions on the unitary time evolution of the total system; \eqref{(A')} and \eqref{(B')} and \eqref{3-14-3eq}, whose $\rho^{(n)'}_{I}$ is substituted by $\rho^{(n)'}_{I,{\cal F}}$. 
When we request other restrictions on the realizable unitary time evolutions, the effect of the catalyst might become larger than Theorem \ref{3-14-3T}.
For example, when we request the following realistic restrictions, the effect of the catalyst might become larger than the order of $O(Q_{n}/n^2)$.
\begin{description}
\item[1]We forbid the direct interaction between each bath and the external system, and permit only the direct interactions between the catalyst and each heat bath and the catalyst and the external system.
\item[2]We impose the request that the total unitary time evolution finishes within a finite time $\tau$.
\end{description}
}
\item[(5) Problem about reusing heat baths: ]~\\{
In the present paper and the previous researches\cite{Jarzynski,tasaki,Kurchan,Car1,Car2,Popescu2014,Xiao,oldresult5,Croocks,Sagawa2010,Ponmurugan2010,Horowitz2010,Horowitz2011,Sagawa2012,Ito2013,sagawa1,jacobs,sagawa2,Funo,Morikuni,Parrondo,IE1,IE2,IE1.5,oneshot2,Horodecki,oneshot1,oneshot3,Egloff,Brandao}, we only consider the first use of the heat baths.
We fix the initial states of the heat baths, and use the baths as a kind of resource.
After the interaction with other systems, the heat baths' states become different from the initial states.
Therefore, we have not known the limit of the performance of heat engines when we use the heat baths repeatedly, yet.
We expect that we can answer this problem by evaluating the disturbance of the baths caused by the work extraction process.
}
\end{description}  
Since they are beyond the range of the present paper, 
we do not discuss them in this paper.
The topics (2) and (3) will be discussed in our next paper \cite{pre}.
The topic (5) also will be discussed in our next paper \cite{pre2}.

\section*{Acknowledgments}
The authors are grateful to  Dr. Kiyoshi Kanazawa, Dr. Jun'ichi Ozaki, and Mr. Kosuke Ito for a helpful comments.
HT was partially supported by the Grants-in-Aid for Japan Society for Promotion of Science (JSPS) Fellows (Grant No. 24.8116).
MH is partially supported by a MEXT Grant-in-Aid for Scientific Research (A) No. 23246071.
The Centre for Quantum Technologies is funded by the
Singapore Ministry of Education and the National Research Foundation
as part of the Research Centres of Excellence programme.

\renewcommand{\refname}{\vspace{-1cm}}

\widetext

\appendix

\section*{Organization of Appendix}

In the present appendix, we give the detail of our information theoretical methods.
We expect it of giving useful informations for the readers who want to use our methods.
The present appendix is organized as follows.
In appendix \ref{Aa3}, we give the prediction of the macroscopic thermodynamics for the optimal efficiency in the thermodynamic, and translate the macroscopic thermodynamic prediction to the statistical mechanical expression.
It corresponds to the derivation of Proposition 1 and
the equality (26) in the main text.
In Appendix \ref{AApB}, we show that $\eta^{(n)}_{T}[\beta_{H},\beta_{L},Q]$ is a general upper bound for the efficiencies $\eta_{C}(\beta_{H},\beta_{L},P_{Z},\T)$, $\eta_{C}(\beta_{H},\beta_{L},\rho_{C},{\cal W})$, $\eta_{C\lnot}(\beta_{H},\beta_{L},P_{Z},\T)$ and $\eta_{C\lnot}(\beta_{H},\beta_{L},\rho_{C},{\cal W})$, and that the differences between $\eta^{(n)}_{T}[\beta_{H},\beta_{L},Q]$ and 
the efficiencies are bounded by the disturbance of the heat baths.
It corresponds to the derivation of Theorem 1 in the main text.
In Appendix \ref{AproofofC=Q}, we derive the relationship between the classical and quantum optimal efficiencies.
It corresponds to the derivation of Proposition 2 in the main text.
In Appendix \ref{AApC}, we derive an asymptotic expansion of $\eta^{(n)}_{T}[\beta_{H},\beta_{L},Q_{n}]$.
It corresponds to the derivation of Proposition 3 in the main text.
In Appendix \ref{Aproof32}, 
we derive the asymptotic expansions of $\eta^{(n)}_{C\lnot}(\beta_{H},\beta_{L},\T^{\mathrm{opt}}_{Q_n})$ and $Q^{(n)}_{H\lnot}(\beta_{H},\beta_{L},\T^{\mathrm{opt}}_{Q_n})$, 
It corresponds to the derivation of Proposition 4 in the main text.
In this proof, we need to asymptotically expand the relative entropy $D_X^n(m_n)$ between the two distributions very close to each other.
The conventional approaches for information theory and quantum information theory do not work well.
In Appendix \ref{As14}, we resolve this problem. We employ strong large deviation by Bahadur-Rao and Blackwell-Hodges \cite{LD1,BH1} that brings us a more detailed evaluation for the tail probability. 
This section is the technical highlight of this supplementary materials.
In Appendix \ref{AApH}, we evaluate the entropy differences of $H$ and $L$ caused by $U_{f_{n}}$
It corresoponds to the derivation of Proposition 5 in the main text.

\section{Optimal efficiency in thermodynamic setup}\Label{Aa3}
In the present section, 
we show Propositions 1, and prepare Lemma \ref{A8-18-1} for the proof of Proposition 2 in the next section.
In the first subsection, we give the prediction of the macroscopic thermodynamics for the optimal efficiency $\eta_{T}[\beta_{H},\beta_{L},\Omega_{R_{H}},\Omega_{R_{L}},Q]$ in the thermodynamic setup.
Then, we show Proposition 1.
In the second subsection, we characterize $\eta^{(n)}_{T}[\beta_{H},\beta_{L},Q]$ by using Carnot's efficiency and the relative entropy.

\subsection{Prediction of the macroscopic thermodynamics}\Label{AApA1}
At first, we write down Proposition 1 in the main text again;
\begin{proposition}\Label{A8-18-0}
The efficiency $\eta_{T}[\beta_{H},\beta_{L},\Omega_{R_{H}},\Omega_{R_{L}},Q]$ satisfies
\begin{align}
&\eta_{T}[\beta_{H},\beta_{L},\Omega_{R_{H}},\Omega_{R_{L}},Q]=1-\frac{E_{R_{L}}(T'^{\mathrm{quasi}}_{R_{L}},\Omega_{R_{L}})-E_{R_{L}}(T_{R_{L}},\Omega_{R_{L}})}{Q}\Label{Athermover2},
\end{align}
where
\begin{align}
\eta_{T}[\beta_{H},\beta_{L},\Omega_{R_{H}},\Omega_{R_{L}},Q]&:=\sup_{\substack{T'_{R_{H}},T'_{R_{L}},T_{W},\Omega_{W}\\:Q_{\mathrm{ad}}=Q,\eqref{Aloei}}}\frac{W_{\mathrm{ad}}}{Q_{\mathrm{ad}}}\Label{Athermover},\\
S_{R_{H}}(T_{R_{H}},\Omega_{R_{H}})+S_{R_{L}}(T_{R_{L}},\Omega_{R_{L}})+S_{W}(T_{W},\Omega_{W})&\le S_{R_{H}}(T'_{R_{H}},\Omega'_{R_{H}})+S_{R_{L}}(T'_{R_{L}},\Omega'_{R_{L}})+S_{W}(T'_{W},\Omega'_{W}),\Label{Aloei}
\end{align}
and where $T'^{\mathrm{quasi}}_{R_{L}}$ is defined by the following two equalities;
\begin{align}
&E_{R_{H}}(T_{R_{H}},\Omega_{R_{H}})-E_{R_{H}}(T'^{\mathrm{quasi}}_{R_{H}},\Omega_{R_{H}})=Q,\Label{Aseigenthermo0}\\
&S_{R_{H}}(T_{R_{H}},\Omega_{R_{H}})+S_{R_{L}}(T_{R_{L}},\Omega_{R_{L}})=S_{R_{H}}(T'^{\mathrm{quasi}}_{R_{H}},\Omega_{R_{H}})+S_{R_{L}}(T'^{\mathrm{quasi}}_{R_{L}},\Omega_{L}).\Label{Aseigenthermo}
\end{align}
\end{proposition}

\begin{proofof}{Proposition \ref{A8-18-0}}
Let us derive \eqref{Athermover2} in Proposition 1. 
Because of \eqref{Aseigenthermo0}, we can rewrite \eqref{Athermover} as follows;
\begin{align}
\eta_{T}[\beta_{H},\beta_{L},\Omega_{R_{H}},\Omega_{R_{L}},Q]=\sup_{T'_{R_{L}};\eqref{Aloei}}\left(1-\frac{E_{R_{L}}(T'_{R_{L}},\Omega_{R_{L}})-E_{R_{L}}(T_{R_{L}},\Omega_{R_{L}})}{Q}\right).\Label{AA1}
\end{align}
Since \eqref{Aloei} implies that,
\begin{align}
S_{R_{H}}(T_{R_{H}},\Omega_{R_{H}})+S_{R_{L}}(T_{R_{L}},\Omega_{R_{L}})
\le 
S_{R_{H}}(T'^{\mathrm{quasi}}_{R_{H}},\Omega_{R_{H}})+S_{R_{L}}(T'_{R_{L}},\Omega_{R_{L}}).\Label{Asecondlaw}
\end{align}
The expressions \eqref{Aseigenthermo} and \eqref{Asecondlaw} guarantee that
\begin{align}
S_{R_{L}}(T'^{\mathrm{quasi}}_{R_{L}},\Omega_{R_{L}})\le 
S_{R_{L}}(T'_{R_{L}},\Omega_{R_{L}}).
\end{align}
Because the thermodynamic entropy is an increasing function of the temperature \cite{Tasakitext}, we have $T'^{\mathrm{quasi}}_{R_{L}}\le T'_{R_{L}}$.
Since the thermodynamic internal energy is an increasing function of the temperature \cite{Tasakitext}, we have
\begin{equation}
E_{R_{L}}(T'^{\mathrm{quasi}}_{R_{L}},\Omega_{R_{L}})\le 
E_{R_{L}}(T'_{R_{L}},\Omega_{R_{L}}).\Label{AA4}
\end{equation}
Combining \eqref{AA1} and \eqref{AA4}, we obtain
\begin{align}
\eta_{T}[\beta_{H},\beta_{L},\Omega_{R_{H}},\Omega_{R_{L}},Q]&\le1-\frac{E_{R_{L}}(T'^{\mathrm{quasi}}_{R_{L}},\Omega_{R_{L}})-E_{R_{L}}(T_{R_{L}},\Omega_{R_{L}})}{Q}.\Label{Aarxthermover2}
\end{align}
The equality of \eqref{Aarxthermover2} is achieved when the equality of \eqref{Asecondlaw} holds.
In the framework of thermodynamics, we assume the thermalization assumption, and thus we can perform the adiabatic quasi-static process which achieves the equality of \eqref{Asecondlaw}.
Hence, we obtain \eqref{Athermover2} in Proposition 1.
\end{proofof}

\subsection{Statistical mechanical expression of $\eta^{(n)}_{T}[\beta_{H},\beta_{L},Q]$}\Label{AApA2}
In the present subsection, we prove the following Lemma which will be used to show Proposition 2.
It corresponds to (26) in the main text.
\begin{lemma}\Label{A8-18-1}
The thermodynamic optimal efficiency $\eta^{(n)}_{T}$ is characterized as \begin{align}
\eta^{(n)}_{T}[\beta_{H},\beta_{L},Q]=\left(1-\frac{\beta_{H}}{\beta_{L}}\right)-\frac{D(\rho^{(n)}_{\beta'_{H}\beta'_{L}}\|\rho^{(n)}_{\beta_{H}\beta_{L}})}{\beta_{L}Q}\Label{ATandD}.
\end{align}
\end{lemma}

\begin{proofof}{Lemma \ref{A8-18-1}}
At first, let us write down the condition (24) assumed in the main text, which means the equivalence between the thermodynamic internal energy and the statistical mechanical internal energy, and the equivalence between the thermodynamic entropy and the statistical mechanical entropy;
\begin{align}
&E_{R_{H}}(T,\Omega^{(n)}_{R_{H}})=\Tr[\rho_{1/T|\hat{H}^{(n)}_{H}}\hat{H}^{(n)}_{H}]=\left<h^{(n)}_{X}\right>_{P_{1/T|h^{(n)}_{X}}},\nonumber\\
&E_{R_{L}}(T,\Omega^{(n)}_{R_{L}})=\Tr[\rho_{1/T|\hat{H}^{(n)}_{L}}\hat{H}^{(n)}_{L}]=\left<h^{(n)}_{Y}\right>_{P_{1/T|h^{(n)}_{Y}}},\nonumber\\
&S_{R_{H}}(T,\Omega^{(n)}_{R_{H}})=S(\rho_{1/T|\hat{H}^{(n)}_{H}})=S(P_{1/T|h^{(n)}_{X}}),\nonumber\\
&S_{R_{L}}(T,\Omega^{(n)}_{R_{L}})=S(\rho_{1/T|\hat{H}^{(n)}_{L}})=S(P_{1/T|h^{(n)}_{Y}}),
\Label{A4-3-1eq}
\end{align}

Then, because of \eqref{A4-3-1eq} and the definition of $\beta'_{L}$, we can convert \eqref{Athermover2} in Proposition 1 as follows;
\begin{align}
\eta^{(n)}_{T}[\beta_{H},\beta_{L},Q]&=1-\frac{E_{R_{L}}(T'^{\mathrm{quasi}}_{R_{L}},\Omega^{(n)}_{R_{L}})-E_{R_{L}}(T_{R_{L}},\Omega^{(n)}_{R_{L}})}{Q},\nonumber\\
&=1-\frac{\Tr[(\rho_{\beta'_{L}|\hat{H}^{(n)}_{L}}-\rho_{\beta_{L}|\hat{H}^{(n)}_{L}})\hat{H}^{(n)}_{L}]}{Q},\nonumber\\
&=1+\frac{\Tr[(\rho_{\beta'_{L}|\hat{H}^{(n)}_{L}}-\rho_{\beta_{L}|\hat{H}^{(n)}_{L}})\log\rho_{\beta_{L}|\hat{H}^{(n)}_{L}}]}{\beta_{L}Q}\nonumber\\
&=1+\frac{-D(\rho_{\beta'_{L}|\hat{H}^{(n)}_{L}}\|\rho_{\beta_{L}|\hat{H}^{(n)}_{L}})-S(\rho_{\beta'_{L}|\hat{H}^{(n)}_{L}})+S(\rho_{\beta_{L}|\hat{H}^{(n)}_{L}})}{\beta_{L}Q}\Label{AA5}
\end{align}
The combination of \eqref{Aseigenthermo}, the definitions of $\beta'_{H}$ and $\beta'_{L}$ and the assumption \eqref{A4-3-1eq} yields that
\begin{align}
S(\rho_{\beta_{H}|\hat{H}^{(n)}_{H}})+S(\rho_{\beta_{L}|\hat{H}^{(n)}_{L}})=S(\rho_{\beta'_{H}|\hat{H}^{(n)}_{H}})+S(\rho_{\beta'_{H}|\hat{H}^{(n)}_{H}}).
\end{align}
Therefore, we can convert the RHS of \eqref{AA5} as follows;
\begin{align}
\mathrm{RHS}\enskip\mathrm{of}\enskip\eqref{AA5}
&=1+\frac{-D(\rho_{\beta'_{L}|\hat{H}^{(n)}_{L}}\|\rho_{\beta_{L}|\hat{H}^{(n)}_{L}})-S(\rho_{\beta_{H}|\hat{H}^{(n)}_{H}})-\Tr[\rho_{\beta'_{H}|\hat{H}^{(n)}_{H}}\log\rho_{\beta_{H}|\hat{H}^{(n)}_{H}}]-D(\rho_{\beta'_{H}|\hat{H}^{(n)}_{H}}\|\rho_{\beta_{H}|\hat{H}^{(n)}_{H}})}{\beta_{L}Q}
\end{align}
Because $\beta'_{H}$ is defined as $1/T'^{\mathrm{quasi}}_{H}$, the relations \eqref{Aseigenthermo0} and \eqref{A4-3-1eq} imply that  
\begin{align}
Q=\Tr[(\rho_{\beta'_{H}|\hat{H}^{(n)}_{H}}-\rho_{\beta_{H}|\hat{H}^{(n)}_{H}})\hat{H}^{(n)}_{H}].
\end{align} 
Hence,
\begin{align}
\eta^{(n)}_{T}[\beta_{H},\beta_{L},Q]&=1+\frac{-D(\rho_{\beta'_{L}|\hat{H}^{(n)}_{L}}\|\rho_{\beta_{L}|\hat{H}^{(n)}_{L}})+\beta_{H}\Tr[(\rho_{\beta'_{H}|\hat{H}^{(n)}_{H}}-\rho_{\beta_{H}|\hat{H}^{(n)}_{H}})\hat{H}^{(n)}_{H}]-D(\rho_{\beta'_{H}|\hat{H}^{(n)}_{H}}\|\rho_{\beta_{H}|\hat{H}^{(n)}_{H}})}{\beta_{L}Q}\nonumber\\
&=1+\frac{-D(\rho_{\beta'_{L}|\hat{H}^{(n)}_{L}}\|\rho_{\beta_{L}|\hat{H}^{(n)}_{L}})-\beta_{H}Q-D(\rho_{\beta'_{H}|\hat{H}^{(n)}_{H}}\|\rho_{\beta_{H}|\hat{H}^{(n)}_{H}})}{\beta_{L}Q}\nonumber\\
&=1-\frac{\beta_{H}}{\beta_{L}}-\frac{D(\rho^{(n)}_{\beta'_{H}\beta'_{L}}\|\rho^{(n)}_{\beta_{H}\beta_{L}})}{\beta_{L}Q}.
\end{align}
So, we obtain Lemma \ref{A8-18-1}.
\end{proofof}

\section{General upper bound}\Label{AApB}
In the present section, we prove Theorem 1 in the main text.
Namely, we show that $\eta^{(n)}_{T}[\beta_{H},\beta_{L},Q]$ is a general upper bound for the efficiencies $\eta_{C}(\beta_{H},\beta_{L},P_{Z},\T)$, $\eta_{C}(\beta_{H},\beta_{L},\rho_{C},{\cal W})$, $\eta_{C\lnot}(\beta_{H},\beta_{L},P_{Z},\T)$ and $\eta_{C\lnot}(\beta_{H},\beta_{L},\rho_{C},{\cal W})$, and that the differences between $\eta^{(n)}_{T}[\beta_{H},\beta_{L},Q]$ and the efficiencies are bounded by the disturbance of the heat baths.

At first, let us write down Theorem 1 again.
\begin{theorem}\Label{A3-14-0T}
Let $Q$ be an arbitrary positive real number, and
$P_{Z}$ be an arbitrary initial distribution of $Z$.
When a classical dynamics ${\cal T}$ satisfies $Q^{(n)}(\beta_{H},\beta_{L},\T)=Q$ and 
\begin{align}
\sum_{x,y,z}\T(x,y,z|x',y',z')&=\sum_{x',y',z'}\T(x,y,z|x',y',z')=1.\Label{A1}\\
\sum_{x,y}P^{(n)'}_{I}(x,y,z)&=P_{Z}(z),\Label{A2}
\end{align}
and when the thermodynamic baths $R_{H}$ and $R_{L}$ satisfy \eqref{A4-3-1eq},
the following inequalities hold;
\begin{align}
\eta^{(n)}_{C}(\beta_{H},&\beta_{L},P_{Z},\T)
\le 
1-\frac{\beta_{H}}{\beta_{L}}-\frac{D(P^{(n)'}_{XY}\|P^{(n)}_{\beta_{H}\beta_{L}})}{\beta_{L}Q} \Label{ACT0}\\
&\le\eta^{(n)}_{T}[\beta_{H},\beta_{L},Q]-\frac{D(P^{(n)'}_{XY}\|P^{(n)}_{\beta'_{H},\beta'_{L}})}{\beta'_{L}Q},\Label{ACT}
\end{align}
where $P^{(n)'}_{XY}:=\sum_{z}P^{(n)'}_{I}(x,y,z)$ is the final state of the classical heat baths, and $\beta'_{H}:=1/T'^{\mathrm{quasi}}_{H}$ is the inverse temperature of $T'^{\mathrm{quasi}}_{H}$ in \eqref{Aseigenthermo0}. 
Similarly, let $Q$ be an arbitrary positive real number, and
$\rho_{C}$ be an arbitrary initial state $\rho_{C}$ of $C$.
When a quantum dynamics ${\cal W}$ satisfies $Q^{(n)}_{H}(\beta_{H},\beta_{L},{\cal W})=Q$ and
\begin{align}
\sum_{j}p_{j}w_{j}&=\Tr[\hat{H}^{(n)}_{I}(\rho^{(n)}_{I}-\rho^{(n)'}_{I})],\enskip\forall\rho^{(n)}_{I}\Label{AE.C.1}\\
\sum_{j}{\cal E}_{j}(\hat{1}_{HL})&=\hat{1}_{HL},\Label{Aunital}
\end{align}
and when the thermodynamic baths $R_{H}$ and $R_{L}$ satisfy \eqref{A4-3-1eq},
the following inequalities hold;
\begin{align}
\eta^{(n)}_{Q}(\beta_{H},\beta_{L},&\rho_{C},{\cal W})
\le1-\frac{\beta_{H}}{\beta_{L}}-\frac{D(\rho^{(n)'}_{HL}\|\rho^{(n)}_{\beta_{H}\beta_{L}})}{\beta_{L}Q} \Label{AQT0}\\
&
\le \eta^{(n)}_{T}[\beta_{H},\beta_{L},Q]-\frac{D(\rho^{(n)'}_{HL}\|\rho_{\beta'_{H},\beta'_{L}})}{\beta'_{L}Q},\Label{AQT}
\end{align}
where $\rho^{(n)'}_{HL}:=\Tr_{C}[\rho^{(n)'}_{I}]$ is the final state of the quantum heat baths.
\end{theorem}

\begin{proofof}{Theorem 1}
We firstly point out that \eqref{ACT0} and \eqref{ACT} follow from \eqref{AQT0} and \eqref{AQT}, because the classical case is included the quantum case. 
Therefore, we only have to prove \eqref{AQT0} and \eqref{AQT}.
Because \eqref{AQT0} is equivalent to 
\begin{align}
D(\rho^{(n)'}_{HL}\|\rho^{(n)}_{\beta_{H}\beta_{L}})\le-\beta_{H}Q-\beta_{L}(W^{(n)}(\beta_{H},\beta_{L},\rho_{C},{\cal W})-Q),\Label{AQT0pre}
\end{align}
we will prove \eqref{AQT0pre} in the following discussion.
The inequality \eqref{AQT0pre} is derived as follows;
\begin{align}
D(\rho^{(n)'}_{HL}\|\rho^{(n)}_{\beta_{H}\beta_{L}})&\le D(\rho^{(n)'}_{HLC}\|\rho^{(n)}_{\beta_{H}\beta_{L}}\otimes\rho_{C})\nonumber\\
&=-S(\rho^{(n)'}_{HLC})-\Tr[\rho^{(n)'}_{HLC}\log\rho^{(n)}_{\beta_{H}\beta_{L}}\otimes\rho_{C}]\nonumber\\
&\stackrel{(a)}{\le} -S(\rho^{(n)}_{\beta_{H}\beta_{L}}\otimes\rho_{C})-\Tr[\rho^{(n)'}_{HLC}\log\rho^{(n)}_{\beta_{H}\beta_{L}}\otimes\rho_{C}]\nonumber\\
&=\Tr[(\rho^{(n)}_{\beta_{H}\beta_{L}}\otimes\rho_{C}-\rho^{(n)'}_{HLC})\log\rho^{(n)}_{\beta_{H}\beta_{L}}\otimes\rho_{C}]\nonumber\\
&\stackrel{(b)}{=}\Tr[(\rho_{\beta_{H}|\hat{H}^{(n)}_{H}}-\rho^{(n)'}_{H})\log\rho_{\beta_{H}|\hat{H}^{(n)}_{H}}]
+\Tr[(\rho_{\beta_{L}|\hat{H}^{(n)}_{L}}-\rho^{(n)'}_{L})\log\rho_{\beta_{H}|\hat{H}^{(n)}_{L}}]+\Tr[(\rho^{(n)}_{C}-\rho^{(n)'}_{HLC})\log\rho^{(n)}_{C}]\nonumber\\
&\stackrel{(c)}{=}\Tr[(\rho_{\beta_{H}|\hat{H}^{(n)}_{H}}-\rho^{(n)'}_{H})\log\rho_{\beta_{H}|\hat{H}^{(n)}_{H}}]
+\Tr[(\rho_{\beta_{L}|\hat{H}^{(n)}_{L}}-\rho^{(n)'}_{L})\log\rho_{\beta_{H}|\hat{H}^{(n)}_{L}}]\nonumber\\
&=-\beta_{H}\Tr[(\rho_{\beta_{H}|\hat{H}^{(n)}_{H}}-\rho^{(n)'}_{H})\hat{H}^{(n)}_{H}]
-\beta_{L}\Tr[(\rho_{\beta_{L}|\hat{H}^{(n)}_{L}}-\rho^{(n)'}_{L})\hat{H}^{(n)}_{L}]\nonumber\\
&=-\beta_{H}Q^{(n)}_{H}(\beta_{H},\beta_{L},\rho_{C},{\cal W})-\beta_{L}(W^{(n)}(\beta_{H},\beta_{L},\rho_{C},{\cal W})-Q^{(n)}_{H}(\beta_{H},\beta_{L},\rho_{C},{\cal W}))\nonumber\\
&=-\beta_{H}Q-\beta_{L}(W^{(n)}(\beta_{H},\beta_{L},\rho_{C},{\cal W})-Q),
\end{align}
where $(a)$ is shown by $S(\rho^{(n)}_{\beta_{H}}\otimes\rho_{C})\le S(\rho^{(n)'}_{HLC})$,
$(b)$ is shown by $\Tr[\rho_{AB}\log\sigma_{A}\otimes\sigma_{B}]=\Tr[\rho_{A}\rho\sigma_{A}]+\Tr[\rho_{B}\log\sigma_{B}]$,
and $(c)$ is shown by \eqref{Aunital}. 

Next, we prove \eqref{AQT}.
We only have to prove 
\begin{align}
D(\rho^{(n)'}_{\beta_{H}\beta_{L}}\|\rho^{(n)}_{\beta_{H}\beta_{L}})-D(\rho^{(n)}_{\beta'_{H}\beta'_{L}}\|\rho^{(n)}_{\beta_{H}\beta_{L}})-\frac{\beta_{L}}{\beta'_{L}}D(\rho^{(n)'}_{HL}\|\rho^{(n)}_{\beta'_{H}\beta'lll_{L}})\ge0.\Label{AQTpre}
\end{align}
In order to show \eqref{AQTpre}, we firstly convert the lefthand side (LHS) of \eqref{AQTpre} as follows;
\begin{align}
\mbox{The}\enskip\mbox{LHS}\enskip\eqref{AQTpre}&=-S(\rho^{(n)'}_{HL})-\Tr[\rho^{(n)'}_{HL}\log\rho^{(n)}_{\beta_{H}\beta_{L}}]+S(\rho^{(n)}_{\beta'_{H}\beta'_{L}})+\Tr[\rho^{(n)}_{\beta'_{H}\beta'_{L}}\log\rho^{(n)}_{\beta_{H}\beta_{L}}]+\frac{\beta_{L}}{\beta'_{L}}S(\rho^{(n)'}_{HL})+\frac{\beta_{L}}{\beta'_{L}}\Tr[\rho^{(n)'}_{HL}\rho^{(n)}_{\beta'_{H}\beta'_{L}}]\nonumber\\
&=\left(\frac{\beta_{L}}{\beta'_{H}}-1\right)S(\rho^{(n)'}_{HL})+\Tr[(\rho^{(n)}_{\beta'_{H}\beta'_{L}}-\rho^{(n)'}_{HL})\log\rho^{(n)}_{\beta_{H}\beta_{L}}]
+S(\rho^{(n)}_{\beta'_{H}\beta'_{L}})
+\frac{\beta_{L}}{\beta'_{L}}\Tr[\rho^{(n)'}_{HL}\log\rho^{(n)}_{\beta'_{H}\beta'_{L}}]\nonumber\\
&\stackrel{(a)}{\ge}\frac{\beta_{L}}{\beta'_{L}}(S(\rho^{(n)}_{\beta'_{H}\beta'_{L}})+\Tr[\rho^{(n)'}_{HL}\log\rho^{(n)}_{\beta'_{H}\beta'_{L}}])
+\Tr[(\rho^{(n)}_{\beta'_{H}\beta'_{L}}-\rho^{(n)'}_{HL})\log\rho^{(n)}_{\beta_{H}\beta_{L}}]\nonumber\\
&=-\frac{\beta_{L}}{\beta'_{L}}(\Tr[(\rho^{(n)}_{\beta'_{H}\beta'_{L}}-\rho^{(n)'}_{HL})\log\rho^{(n)}_{\beta'_{H}\beta'_{L}}])
+\Tr[(\rho^{(n)}_{\beta'_{H}\beta'_{L}}-\rho^{(n)'}_{HL})\log\rho^{(n)}_{\beta_{H}\beta_{L}}],\Label{AQTpre2}
\end{align}
where $(a)$ is shown by $S(\rho^{(n)'}_{HL})\ge S(\rho^{(n)}_{\beta'_{H}\beta'_{L}})$.
Let us show that the first term and the second term of the righthand side (RHS) of \eqref{AQTpre2} offset each others.
The first term is converted as follows;
\begin{align}
-\frac{\beta_{L}}{\beta'_{L}}(\Tr[(\rho^{(n)}_{\beta'_{H}\beta'_{L}}-\rho^{(n)'}_{HL})\log\rho^{(n)}_{\beta'_{H}\beta'_{L}}])
&=-\frac{\beta_{L}}{\beta'_{L}}(\Tr[(\rho_{\beta'_{H}|\hat{H}^{(n)}_{H}}-\rho^{(n)'}_{H})\log\rho_{\beta'_{H}|\hat{H}^{(n)}_{H}}]
+\Tr[(\rho_{\beta'_{L}|\hat{H}^{(n)}_{L}}-\rho^{(n)'}_{L})\log\rho_{\beta'_{L}|\hat{H}^{(n)}_{L}}])\nonumber\\
&=-\frac{\beta_{L}}{\beta'_{L}}(-\beta'_{H}\Tr[(\rho_{\beta'_{H}|\hat{H}^{(n)}_{H}}-\rho^{(n)'}_{H})\hat{H}^{(n)}_{H}]
-\beta'_{L}\Tr[(\rho_{\beta'_{L}|\hat{H}^{(n)}_{L}}-\rho^{(n)'}_{L})\hat{H}^{(n)}_{L}])\nonumber\\
&\stackrel{(a)}{=}\beta_{L}\Tr[(\rho_{\beta'_{L}|\hat{H}^{(n)}_{L}}-\rho^{(n)'}_{L})\hat{H}^{(n)}_{L}]),\Label{AQTpre4}
\end{align}
where $(a)$ follows from
\begin{align}
Q=\Tr[(\rho_{\beta_{H}|\hat{H}^{(n)}_{H}}-\rho_{\beta'_{H}|\hat{H}^{(n)}_{H}})\hat{H}^{(n)}_{H}]
=\Tr[(\rho_{\beta_{H}|\hat{H}^{(n)}_{H}}-\rho^{(n)'}_{H})\hat{H}^{(n)}_{H}].\Label{AQTpre3}
\end{align}
On the other hand, the first term is converted as follows;
\begin{align}
\Tr[(\rho^{(n)}_{\beta'_{H}\beta'_{L}}-\rho^{(n)'}_{HL})\log\rho^{(n)}_{\beta_{H}\beta_{L}}]
&=\Tr[(\rho_{\beta'_{H}|\hat{H}^{(n)}_{H}}-\rho^{(n)'}_{H})\log\rho_{\beta_{H}|\hat{H}^{(n)}_{H}}]
+\Tr[(\rho_{\beta'_{L}|\hat{H}^{(n)}_{L}}-\rho^{(n)'}_{L})\log\rho_{\beta_{L}|\hat{H}^{(n)}_{L}}])\nonumber\\
&=-\beta_{H}\Tr[(\rho_{\beta'_{H}|\hat{H}^{(n)}_{H}}-\rho^{(n)'}_{H})\hat{H}^{(n)}_{H}]
-\beta_{L}\Tr[(\rho_{\beta'_{L}|\hat{H}^{(n)}_{L}}-\rho^{(n)'}_{L})\hat{H}^{(n)}_{L}])\nonumber\\
&\stackrel{(a)}{=}-\beta_{L}\Tr[(\rho_{\beta'_{L}|\hat{H}^{(n)}_{L}}-\rho^{(n)'}_{L})\hat{H}^{(n)}_{L}]),\Label{AQTpre5}
\end{align}
where $(a)$ follows from \eqref{AQTpre3}.
Due to \eqref{AQTpre4} and \eqref{AQTpre5}, the RHS of \eqref{AQTpre2} is zero, and thus \eqref{AQTpre} holds. 
\end{proofof}

\section{The relationship between the classical and quantum optimal efficiencies}\Label{AproofofC=Q}
In the present section, we prove Proposition 2 in the main text.
Let us write down Proposition 2 again;
\begin{proposition}\Label{AC=Q}
\begin{align}
\eta^{(n)}_{C\lnot}[\beta_{H},\beta_{L},Q]&=\eta^{(n)}_{Q\lnot}[\beta_{H},\beta_{L},Q],\Label{AC=Q1}\\
\eta^{(n)}_{C}[\beta_{H},\beta_{L},Q]&\le\eta^{(n)}_{Q}[\beta_{H},\beta_{L},Q]\Label{AC=Q2}
\end{align}
\end{proposition}

\begin{proofof}{Proposition \ref{AC=Q}}
We firstly prove \eqref{AC=Q2}.
In order to show \eqref{AC=Q2}, we only have to show that when $P_{Z}$ and $\T$  satisfy \eqref{A1} and \eqref{A2}, we can take a state $\rho^{P_{Z}}_{C}$ and a work extration ${\cal W}_{\T}$ that satisfy the following relations;
\begin{align}
W^{(n)}(\beta_{H},\beta_{L},P_{Z},\T)&= W^{(n)}(\beta_{H},\beta_{L},\rho^{P_{Z}}_{C},{\cal W}_{T}).\Label{ACQP1}\\
Q^{(n)}_{H}(\beta_{H},\beta_{L},P_{Z},\T)&= Q^{(n)}_{H}(\beta_{H},\beta_{L},\rho^{P_{Z}}_{C},{\cal W}_{T}).\Label{ACQP2}
\end{align}
In order to find ${\cal W}_{\T}$, we notice that the bi-stochastic $\T$ can be written as
\begin{align}
\T(x,y,z|x',y',z')=\sum_{m}p_{m}\delta_{(x,y,z),f_{m}(x',y',z')},
\end{align}
by using a proper probability $\{p_{m}\}$, invertible functions $\{f_{m}:=(f^{X}_{m}(x,y,z),f^{Y}_{m}(x,y,z),f^{Z}_{m}(x,y,z))\}$, and the Kronecker delta $\delta$. 
Then, we choose the state $\rho^{P_{Z}}_{C}$ and the work extraction ${\cal W}_{\T}$ as follows;
\begin{align}
\rho^{P_{Z}}_{C}&:=\sum_{z}P_{Z}(z)|z\rangle_{C}~_{C}\langle z|,\\
{\cal E}_{x,y,z}(\rho_{I})&:=\sum_{m}p_{m}U_{f_{m}}|x,y,z\rangle\langle x,y,z|\rho_{I}|x,y,z\rangle\langle x,y,z|U^{\dagger}_{f_{m}},\\
w_{x,y,z}&:=h_{I}(x,y,z)-\sum_{m}p_{m}h_{I}(f_{m}(x,y,z)).
\end{align}
where 
we introduce an energy preserving unitary $U_{f}$ on $HLE_{X}$
for an invertible function $f$ on $({\cal X} \times {\cal Y})^n$ as follows;
\begin{align}
U_{f}:=\sum_{x,y}\sum_{e\in \Lambda}|&f_{n}(x,y)\rangle_{HL}
|e'(x,y)\rangle_{E_{X}}~_{HL}\langle x,y|~ _{E_{X}}\langle e|,\Label{AUfn}
\end{align}
where $|e'(x,y)\rangle_{E_{X}}:=|e+h^{(n)}_{XY}(x,y)-h^{(n)}_{XY}(f_{n}(x,y))\rangle_{E_{X}}$.

Let us prove that $\rho^{P_{Z}}_{C}$ and ${\cal W}_{\T}$ satisfy \eqref{ACQP1} and \eqref{ACQP2}.
\begin{align}
W^{(n)}(\beta_{H},\beta_{L},\rho^{P_{Z}}_{C},{\cal W}_{T})
&=\sum_{x,y,z} w_{x,y,z} \Tr[{\cal E}_{x,y,z}(\rho^{(n)}_{\beta_{H},\beta_{L}}\otimes\rho^{P_{Z}}_{C})]\nonumber\\
&=\sum_{x,y,z} (h_{I}(x,y,z)-\sum_{m}p_{m}h_{I}(f_{m}(x,y,z)))P^{(n)}_{\beta_{H},\beta_{L}}(x,y)P_{Z}(z)\nonumber\\
&=\left<h_{I}\right>_{P^{(n)}_{\beta_{H},\beta_{L}}P_{Z}}-\left<h_{I}\right>_{\T(P^{(n)}_{\beta_{H},\beta_{L}}P_{Z})}\nonumber\\
&=W^{(n)}(\beta_{H},\beta_{L},P_{Z},\T)\nonumber\\
Q^{(n)}_{H}(\beta_{H},\beta_{L},\rho^{P_{Z}}_{C},{\cal W}_{T})
&=\Tr[(\rho^{(n)}_{\beta_{H},\beta_{L}}\otimes\rho^{P_{Z}}_{C}-\sum_{x,y,z}{\cal E}_{x,y,z}(\rho^{(n)}_{\beta_{H},\beta_{L}}\otimes\rho^{P_{Z}}_{C}))\hat{H}^{(n)}_{H}]\nonumber\\
&=\Tr[(\rho^{(n)}_{\beta_{H}\beta_{L}}-\sum_{j}{\cal E}_{j}(\rho^{(n)}_{\beta_{H}\beta_{L}}))\hat{H}^{(n)}_{H}] \nonumber\\
&=\sum_{x,y,z} (h_{X}(x)-\sum_{m}p_{m}h_{X}(f^{X}_{m}(x,y,z)))P^{(n)}_{\beta_{H},\beta_{L}}(x,y)P_{Z}(z)\nonumber\\
&=\left<h_{X}\right>_{P^{(n)}_{\beta_{H},\beta_{L}}P_{Z}}-\left<h_{X}\right>_{\T(P^{(n)}_{\beta_{H},\beta_{L}}P_{Z})}\nonumber\\
&=Q^{(n)}_{H}(\beta_{H},\beta_{L},P_{Z},\T).
\end{align}
Therefore, we obtain \eqref{AC=Q2}.

Next, we prove \eqref{AC=Q1}.
Because $\eta^{(n)}_{C\lnot}[\beta_{H},\beta_{L},Q_n]\le\eta^{(n)}_{Q\lnot}[\beta_{H},\beta_{L},Q_n]$ can be shown in the same way of \eqref{AC=Q2}, we only have to prove $\eta^{(n)}_{Q\lnot}[\beta_{H},\beta_{L},Q_n]
\le\eta^{(n)}_{C\lnot}[\beta_{H},\beta_{L},Q_n]$.
Let any CP-instrument work extraction $\mathscr{W}_{n}=\{{\cal E}_j,w_j\}$ satisfy \eqref{Aunital} and 
\begin{align}
{\cal E}_{j}(\Pi_{x}) 
=P_{h_{x}-w_{j}}{\cal E}_{j}(\Pi_{x})P_{h_{x}-w_{j}}
\Label{ACP2}
\end{align}
for any initial eigenstate $\Pi_{x}:=|x\rangle\langle x|$ of $\hat{H}_I^{(n)}$,
where $P_{h}$ is the projection to the eigenspace of the Hamiltonian 
$\hat{H}_I^{(n)}$ with eigenvalue $h$.
For $\mathscr{W}_{n}$, we give the bi-stochastic matrix $\T_{\mathscr{W}_{n}}$ as follows;
\begin{align}
\T_{\mathscr{W}_{n}}(x',y'|x,y)=\langle x',y'|{\cal E}_{j}(|x,y\rangle \langle x,y|) |x',y'\rangle.
\end{align}
Then, we have
\begin{align}
W^{(n)}_{\lnot}(\beta_{H},\beta_{L},\mathscr{W}_{n})
&=\sum_{j} w_j \Tr {\cal E}_{j}(\rho_I)
=\Tr[(\rho^{(n)}_{\beta_{H}\beta_{L}}-\sum_{j}{\cal E}_{j}(\rho^{(n)}_{\beta_{H}\beta_{L}}))\hat{H}^{(n)}_{HL}] \nonumber\\
&=\sum_{x',y'}\langle x',y'|  (\rho_{\beta_{H}\beta_{L}}-\sum_{j}{\cal E}_{j}(\rho^{(n)}_{\beta_{H}\beta_{L}}))\hat{H}^{(n)}_{HL}  |x',y'\rangle
\nonumber\\
&=\sum_{x',y',x,y}(P^{(n)}_{\beta_{H}\beta_{L}}(x',y')-\T_{\mathscr{W}_{n}}(x',y'|x,y)P^{(n)}_{\beta_{H}\beta_{L}}(x,y))(h^{(n)}_{X}(x')+h^{(n)}_{Y}(y'))\nonumber\\
&=W^{(n)}_{\lnot}(\beta_{H},\beta_{L},\T_{\mathscr{W}_{n}})\nonumber\\
Q^{(n)}_{H\lnot}(\beta_{H},\beta_{L},\mathscr{W}_{n})
&=\Tr[(\rho^{(n)}_{\beta_{H}\beta_{L}}-\sum_{j}{\cal E}_{j}(\rho^{(n)}_{\beta_{H}\beta_{L}}))\hat{H}^{(n)}_{H}] 
=\sum_{x',y'}\langle x',y'|  (\rho_{\beta_{H}\beta_{L}}-\sum_{j}{\cal E}_{j}(\rho^{(n)}_{\beta_{H}\beta_{L}}))\hat{H}^{(n)}_{H}  |x',y'\rangle
\nonumber\\
&=\sum_{x',y',x,y}(P^{(n)}_{\beta_{H}\beta_{L}}(x',y')-\T_{\mathscr{W}_{n}}(x',y'|x,y)P^{(n)}_{\beta_{H}\beta_{L}}(x,y))(h^{(n)}_{X}(x'))\nonumber\\
&=Q^{(n)}_{H\lnot}(\beta_{H},\beta_{L},\T_{\mathscr{W}_{n}}).
\end{align}
Therefore,
we obtain $\eta^{(n)}_{Q\lnot}[\beta_{H},\beta_{L},Q_n]
\le\eta^{(n)}_{C\lnot}[\beta_{H},\beta_{L},Q_n]$, which implies \eqref{AC=Q1}.
\end{proofof}

\section{Asymptotic behavior of thermodynamic efficiency}\Label{AApC}
In the present section, 
using Lemma \ref{A8-18-1},
we prove Proposition 3 in the main text.
At first, we write down Proposition 3 again;
\begin{proposition}\Label{A8-18-2}
Let $\{Q_{n}\}^{\infty}_{n=1}$ be arbitrary positive real numbers which satisfy
$\lim_{n\rightarrow\infty}Q_{n}/n=0$.
When the thermodynamic baths $R_{H}$ and $R_{L}$ satisfy \eqref{A4-3-1eq},
the thermodynamic optimal efficiency $\eta^{(n)}_{T}$ is asymptotically calculated as
\begin{align}
\eta^{(n)}_{T}[\beta_{H},\beta_{L},Q_{n}]
=1-\frac{\beta_{H}}{\beta_{L}}-\sum^{2}_{k=1}c^{(k)}_{\beta_{H},\beta_{L}}\frac{Q^{k}_{n}}{n^{k}}+O\left(\frac{Q^{3}_{n}}{n^3}\right),\Label{AB1}
\end{align}
where $c^{(1)}_{\beta_{H},\beta_{L}}$ 
and $c^{(2)}_{\beta_{H},\beta_{L}}$ are given by
\begin{align}
c^{(1)}_{\beta_H,\beta_L}
:=&\left(\frac{1}{2\beta^2_{H}\sigma^{2}_{\hat{H}_{1}}(\beta_{H})}+\frac{1}{2\beta^2_{L}\sigma^{2}_{\hat{H}_{2}}(\beta_{L})}\right)\frac{\beta^{2}_{H}}{\beta_{L}},\Label{Ac1} \\
c^{(2)}_{\beta_H,\beta_L}
:=&
\biggl(-\frac{\gamma_{\hat{H}_{1}}(\beta_H)}{6 \beta_H^3 \sigma^{3}_{\hat{H}_{1}}(\beta_H)}
+\frac{\gamma_{\hat{H}_{2}}(\beta_L)}{6 \beta_L^3 \sigma^{3}_{\hat{H}_{2}}(\beta_L)}
+\frac{1}{2 \beta_L^4 \sigma^{4}_{\hat{H}_{2}}(\beta_L)}+\frac{1}{2 \beta_H^2 \beta_L^2 \sigma^{2}_{\hat{H}_{1}}(\beta_H)\sigma^{2}_{\hat{H}_{2}}(\beta_L)}\biggl)\frac{\beta^{3}_{H}}{\beta_{L}}.\Label{Ac2}
\end{align}
\end{proposition}

\subsection{Preliminary: Relation with cumulant generating function}\Label{As2'}
In the present subsection, we prepare several useful notations 
based on the cumulant generating function for latter discussions.
Let us introduce the cumulant generating functions
$\phi_A(1+s):= \log \Tr \rho_{\beta_{\tilde{A}}|\hat{H}_A}^{1+s}$ for $A=H,L$
and 
$\phi_A(1+s):= \log \sum_{a}P_{\beta_{\tilde{A}}|\hat{H}_A}(a)^{1+s}$ for $A=X,Y$.
Since the classical case can be discussed in the same way as the quantum case,
we treat only the quantum case in this section.
The corresponding notation in the classical case will be applied in 
Appendices 
\ref{Aproof32} and \ref{As14}.
Then, the variance $\sigma^{2}_{A}(\beta_{\tilde{A}})$ and the skewness $\gamma_{A}(\beta_{\tilde{A}})$ of energy 
satisfy
\begin{align}
\phi_A''(1)&= \beta_{\tilde{A}}^2  \sigma^{2}_{A}(\beta_{\tilde{A}}) \\
\phi_A'''(1)&= \beta_{\tilde{A}}^3  \gamma_{A}(\beta_{\tilde{A}}) \sigma^{3}_{A}(\beta_{\tilde{A}}).
\end{align}
The relative entropy and the entropy can be written by using 
the cumulant generating function, i.e., 
they are written as \cite{AN}
\begin{align}
D(\rho_{\beta_{\tilde{A}} (1+s)|\hat{H}_A}\|\rho_{\beta_{\tilde{A}}|\hat{H}_A})
&=
\phi_A'(1+s)s- \phi_A(1+s) \\
S(\rho_{\beta_{\tilde{A}} (1+s)|\hat{H}_A})
&=-(1+s) \phi_A'(1+s)+ \phi_A(1+s)
= - \phi_A'(1+s)- 
D(\rho_{\beta_{\tilde{A}} (1+s)|\hat{H}_A} \| \rho_{\beta_{\tilde{A}}|\hat{H}_A}).
\end{align}
In the following, we focus on the entropy $S(\rho_{\beta_{\tilde{A}} |\hat{H}_A}) $, which is denoted by $S_A$.

In the following section, we need to discuss the above two quantities.
To handle them effectively,
we introduce a method to characterize them by a function with a single input variable. 
Since $\phi_A$ is strictly convex, 
we can define the inverse function of $\phi_A'$, which is denoted by $\psi_A$.
The function $\psi_A$ is characterized as follows.
\begin{align}
\psi'_A(-S_A)
&=\phi_A''(1)^{-1}
=\frac{1}{\beta_{\tilde{A}}^2 \sigma^{2}_{A}(\beta_{\tilde{A}})} \Label{A6-6-14}\\
\psi''_A(-S_A)
&=-\phi_A'''(1) \phi_A''(1)^{-3}
=-\frac{\gamma_{A}(\beta_{\tilde{A}})}{\beta_{\tilde{A}}^3 \sigma^{3}_{A}(\beta_{\tilde{A}})} .\Label{A6-6-15}
\end{align}
When $s$ is close to $0$, 
$\alpha:=\phi_A'(1+s)$ is also close to $-S_A$.
Hence, using the function $\psi_A$, we can characterize the relative entropy as
\begin{align}
D(\rho_{\beta_{\tilde{A}} (1+s)|\hat{H}_A} \|\rho_{\beta_{\tilde{A}} |\hat{H}_A} )
=&\phi_A'(1+s)s- \phi_A(1+s)
=\phi_A'(1+s)(\psi_A(\alpha) -1)- \phi_A(\psi_A(\alpha)) \nonumber \\
=&
(\psi_A(-S_A) -1) (\alpha+S_A) 
+\frac{1}{2}\psi_A'(-S_A) (\alpha+S_A)^2
+\frac{1}{6}\psi_A''(-S_A) (\alpha+S_A)^3+O((\alpha+S_A)^4) \nonumber \\
=&
\frac{1}{2}\psi_A'(-S_A) (\alpha+S_A)^2
+\frac{1}{6}\psi_A''(-S_A) (\alpha+S_A)^3+O((\alpha+S_A)^4) .
\end{align}
The entropy is characterized as 
\begin{align}
S(\rho_{\beta_{\tilde{A}} (1+s)|\hat{H}_A})
= - \phi_A'(1+s)- D(\rho_{\beta_{\tilde{A}} (1+s)|\hat{H}_A} \|\rho_{\beta_{\tilde{A}} |\hat{H}_A} )
= -\alpha  -\frac{1}{2}\psi_A'(-S_A) (\alpha+S_A)^2+O((\alpha+S_A)^3) .
\end{align}
Next, we focus on the difference 
$\Delta_A S:=S(\rho_{\beta_{\tilde{A}} (1+s)|\hat{H}_A})-S_A$ 
from the entropy of the state $\rho_{\beta_{\tilde{A}}|\hat{H}_A}$.
This difference is characterizes as
\begin{align}
\Delta_A S
= -(\alpha+S_A)  -\frac{1}{2}\psi_A'(-S_A) (\alpha+S_A)^2+O((\alpha+S_A)^3) ,
\end{align}
which implies
\begin{align}
-(\alpha+S_A)
=\Delta_A S
 +\frac{1}{2}\psi_A'(-S_A) (\Delta_A S)^2+O((\Delta_A S)^3) .
\end{align}
Thus, the relative entropy is characterized by the difference $\Delta_A S$ as follows.
\begin{align}
D(\rho_{\beta_{\tilde{A}} (1+s)|\hat{H}_A} \|\rho_{\beta_{\tilde{A}}|\hat{H}_A} )
=&
\frac{1}{2}\psi_A'(-S_A) (\Delta_A S)^2
+(\frac{1}{2}\psi_A'(-S_A)^2-\frac{1}{6}\psi_A''(-S_A) )
(\Delta_A S)^3+O((\Delta_A S)^4) \nonumber  \\
=&
\frac{1}{2\beta_{\tilde{A}}^2 \sigma^{2}_{A}(\beta_{\tilde{A}})} 
(\Delta_A S)^2
+
(\frac{1}{2\beta_{\tilde{A}}^4 \sigma^{4}_{A}(\beta_{\tilde{A}})} 
+
\frac{\gamma_{A}(\beta_{\tilde{A}})}{6\beta_{\tilde{A}}^3 \sigma^{3}_{A}(\beta_{\tilde{A}})} 
)
(\Delta_A S)^3+O((\Delta_A S)^4) .\Label{A6-6-3}
\end{align}
This relation will be used in the following sections.

\subsection{Asymptotic expansion of thermodynamic efficiency}\Label{As12} 
Now, let us derive the asymptotic expansion \eqref{AB1}
in Proposition 2 by using Lemma \ref{A8-18-1}.
Because now we assume that the heat baths are composed of identical and uncorrelated particles, the Gibbs states of the baths $H$ and $L$ are given as 
$\rho_{\beta_{H}|\hat{H}_{1} }^{\otimes n}$ and $\rho_{\beta_{L}|\hat{H}_{2}}^{\otimes n}$.
Therefore, with using $D(\rho^{\otimes}\|\sigma^{\otimes n})=nD(\rho\|\sigma)$, we can transform the equality \eqref{ATandD} in Lemma \ref{A8-18-1} into 
\begin{align} 
\eta_{T}[\beta_{H},\beta_{L},Q_{n}]
=
\left(1-\frac{\beta_{H}}{\beta_{L}}\right)
- \frac{n(D(\rho_{\beta'_{H}|\hat{H}_{H}}\|\rho_{\beta_{H}|\hat{H}_{H}})+
D(\rho_{\beta'_{L}|\hat{H}_{L}}\|\rho_{\beta_{L}|\hat{H}_{L}}))}
{\beta_{L}Q_n} ,\Label{Aarxupperboundz}
\end{align}
where the real numbers $\beta'_{H}$ and $\beta'_{L}$ are determined by
\begin{align}
Q_n
&=\Tr (\rho_{\beta_{H}|\hat{H}_{H}}^{\otimes n}
-\rho_{\beta'_{H}|\hat{H}_{H}}^{\otimes n}) \hat{H}_{H}^{(n)}
\\
S(\rho_{\beta'_{H}|\hat{H}_{H}}^{\otimes n} )
+S(\rho_{\beta'_{L}|\hat{H}_{L}}^{\otimes n})
&=
S(\rho_{\beta_{H}|\hat{H}_{H}}^{\otimes n} )
+S(\rho_{\beta_{L}|\hat{H}_{L}}^{\otimes n} ),
\end{align}
which are equivalent with
\begin{align}
\frac{Q_n}{n}
&= \Tr (\rho_{\beta_{H}|\hat{H}_{H}}-\rho_{\beta'_{H}|\hat{H}_{H}}) \hat{H}_{H}
\Label{Aarxseigen1z}
\\
S(\rho_{\beta'_{H}|\hat{H}_{H}})+ S(\rho_{\beta'_{L}|\hat{H}_{L}})
&= S(\rho_{\beta_{H}|\hat{H}_{H}})+ S(\rho_{\beta_{L}|\hat{H}_{L}}).
\Label{Aarxseigen2z}
\end{align}
Then, \eqref{Aarxseigen1z} implies that
\begin{align}
-\frac{\beta_H Q_H}{n}
= \Delta_H S + D(\rho_{\beta'_H|\hat{H}_{H}}\|\rho_{\beta_H|\hat{H}_{H}}) \Label{A4-8-11eq}.
\end{align}
Now, we employ the differences of entropies 
$\Delta_H S=S(\rho_{\beta_H'|\hat{H}_H})-S_H$ and
$\Delta_L S=S(\rho_{\beta_L'|\hat{H}_L})-S_L$,
which are introduced in Appendix \ref{As2'}.
Then, 
\eqref{A6-6-3} and \eqref{Aarxseigen2z} yield that
\begin{align}
\Delta_H S + D(\rho_{\beta'_H|\hat{H}_{H}}\|\rho_{\beta_H|\hat{H}_{H}})
&= \Delta_H S 
+ \frac{1}{2\beta_H^2 \sigma^2(\beta_H)} (\Delta_H S)^2
+O((\Delta_H S)^3) \Label{A4-8-10eq}\\
\Delta_L S 
&=-\Delta_H S .
\end{align}
Thus, \eqref{A4-8-11eq} and \eqref{A4-8-10eq} imply
\begin{align}
\Delta_H S 
=
-\frac{\beta_H Q_H}{n}
-\frac{1}{2\beta_H^2 \sigma^2(\beta_H)} (\frac{\beta_H Q_H}{n})^2
+O((\frac{\beta_H Q_H}{n})^3) ,
\end{align}
which yields that
\begin{align}
\Delta_H S ^2
=&
(\frac{\beta_H Q_H}{n})^2
+\frac{1}{\beta_H^2 \sigma^2(\beta_H)} (\frac{\beta_H Q_H}{n})^3
+O((\frac{\beta_H Q_H}{n})^4)  \\
\Delta_H S ^3
=&
-(\frac{\beta_H Q_H}{n})^3
+O((\frac{\beta_H Q_H}{n})^4).
\end{align}
Therefore,
\begin{align}
&n(D(\rho_{\beta'_{H}|\hat{H}_{H}}\|\rho_{\beta_{H}|\hat{H}_{H}})+D(\rho_{\beta'_{L}|\hat{H}_{L}}\|\rho_{\beta_{L}|\hat{H}_{L}})) 
\nonumber\\
=&
\frac{1}{2\beta_H^2 \sigma^2(\beta_H)} 
(\Delta_H S)^2
+
(\frac{1}{2\beta_H^4 \sigma^4(\beta_H)} 
+
\frac{\gamma_1(\beta_H)}{6\beta_H^3 \sigma^3(\beta_H)} 
)
(\Delta_H S)^3 \nonumber \\
&+
\frac{1}{2\beta_L^2 \sigma^2(\beta_L)} 
(\Delta_L S)^2
+
(\frac{1}{2\beta_L^4 \sigma^4(\beta_L)} 
+
\frac{\gamma_1(\beta_L)}{6\beta_L^3 \sigma^3(\beta_L)} 
)
(\Delta_L S)^3+O((\Delta_H S)^4) \nonumber \\
=&
(\frac{1}{2\beta_H^2 \sigma^2(\beta_H)} 
+\frac{1}{2\beta_L^2 \sigma^2(\beta_L)} )
(\Delta_H S)^2
+
(\frac{1}{2\beta_H^4 \sigma^4(\beta_H)} 
-\frac{1}{2\beta_L^4 \sigma^4(\beta_L)} 
+\frac{\gamma_1(\beta_H)}{6\beta_H^3 \sigma^3(\beta_H)} 
-\frac{\gamma_1(\beta_L)}{6\beta_L^3 \sigma^3(\beta_L)} 
)
(\Delta_H S)^3+O((\Delta_H S)^4) 
\nonumber \\
=&
(\frac{1}{2\beta_H^2 \sigma^2(\beta_H)} 
+\frac{1}{2\beta_L^2 \sigma^2(\beta_L)} )
(\frac{\beta_H Q_H}{n})^2 \nonumber \\
&+
(-\frac{\gamma_1(\beta_H)}{6\beta_H^3 \sigma^3(\beta_H)} 
+\frac{\gamma_1(\beta_L)}{6\beta_L^3 \sigma^3(\beta_L)} 
+\frac{1}{2\beta_H^2 \beta_L^2 \sigma^2(\beta_H)\sigma^2(\beta_L)} 
+\frac{1}{2\beta_L^4 \sigma^4(\beta_L)} 
)
(\frac{\beta_H Q_H}{n})^3+O((\Delta_H S)^4) .\Label{A6-5-9}
\end{align}
Therefore, substituting \eqref{A6-5-9} into \eqref{Aarxupperboundz}, we obtain
\eqref{AB1} in Proposition 2.



\section{Asymptotic expansions of $\eta^{(n)}_{C\lnot}(\beta_{H},\beta_{L},\T^{\mathrm{opt}}_{Q_n})$ and $Q^{(n)}_{H\lnot}(\beta_{H},\beta_{L},\T^{\mathrm{opt}}_{Q_n})$}\Label{Aproof32}
In the present section and the next section, 
we show Proposition \ref{A3-14-2T} in the main text.
As we have mentioned in the main text, hereafter, we treat the case that the baths ${\cal X}$, ${\cal Y}$, $H$ and $L$ are the composed of $n$ uncorrelated identical particle, i.e., the following equalities hold;
\begin{eqnarray}
&h^{(n)}_{X}(x)=\sum^{n}_{k=1}h_{1}(x_{k}),\enskip h^{(n)}_{Y}(y)=\sum^{n}_{k=1}h_{1}(y_{k}),\Label{Aid1}\\
&\hat{H}^{(n)}_{H}=\sum^{n}_{k=1}\hat{H}^{[k]}_{H},\enskip
\hat{H}^{(n)}_{L}=\sum^{n}_{k=1}\hat{H}^{[k]}_{L},\Label{Aid2}
\end{eqnarray}
where $\hat{H}^{[k]}_{H}=\hat{H}_{1}:=\sum^{d}_{s=1}h_{1}(s)|s\rangle\langle s|$ and $\hat{H}^{[k]}_{L}=\hat{H}_{2}:=\sum^{d}_{s=1}h_{2}(s)|s\rangle\langle s|$ are the Hamiltonians of the $k$th particle of $H$ and $L$, respectively.

Let us write down Proposition \ref{A3-14-2T} and Protocol \ref{Aprotocol1} again;
\begin{proposition}\Label{A3-14-2T}
Assume that the baths ${\cal X}$ and ${\cal Y}$ are composed of $n$ uncorrelated identical particles, i.e., the equality \eqref{Aid1} holds. 
Then, the classical work extraction $\T^{\mathrm{opt}}_{Q_n}$ given in Protocol \ref{Aprotocol1} satisfies
\begin{align} 
\eta^{(n)}_{C\lnot}(\beta_{H},\beta_{L},\T^{\mathrm{opt}}_{Q_n})
=
1-\frac{\beta_{H}}{\beta_{L}}
-
c^{(1)}_{\beta_{H},\beta_{L}} \frac{Q_{n}}{n}
+O(\frac{1}{n})
+O(\frac{Q_{n}^2}{n^2})\Label{Aarxsecondorder2x} ,
\end{align}
and
\begin{align} 
Q^{(n)}_{H\lnot}(\beta_{H},\beta_{L},\T^{\mathrm{opt}}_{Q_n})
=Q_{n}+ o(\frac{Q^{2}_{n}}{n^3}).\Label{Aarxsecondorder2.2x}
\end{align}

Further,
when $h_X$ and $h_Y$ are non-lattice,
i.e., when there does not exist a positive number $t$
such that $\{h(z)-h(z')\}_{z,z'} \subset t\mathbb{Z}$ for $h^{(n)}_{X}(x)$ nor $h^{(n)}_{Y}(y)$, 
the classical work extraction $\T^{\mathrm{opt}}_{Q_n}$ has more detailed asymptotic expansions as
\begin{align} 
\eta^{(n)}_{C\lnot}(\beta_{H},\beta_{L},\T^{\mathrm{opt}}_{Q_n})
=
1-\frac{\beta_{H}}{\beta_{L}}
-\sum^{2}_{k=1}
c^{(k)}_{\beta_{H},\beta_{L}} \frac{Q^{k}_{n}}{n^{k}}
-
d^{(1)}_{\beta_{H},\beta_{L}} \frac{Q_{n}}{n^2}
+O\left(\frac{Q^2_{n}}{n^{5/2}}\right)
+O\left(\frac{Q^{3}_{n}}{n^{3}}\right)\Label{Aarxsecondorder2} ,
\end{align}
and
\begin{align} 
Q^{(n)}_{H\lnot}(\beta_{H},\beta_{L},\T^{\mathrm{opt}}_{Q_n})
=Q_{n}+ O(\frac{Q^{3}_{n}}{n^4}),\Label{Aarxsecondorder2.2}
\end{align}
where  
\begin{align}
d^{(1)}_{\beta_{H},\beta_{L}}
:=
\Biggl((\frac{\gamma_1(\beta_{H})}{2 \beta_{H} \sigma(\beta_{H})}
+\frac{1}{\beta_{H}^2 \sigma^2(\beta_{H})})^2
+(\frac{\gamma_1(\beta_{L})}{2 \beta_{L} \sigma(\beta_{L})}
+\frac{1}{\beta_{L}^2 \sigma^2(\beta_{L})})^2
\Biggr)\frac{\beta^2_{X}}{\beta_{L}}.\Label{Adbeta}
\end{align}
\end{proposition}

\begin{Protocol}[t]                  
\caption{Classical optimal work extraction}         
\Label{Aprotocol1}      
\begin{algorithmic}
\LECTURE  We consider the combination of the following three steps $g_{1n}$, $g_{2n}$, and $g_{1n}^{-1}$, i.e., 
$f_n := g_{1n}^{-1} \circ g_{2n}\circ g_{1n}$.
Hence, the classical optimal work extraction 
${\cal T}^{opt}_{Q_n}$
is given as the following deterministic and invertible dynamics;
\begin{align}
\T^{\mathrm{opt}}_{Q_n}(x,y|x',y'):=\delta_{(x,y),f_{n}(x',y')}.\Label{ATopt} 
\end{align}             
\STEPONE  First, we convert the elements $x$ and $y$ in ${\cal X}$ and ${\cal Y}$, which are described as the pair of integers $(i,j)$ in the set $\mathbb{Z}_{d^n}^2$,
by using the following two functions $g_{1nX}$ and $g_{1nY}$ 
from $\mathbb{Z}_{d}^n$ to $\mathbb{Z}_{d^n}$
as 
\begin{align}
P^{n}_{\beta_{H}|h^{(n)}_{X}}(x)
=P^{n\downarrow}_{\beta_{H}|h^{(n)}_{X}}(g_{1nX}(x)),
\enskip
P^{n}_{\beta_{L}|h^{(n)}_{Y}}(y)
=P^{n\downarrow}_{\beta_{L}|h^{(n)}_{Y}}(g_{1nY}(y))
\Label{Ag1ndef},
\end{align}
where $P^{n\downarrow}_{\beta_{H}|h^{(n)}_{X}}$ and $P^{n\downarrow}_{\beta_{L}|h^{(n)}_{Y}}$ are the descending reordered distributions of the distributions
$P^{n}_{\beta_{H}|h^{(n)}_{X}}$ and $P^{n}_{\beta_{L}|h^{(n)}_{Y}}$,
respectively. 
That is, we apply the function $g_{1n}(x,y):=(g_{1nX}(x),g_{1nY}(y))$
to the classical system ${\cal X}\times{\cal Y}$.
\STEPTWO Next, we apply the function $g_{2n}$ defined as
\begin{align}
g_{2n}:&
(i_A d^{m_n} + j_B ,i_B d^{n-m_n}+ j_A )
\mapsto 
(i_A +i_B d^{m_n},j_A +j_B d^{n-m_n}) 
\Label{Ag2ndef}
\end{align}
for $i_A , j_B\in \mathbb{Z}_{d^{m_n}} $ and $i_B , j_A\in \mathbb{Z}_{d^{n-m_n}} $,
where
\begin{align}
m_{n}&:=\left\lfloor 
\frac{\beta_{H} Q_{n}
+ \frac{Q_{n}^2}{2 n \sigma_X^2(\beta_{H})}}{\log d}
\right\rfloor.
\end{align}
\STEPTHREE Finally, we apply the inverse function $g_{1n}^{-1}$.
\end{algorithmic}
\end{Protocol}

In this section, 
we derive the asymptotic expansions \eqref{Aarxsecondorder2x}--\eqref{Aarxsecondorder2.2} by using Lemma \ref{Al5}, which is given in this section and will be shown in the next section.
The start point is the following equality;
\begin{align} 
\eta^{(n)}_{C\lnot}(\beta_{H},\beta_{L},\T^{\mathrm{opt}}_{Q_n})
&=
\left(1-\frac{\beta_{H}}{\beta_{L}}\right)
-
\frac{D(
f_{n*}(P^{(n)}_{\beta_{H}\beta_{L}})
\|
P^{(n)}_{\beta_{H}\beta_{L}})}{\beta_{L}Q^{(n)}_{H\lnot}(\beta_{H},\beta_{L},\T^{\mathrm{opt}}_{Q_n})},\Label{Atoushiki2}
\end{align}
where for an arbitrary invertible function $f$ we give
\begin{align}
f_{*}(P)(x)=P(f^{-1}(x)).\Label{A4-19-4eq}
\end{align}
Notice the relation 
\begin{align}
(g\circ f)_{*}(P)=g_{*}(f_{*}(P)).\Label{A4-19-5eq}
\end{align}

\begin{proofof}{\eqref{Atoushiki2}}
Because $f_{n*}$ is invertible and deterministic, 
it preserves the entropy; 
\begin{align}
S(f_{n*}(P^{(n)}_{\beta_{H}\beta_{L}})
)
=S(P^{(n)}_{\beta_{H}\beta_{L}}).\Label{A4-7-1eq}
\end{align}
Thus, similarly as the derivation of \eqref{AQT0pre}, 
using the relation \eqref{A4-7-1eq}, we have 
\begin{eqnarray}
-D(
f_{n*}(P^{(n)}_{\beta_{H}\beta_{L}})
\|
P^{(n)}_{\beta_{H}\beta_{L}})
&=&S(f_{n*}(P^{(n)}_{\beta_{H}\beta_{L}}))
+\sum_{x,y}[P^{(n)}_{\beta_{H}\beta_{L}}
(f_{n}^{-1}(x,y))
\log P^{(n)}_{\beta_{H}\beta_{L}}(x,y)]
\nonumber\\
&=&S(P^{(n)}_{\beta_{H}\beta_{L}})
+\sum_{x,y}[P^{(n)}_{\beta_{H}\beta_{L}}(f_{n}^{-1}(x,y))
\log P^{(n)}_{\beta_{H}\beta_{L}}(x,y)]
\nonumber\\
&=&-\sum_{x}[(P_{\beta_{H}|h_X}^{n}(x)-\sum_{y}P_{\beta_{H}|h_X}^{n}
P_{\beta_{L}|h_Y}^{n}(f_{n}^{-1}(x,y))
\log P_{\beta_{H}|h_X}^{n}(x))]
\nonumber\\
&&-\sum_{y}[(P_{\beta_{L}|h_Y}^{n}(y)-\sum_{x}P^{(n)}_{\beta_{H}\beta_{L}}(f_{n}^{-1}(x,y))
\log P_{\beta_{L}|h_y}^{n}(y))]\nonumber\\
&=&\beta_{H}\sum_{x}[(P_{\beta_{H}|h_X}^{n}(x)-\sum_{y}P^{(n)}_{\beta_{H}\beta_{L}}(g_{2n}^{-1}(x,y)))h_{X}(x)] \nonumber\\
&&+\beta_{L}\sum_{y}[(P_{\beta_{H}|h_Y}^{n}(y)-\sum_{x}P^{(n)}_{\beta_{H}\beta_{L}}(g_{2n}^{-1}(x,y)))h_{Y}(y)]\nonumber\\
&=&(\beta_{H}-
\beta_{L})Q^{(n)}_{H\lnot}(\beta_{H},\beta_{L},\T^{\mathrm{opt}}_{Q_n})+\beta_{L}W^{(n)}_{\lnot}(\beta_{H},\beta_{L},\T^{\mathrm{opt}}_{Q_n}).
\end{eqnarray}
Therefore, we obtain \eqref{Atoushiki2}.
\end{proofof}

Due to the equality \eqref{Atoushiki2}, in order to obtain the asymptotic expansion of $\eta^{(n)}_{C\lnot}(\beta_{H},\beta_{L},\T^{\mathrm{opt}}_{Q_n})$, it is enough to asymptotically expand the relative entropy 
$D(
f_{n*}(P^{(n)}_{\beta_{H}\beta_{L}})
\|P^{(n)}_{\beta_{H}\beta_{L}})$ 
and the endothermic amount $Q^{(n)}_{H\lnot}(\beta_{H},\beta_{L},\T^{\mathrm{opt}}_{Q_n})$.
We do not calculate these two asymptotic expansions directly, but we calculate them with using an approximation method.
To be concrete, we show the following lemma;
\begin{lemma}\Label{Al4}
When the sequence $\{m_{n}\}$ satisfies
\begin{equation}
m_{n}< n 
\left(
\min\left\{1-
\frac{S(P_{\beta_{L}|h_Y})}{\log d},
\enskip-
\frac{\log \max_{x} P_{\beta_{H}|h_X}(x)}{\log d}
\right\}
-\epsilon\right)
\Label{Akairyou2}
\end{equation}
with a small $\epsilon>0$,
the following equalities satisfies;
\begin{eqnarray}
D(
f_{n*}(P^{(n)}_{\beta_{H}\beta_{L}})
\| P^{(n)}_{\beta_{H}\beta_{L}})
&=&D_{X}^n(m_{n})+D_{Y}^n(m_{n})+O(e^{-n\alpha_{2}})\Label{Akinkin}\\
\beta_{H}Q^{(n)}_{H\lnot}(\beta_{H},\beta_{L},\T^{\mathrm{opt}}_{Q_n})
&=&m_{n}\log d-D_{X}^n(m_{n})+O(e^{-n\alpha_{3}})\Label{Anobu}
\end{eqnarray}
hold with real constants $\alpha_{2}>0$ and $\alpha_{3}>0$, 
where
\begin{align}
D_{X}^n(m_{n})
&:=
\sum^{d^{n}-1}_{i=0}
P_{\beta_{H}|h_X}^{n\downarrow}(i)
\log \frac{d^{m_n} P_{\beta_{H}|h_X}^{n\downarrow}(i)}
{P_{\beta_{H}|h_X}^{n\downarrow}(\lceil d^{-m_{n}} i\rceil)}
\Label{Akinji}\\
D_{Y}^n(m_{n})&:=
\sum^{d^{n}-1}_{j=0}
P_{\beta_{L}|h_Y}^{n\downarrow}(j)
\log \frac{ P_{\beta_{L}|h_Y}^{n\downarrow}(j)}
{d^{m_n} P_{\beta_{L}|h_Y}^{n\downarrow}(d^{m_{n}} j)}
\Label{Akinji2} ,
\end{align}
where $\lceil a\rceil$ is the ceiling function of a positive number $a$. 
\end{lemma}

Due to Lemma \ref{Al4}, in order to calculate $D(
f_{n*}(P^{(n)}_{\beta_{H}\beta_{L}})
\| P^{(n)}_{\beta_{H}\beta_{L}})$ 
and  $Q^{(n)}_{H\lnot}(\beta_{H},\beta_{L},\T^{\mathrm{opt}}_{Q_n})$, 
we only have to calculate $D_{X}^n(m_{n})$ and $D_{Y}^n(m_{n})$.
Lemma \ref{Al4} will be shown after the following Lemma \ref{AL-H}.

Now, we simulate the distribution
$g_{2n*}(
P_{\beta_{H}|h_X}^{n\downarrow}
P_{\beta_{L}|h_Y}^{n\downarrow})$ by a product distribution,
where $g_{2n}$ is defined in \eqref{Ag2ndef}.
For this purpose,
we introduce the two distributions
$\tilde{P}_{\beta_{H}|h_X}^{n\downarrow}$ and $\tilde{P}_{\beta_{L}|h_Y}^{n\downarrow}$ by
\begin{eqnarray}
\tilde{P}_{\beta_{H}|h_X}^{n\downarrow}(i_A+ i_B d^{m_n})
&:=&d^{m_{n}}P^{n\downarrow}_{\beta_{H}|h_{X}}(i_{A}d^{m_{n}})\delta_{i_{B},0},\Label{AD11}\\
\tilde{P}_{\beta_{L}|h_Y}^{n\downarrow}(j_A+j_B d^{n-m_n})
&:=&d^{-m_{n}}P^{n\downarrow}_{\beta_{L}|h_{Y}}(j_{A})\Label{AD12}
\end{eqnarray}
for $i_A , j_B\in \mathbb{Z}_{d^{m_n}} $ and $i_B , j_A\in \mathbb{Z}_{d^{n-m_n}} $.
As shown in Lemma \ref{AL-H}, the product 
$\tilde{P}_{\beta_{H}|h_X}^{n\downarrow}
\tilde{P}_{\beta_{L}|h_Y}^{n\downarrow}$ 
approximates the true distribution
$g_{2n*}(
P_{\beta_{H}|h_X}^{n\downarrow}
P_{\beta_{L}|h_Y}^{n\downarrow})$.

\begin{lemma}\Label{AL-H}
When a sequence $\{m_{n}\}$ satisfies \eqref{Akairyou2} with a small $\epsilon>0$,
the relation
\begin{eqnarray}
\|
g_{2n*}(
P_{\beta_{H}|h_X}^{n\downarrow}P_{\beta_{L}|h_Y}^{n\downarrow})
-\tilde{P}_{\beta_{H}|h_X}^{n\downarrow} \tilde{P}_{\beta_{L}|h_Y}^{n\downarrow} 
\|_{1}
=O(e^{-n\alpha_{1}})
\Label{A(A)}
\end{eqnarray}
holds, where 
\begin{eqnarray}
\|P-Q\|_{1}&:=&\sum_{j}|P(j)-Q(j)|.
\end{eqnarray}
\end{lemma}

\begin{proofof}{Lemma \ref{AL-H}}
Due to \eqref{A4-19-4eq} and \eqref{Ag2ndef},
we can evaluate the LHS of \eqref{A(A)} as
\begin{align}
&\|
g_{2n*}(P_{\beta_{H}|h_X}^{n\downarrow}P_{\beta_{L}|h_Y}^{n\downarrow})
-\tilde{P}_{\beta_{H}|h_X}^{n\downarrow} \tilde{P}_{\beta_{L}|h_Y}^{n\downarrow})
\|_{1}\nonumber\\
&=
\sum_{i_{A},i_{B},j_{A},j_{B}}
|g_{2n*}(P_{\beta_{H}|h_X}^{n\downarrow}P_{\beta_{L}|h_Y}^{n\downarrow})
(i_A +i_B d^{m_n},j_A +j_B d^{n-m_n}) 
-\tilde{P}_{\beta_{H}|h_X}^{n\downarrow} 
(i_A +i_B d^{m_n})
\tilde{P}_{\beta_{L}|h_Y}^{n\downarrow}
(j_A +j_B d^{n-m_n}) 
|\nonumber\\
&=
\sum_{i_{A},i_{B},j_{A},j_{B}}
|P_{\beta_{H}|h_X}^{n\downarrow}(i_{A}d^{m_{n}}+j_{B})P_{\beta_{L}|h_Y}^{n\downarrow}(i_{B}d^{n-m_{n}}+j_{A})
-P_{\beta_{H}|h_X}^{n\downarrow}(i_{A}d^{m_{n}}) P_{\beta_{L}|h_Y}^{n\downarrow}(j_{A})\delta_{i_{B},0}|\nonumber\\
&=-\sum_{i_{A},j_{A},j_{B}}
\left|
P_{\beta_{H}|h_X}^{n\downarrow}(i_{A}d^{m_{n}}+j_{B})
-P_{\beta_{H}|h_X}^{n\downarrow}(i_{A}d^{m_{n}}) 
\right|
P_{\beta_{L}|h_Y}^{n\downarrow}(j_{A})
\nonumber\\
&\enskip\enskip\enskip\enskip\enskip
+\sum^{d^{n-m_{n}}-1}_{i_{B}=1}\sum_{i_{A},j_{A},j_{B}}
P_{\beta_{H}|h_X}^{n\downarrow}(i_{A}d^{m_{n}}+j_{B})P_{\beta_{L}|h_Y}^{n\downarrow}(i_{B}d^{n-m_{n}}+j_{A})
\nonumber\\
&
\stackrel{(a)}{\le}
\sum^{d^{n-m_{n}}-1}_{i_{A}=0}\sum^{d^{m_{n}}-1}_{j_{B}=0}
\left(
P_{\beta_{H}|h_X}^{n\downarrow}(i_{A}d^{m_{n}})
-P_{\beta_{H}|h_X}^{n\downarrow}(i_{A}d^{m_{n}}+j_{B})
\right)
+\sum^{d^{n}-1}_{j'=d^{n-m_{n}}}\sum^{d^{n}-1}_{i'=0}
P_{\beta_{H}|h_X}^{n\downarrow}(i')P_{\beta_{L}|h_Y}^{n\downarrow}(j')\nonumber\\
&=
\sum^{d^{n}-1}_{j'=d^{n-m_{n}}}P_{\beta_{L}|h_Y}^{n\downarrow}(j')
+\sum^{d^{n-m_{n}}-1}_{i_{A}=0}\sum^{d^{m_{n}}-1}_{j_{B}=0}
\left(
P_{\beta_{H}|h_X}^{n\downarrow}(i_{A}d^{m_{n}})
-
P_{\beta_{H}|h_X}^{n\downarrow}(i_{A}d^{m_{n}}+j_{B})
\right),
\end{align}
where $i':=i_{A}d^{m_{n}}+j_{B}$ and $j':=i_{B}d^{n-m_{n}}+j_{A}$,
and $(a)$ follows from 
$\sum_{j_A}P_{\beta_{L}|h_Y}^{n\downarrow}(j_{A})\le 1$.
We notice that
the distribution $P_{\beta_{L}|h_Y}^{n\downarrow}(j')$
has the probability 
only from $j'=1$ to $j'=e^{n (S(P_{\beta_{L}|h_Y})+\epsilon)}$
except for exponentially small probability
when $m_{n}< n((1-\frac{S(
P_{\beta_{L}|h_Y})}{\log d})-\epsilon)$.
Since this condition holds for the choice of $m_n$,
the probability
$\sum^{d^n-1}_{j'=d^{n-m_{n}}} P_{\beta_{L}|h_Y}^{n\downarrow}(j')$
goes to zero exponentially.
Although the distribution $P_{\beta_{H}|h_X}^{n\downarrow}$ has $d^n$ events, the probabilities $P_{\beta_{H}|h_X}^{n\downarrow}(k)$
take at most $(n+1)^{d-1}$ distinct values due to the combinatorics.
That is, for an fixed integer $j_{B}$,
at most $(n+1)^{d-1}$ integers $i_{A}$ satisfy the following condition;
the difference 
$P_{\beta_{H}|h_X}^{n\downarrow}(i_{A}d^{m_{n}})
-P_{\beta_{H}|h_X}^{n\downarrow}(i_{A}d^{m_{n}}+j_{B})$
takes non-zero values. 
Therefore,
\begin{align}
\sum^{d^{n-m_{n}}-1}_{i_{A}=0}\sum^{d^{m_{n}}-1}_{j_{B}=0}
\left(
P_{\beta_{H}|h_X}^{n\downarrow}(i_{A}d^{m_{n}})
-
P_{\beta_{H}|h_X}^{n\downarrow}(i_{A}d^{m_{n}}+j_{B})
\right)
&\le
d^{m_{n}} (n+1)^{d-1}P_{\beta_{H}|h_X}^{n\downarrow}(i_{A}d^{m_{n}})
\nonumber\\
&=
d^{m_n} (n+1)^{d-1} (\max_x P_{\beta_{H}|h_{X}}(x))^n,\Label{Aeq43}
\end{align}
which goes to zero exponentially because 
$m_{n}<-n
\left(\frac{\log \max_{x} P_{\beta_{H}|h_{X}}(x)
}{\log d}
+\epsilon \right)$.
Thus, when \eqref{Akairyou2} holds, the equation \eqref{A(A)} is valid.
\end{proofof}

\begin{proofof}{Lemma \ref{Al4}}
We firstly transform 
$D(f_{n*}(
P^{(n)}_{\beta_{H}\beta_{L}})
\|P^{(n)}_{\beta_{H}\beta_{L}})$ 
and $\beta_{H}Q^{(n)}_{H\lnot}(\beta_{H},\beta_{L},\T^{\mathrm{opt}}_{Q_n})$ into forms that we can more easily compare with $D_{X}^n(m_{n})$ and $D_{Y}^n(m_{n})$.
Because $g_{1n}$ is invertible and deterministic, the relative entropy 
$D(f_{n*}(P^{(n)}_{\beta_{H}\beta_{L}})
\|P^{(n)}_{\beta_{H}\beta_{L}})$ 
turns into 
\begin{align}
D(f_{n*}(P^{(n)}_{\beta_{H}\beta_{L}})
\|P^{(n)}_{\beta_{H}\beta_{L}})
&=
D(g_{1n*}(f_{n*}(P^{(n)}_{\beta_{H}\beta_{L}}))
\|
g_{1n*}( P^{(n)}_{\beta_{H}\beta_{L}}))
\\
&=D(
g_{2n*}(
P_{\beta_{H}|h_X}^{n\downarrow}P_{\beta_{L}|h_Y}^{n\downarrow})
\|P_{\beta_{H}|h_X}^{n\downarrow}P_{\beta_{L}|h_Y}^{n\downarrow}).
\end{align}
Now, we introduce the marginal distributions:
\begin{align}
{P_X}''(x)&:=
\sum_y f_{n*}(P^{(n)}_{\beta_{H}\beta_{L}})(x,y) \\
{P_X}'(i)&:=
\sum_j g_{2n*}(
P_{\beta_{H}|h_X}^{n\downarrow}P_{\beta_{L}|h_Y}^{n\downarrow}
)(i,j) .
\end{align}
Hence, we have
\begin{align}
S({P_X}')
&=S(g_{1nX*}({P_X}'))=S({P_X}'') \Label{AStrans}\\
D({P_X}'\|P_{\beta_{H}|h_X}^{n} )
&=D(g_{1nX*}({P_X}')
\|g_{1nX*}(P_{\beta_{H}|h_X}^{n}))
=D({P_X}''\| P_{\beta_{H}|h_X}^{n\downarrow}) .
\Label{ADtrans}
\end{align}
With using \eqref{AStrans}, and \eqref{ADtrans}, 
we convert the quantity $\beta_{H}Q^{(n)}_{H\lnot}(\beta_{H},\beta_{L},\T^{\mathrm{opt}}_{Q_n})$ as follows;
\begin{align}
\beta_{H}Q^{(n)}_{H\lnot}(\beta_{H},\beta_{L},\T^{\mathrm{opt}}_{Q_n})
&=\beta_{H}\sum_{X} (P_{\beta_{H}|h_{X}}(x)-P'_{X}(x))h_{X}(x)
=-\sum_{X} (P_{\beta_{H}|h_{X}}(x)-P'_{X}(x))\log P_{\beta_{H}|h_{X}}(x)\Label{Ah2}\\
&=S(P_{\beta_{H}|h_X}^n)-S({P_X}'')
-D({P_X}''\| P_{\beta_{H}|h_X}^{n})
\nonumber\\
&=S(P_{\beta_{H}|h_X}^{n\downarrow})
-S({P_X}')
-D({P_X}'\|P_{\beta_{H}|h_X}^{n\downarrow}).
\end{align}

Second, we define
\begin{align}
\tilde{D}_{X}^n(m_{n}):=D(\tilde{P}_{\beta_{H}|h_X}^{n\downarrow}\|P^{n\downarrow}_{\beta_{H}|h_{X}})
&=\sum^{d^{n}-1}_{i=0}
\tilde{P}_{\beta_{H}|h_X}^{n\downarrow}(i)
\log \frac{\tilde{P}_{\beta_{H}|h_X}^{n\downarrow}(i)}{P^{n\downarrow}_{\beta_{H}|h_{X}}(i)}\nonumber\\
&=\sum^{d^{n-m_{n}}-1}_{i_A=0}
d^{m_{n}}P^{n\downarrow}_{\beta_{H}|h_{X}}(d^{m_{n}}i_A)
\log \frac{d^{m_{n}}P^{n\downarrow}_{\beta_{H}|h_{X}}(d^{m_{n}}i_A)}
{P^{n\downarrow}_{\beta_{H}|h_{X}}(i_A)}\nonumber\\
&
=\sum^{d^{n}-1}_{i=0}
 P^{n\downarrow}_{\beta_{H}|h_{X}}(d^{m_{n}} 
\lceil d^{-m_{n}} i \rceil)
\log \frac{d^{m_{n}}P^{n\downarrow}_{\beta_{H}|h_{X}}(d^{m_{n}} \lceil d^{-m_{n}} i \rceil)}
{P^{n\downarrow}_{\beta_{H}|h_{X}}(\lceil d^{-m_{n}} i \rceil)}.\enskip\enskip\enskip\enskip\enskip
\end{align}
Since 
$\sum_{j=i}^{d^{m_n}-1}
| P^{n\downarrow}_{\beta_{H}|h_{X}}(d^{m_{n}} 
\lceil d^{-m_{n}} i \rceil)-P^{n\downarrow}_{\beta_{H}|h_{X}}(i)|$
is exponentially small
and the difference between 
$\log \frac{d^{m_{n}} P^{n\downarrow}_{\beta_{H}|h_{X}}(d^{m_{n}} \lceil d^{-m_{n}} i \rceil)}
{P^{n\downarrow}_{\beta_{H}|h_{X}}(\lceil d^{-m_{n}} i \rceil)}$
and
$\log \frac{d^{m_n} P^{n\downarrow}_{\beta_{H}|h_{X}}(x)}{P^{n\downarrow}_{\beta_{H}|h_{X}}
(\lceil d^{-m_{n}} i\rceil)}
$ is linear with respect to $n$ at most,
the difference 
$\tilde{D}_{X}^n(m_{n})
-{D}_{X}^n(m_{n})$
is exponentially small.
Similarly, we define
\begin{align}
\tilde{D}_{Y}^n(m_{n})&:=D(\tilde{P}_{\beta_{L}|h_Y}^{n\downarrow}
\|P^{n\downarrow}_{\beta_{L}|h_{Y}}).
\end{align}
Then, the difference 
$\tilde{D}_{Y}^n(m_{n})- {D}_{Y}^n(m_{n})$
is exponentially small.
Hence, in order to prove \eqref{Akinkin}, we only have to show that
\begin{eqnarray}
D(
g_{2n*}(P_{\beta_{H}|h_X}^{n\downarrow}P_{\beta_{L}|h_Y}^{n\downarrow})
\|P_{\beta_{H}|h_X}^{n\downarrow}P_{\beta_{L}|h_Y}^{n\downarrow})
&=&\tilde{D}_{X}^n(m_{n})+\tilde{D}_{Y}^n(m_{n})+O(e^{-n\alpha_{2}})\Label{Akinkin3}.
\end{eqnarray}

Using Lemma \ref{AL-H}, we show \eqref{Akinkin3} as follows.
Because of Fannes's theorem \cite{Fannes}, the equation 
$S(g_{2n*} (P_{\beta_{H}|h_X}^{n\downarrow}P_{\beta_{L}|h_Y}^{n\downarrow}
))
-S(\tilde{P}_{\beta_{H}|h_X}^{n\downarrow}\tilde{P}_{\beta_{L}|h_Y}^{n\downarrow})=O(ne^{-n\alpha_{1}})$ follows from \eqref{A(A)}.
The maximum $h_{\mathrm{max}}:=\max\{ h_{X}(x), h_{Y}(y)\}$ satisfies 
\begin{align}
& \sum_{i,j}|(\tilde{P}_{\beta_{H}|h_X}^{n\downarrow}(i)
\tilde{P}_{\beta_{L}|h_Y}^{n\downarrow}(j)-
P_{\beta_{H}|h_X}^{n\downarrow}P_{\beta_{L}|h_Y}^{n\downarrow}(g_{2n}^{-1}(i,j)))
\log P_{\beta_{H}|h_X}^{n\downarrow}P_{\beta_{L}|h_Y}^{n\downarrow}(i,j)|
\nonumber \\
\le & 
2n
\|
\tilde{P}_{\beta_{H}|h_X}^{n\downarrow} \tilde{P}_{\beta_{L}|h_Y}^{n\downarrow}
-
g_{2n*}(P_{\beta_{H}|h_X}^{n\downarrow}P_{\beta_{L}|h_Y}^{n\downarrow})
\|_{1}
\beta_{L}h_{\mathrm{max}}\nonumber \\
=& O(ne^{-n\alpha_{1}}).
\end{align}
Thus, we obtain \eqref{Akinkin3} as follows;
\begin{align}
&D(
g_{2n*}(P_{\beta_{H}|h_X}^{n\downarrow}P_{\beta_{L}|h_Y}^{n\downarrow})
\|P_{\beta_{H}|h_X}^{n\downarrow}P_{\beta_{L}|h_Y}^{n\downarrow})
-\tilde{D}_{X}^n(m_{n})-\tilde{D}_{Y}^n(m_{n})\nonumber\\
=&
-S(
g_{2n*}(P_{\beta_{H}|h_X}^{n\downarrow}P_{\beta_{L}|h_Y}^{n\downarrow})
)
+S(\tilde{P}_{\beta_{H}|h_X}^{n\downarrow}\tilde{P}_{\beta_{L}|h_Y}^{n\downarrow})
\nonumber \\
&-\sum_{i,j}
|(\tilde{P}_{\beta_{H}|h_X}^{n\downarrow}(i)\tilde{P}_{\beta_{L}|h_Y}^{n\downarrow}(j)
-P_{\beta_{H}|h_X}^{n\downarrow}P_{\beta_{L}|h_Y}^{n\downarrow}(g_{2n}^{-1}(i,j)))\log P_{\beta_{H}|h_X}^{n\downarrow}P_{\beta_{L}|h_Y}^{n\downarrow}(i,j)|\nonumber \\
=&O(ne^{-n\alpha_{1}}).
\end{align}
Thus, we have shown \eqref{Akinkin3}, which implies \eqref{Akinkin}.

Next, we will show \eqref{Anobu}.
Using \eqref{Ah2}, we deform $\beta_{H}Q^{(n)}_{H\lnot}(\beta_{H},\beta_{L},\T^{\mathrm{opt}}_{Q_n})-m_{n}\log d+\tilde{D}_{X}^n(m_{n})$ as follows;
\begin{align}
&\beta_{H}Q^{(n)}_{H\lnot}(\beta_{H},\beta_{L},\T^{\mathrm{opt}}_{Q_n})-m_{n}\log d+\tilde{D}_{X}^n(m_{n})
\nonumber\\
&=S({P_X}')-S(\tilde{P}_{\beta_{H}|h_X}^{n\downarrow})
+
\sum_{i}[(\sum_{j}P_{\beta_{H}|h_X}^{n\downarrow}P_{\beta_{L}|h_Y}^{n\downarrow}(g_{2n}^{-1}(i,j))-\tilde{P}_{\beta_{H}|h_X}^{n\downarrow}(i))\log P^{n\downarrow}_{\beta_{H}|h_{X}}(i)].
\end{align}
Therefore, similar to \eqref{Akinkin3}, 
we find that $\beta_{H}Q_{n}-m_{n}\log d-\tilde{D}_{X}^n(m_{n})$
is exponentially small.
Since the difference $\tilde{D}_{X}^n(m_{n})-{D}_{X}^n(m_{n})$
is also exponentially small, we obtain \eqref{Anobu}.
\end{proofof}


As shown in the next section, we have the following lemma. 
\begin{lemma}\Label{Al5}
When $m_n=o(n)$,
\begin{align}
D_{X}^n(m_n)=&
(m_n \log d)^2 n^{-1} \frac{\psi_X'(-S_X)}{2}
+O((m_n \log d) n^{-1})
+O((m_n \log d)^3 n^{-2})
\Label{A2keyofkey'}.\\
D_{Y}^n(m_n)=&
(m_n \log d)^2 n^{-1} \frac{\psi_Y'(-S_Y)}{2}
+O((m_n \log d) n^{-1})
+O((m_n \log d)^3 n^{-2})
\Label{A2keyofkey2'}
\end{align}
Furthermore, 
when $h_X$ and $h_Y$ are non-lattice,
\begin{align}
D_{X}^n(m_n)=&
(m_n \log d)^2 n^{-1} \frac{\psi_X'(-S_X)}{2}
+(m_n \log d)^3 n^{-2} (\frac{\psi_X''(-S_X)}{6} -\frac{\psi_X'(-S_X)^2}{2} ) 
\nonumber \\
&
+(m_n \log d)^2
 n^{-2} 
(\frac{\psi_X''(-S_X)}{2\psi_X'(-S_X)}-\psi_X'(-S_X))^2
+O(m_n^3 n^{-5/2}) 
+O(m_n^4 n^{-3}),
\Label{Akeyofkey'}.\\
D_{Y}^n(m_n)=&
(m_n \log d)^2 n^{-1} \frac{\psi_Y'(-S_Y)}{2}
+(m_n \log d)^3 n^{-2} (-\frac{\psi_Y''(-S_Y)}{6} +\frac{\psi_Y'(-S_Y)^2}{2} ) 
\nonumber \\
&
+(m_n \log d)^2
 n^{-2} 
(\frac{\psi_Y''(-S_Y)}{2\psi_Y'(-S_Y)}-\psi_Y'(-S_Y))^2
+O(m_n^3 n^{-5/2}) 
+O(m_n^4 n^{-3})
\Label{Akeyofkey2'}
\end{align}
\end{lemma}

We can obtain Proposition \ref{A3-14-2T} by combining the equality \eqref{Atoushiki2}, Lemma \ref{Al4} and Lemma \ref{Al5} as follows;

\begin{proofof}{Proposition \ref{A3-14-2T}}
Now, we show Proposition \ref{A3-14-2T} when $P_X$ and $P_Y$ are non-lattice.
We can rewrite the integer $m_n$ as
\begin{align}
m_{n}=\left\lfloor 
\frac{\beta_{H} Q_{n}+ \frac{\psi_X'(-S_X)}{2 n }\beta_{H}^2 Q_{n}^2}{\log d}\right\rfloor.
\Label{A6-5-11}
\end{align} 
Then, we obtain \eqref{Aarxsecondorder2.2} as follows;
\begin{align}
Q^{(n)}_{H\lnot}(\beta_{H},\beta_{L},\T^{\mathrm{opt}}_{Q_n})
&
\stackrel{(a)}{=}
\frac{m_{n} \log d }{\beta_{H}}-\frac{D_{X}^n(m_{n})}{\beta_{H}}+O(e^{-n\alpha_{3}})\nonumber\\
&\stackrel{(b)}{=}
\frac{m_{n} \log d }{\beta_{H}}
-
(m_n \log d)^2 n^{-1} \frac{\psi_X'(-S_X)}{2 \beta_{H}}
+O(\frac{m_n^3}{n^2}) 
\nonumber\\
&\stackrel{(c)}{=}
Q_{n}
+\frac{\beta_{H} Q_{n}^2 \psi_X'(-S_X)}{2 n}
-\frac{\beta_{H} Q_{n}^2 \psi_X'(-S_X)}{2 n}
+O(\frac{Q_{n}^3}{n^2})+o(1)
=
Q_{n}
+O(\frac{Q_{n}^3}{n^2})+o(1),\Label{AQA}
\end{align}
where 
$(a)$, $(b)$, and $(c)$
follow from \eqref{Anobu}, \eqref{Akeyofkey'}, and \eqref{A6-5-11}, respectively.

Substituting the relation \eqref{A6-5-11}
into \eqref{Akeyofkey'} and \eqref{Akeyofkey2'}, we have
\begin{align}
D_{X}^n(m_n)=&
(\beta_{H} Q_{n})^2 n^{-1} \frac{\psi_X'(-S_X)}{2}
+(\beta_{H} Q_{n})^3 n^{-2} 
\frac{\psi_X''(-S_X)}{6}
+(m_n \log d)^2
 n^{-2} 
(\frac{\psi_X''(-S_X)}{2\psi_X'(-S_X)}-\psi_X'(-S_X))^2
\nonumber \\
&
+O(Q_{n}^3 n^{-5/2}) 
+O(Q_{n}^4 n^{-3}),
\Label{Akeyofkey-1} \\
D_{Y}^n(m_n)=&
(\beta_{H} Q_{n})^2 n^{-1} \frac{\psi_Y'(-S_Y)}{2}
+(\beta_{H} Q_{n})^3 n^{-2} 
(-\frac{\psi_Y''(-S_Y)}{6} 
+\frac{\psi_X'(-S_X) \psi_Y'(-S_Y)}{2} 
+\frac{\psi_Y'(-S_Y)^2}{2} ) \nonumber\\
&+(\beta_{H} Q_{n})^2
 n^{-2} 
(\frac{\psi_Y''(-S_Y)}{2\psi_Y'(-S_Y)}-\psi_Y'(-S_Y))^2
+O(Q_{n}^3 n^{-5/2}) 
+O(Q_{n}^4 n^{-3}).
\Label{Akeyofkey-2}
\end{align}
Therefore, 
\eqref{A6-6-14} and \eqref{A6-6-15} imply that
\begin{align}
D_{X}^n(m_n)+ D_{Y}^n(m_n)
=&
c^{(1)}_{\beta_{H},\beta_{L}}
(\beta_{H} Q_{n})^2 n^{-1} 
+
c^{(2)}_{\beta_{H},\beta_{L}}
(\beta_{H} Q_{n})^3 n^{-2} 
+
d^{(1)}_{\beta_{H},\beta_{L}}
(\beta_{H} Q_{n})^2 n^{-2} 
\nonumber \\
&
+O(Q_{n}^3 n^{-5/2}) 
+O(Q_{n}^4 n^{-3}).
\Label{Akeyofkey-3}
\end{align}
Combining \eqref{Atoushiki2}, \eqref{Akinkin} and \eqref{Akeyofkey-3}, 
we obtain the following expansion
\begin{align*} 
\eta^{(n)}_{C\lnot}(\beta_{H},\beta_{L},\T^{\mathrm{opt}}_{Q_n})
&=
\left(1-\frac{\beta_{H}}{\beta_{L}}\right)
-
\sum^{2}_{k=1}c^{(k)}_{\beta_{H},\beta_{L}} \frac{Q^{k}_{n}}{n^{k}}
-
d^{(1)}_{\beta_{H},\beta_{L}} \frac{Q_{n}}{n^2}
+O\left(\frac{Q^2_{n}}{n^{5/2}}\right)
+O\left(\frac{Q^{3}_{n}}{n^{3}}\right),
\end{align*}
which is the same as \eqref{Aarxsecondorder2}. 

Now, we consider the case when $P_X$ or $P_Y$ is lattice.
Using \eqref{A2keyofkey'} and \eqref{A2keyofkey2'}
instead of \eqref{Akeyofkey'} and \eqref{Akeyofkey2'}, we obtain
\begin{align*} 
\eta^{(n)}_{C\lnot}(\beta_{H},\beta_{L},\T^{\mathrm{opt}}_{Q_n})
&=
\left(1-\frac{\beta_{H}}{\beta_{L}}\right)
-
c^{(1)}_{\beta_{H},\beta_{L}} \frac{Q_{n}}{n}
+O(\frac{1}{n})
+O(\frac{Q_{n}^2}{n^2}),
\end{align*}
which is the same as \eqref{Aarxsecondorder2x}. 
Also, we obtain \eqref{Aarxsecondorder2.2x} as follows;
\begin{align}
Q^{(n)}_{H\lnot}(\beta_{H},\beta_{L},\T^{\mathrm{opt}}_{Q_n})
&
\stackrel{(a)}{=}
\frac{m_{n} \log d }{\beta_{H}}-\frac{D_{X}^n(m_{n})}{\beta_{H}}+O(e^{-n\alpha_{3}})\nonumber\\
&\stackrel{(b)}{=}
\frac{m_{n} \log d }{\beta_{H}}
-
(m_n \log d)^2 n^{-1} \frac{\psi_X'(-S_X)}{2 \beta_{H}}
+O((m_n \log d) n^{-1})
+O((m_n \log d)^3 n^{-2})
\nonumber\\
&\stackrel{(c)}{=}
Q_{n}
+\frac{\beta_{H} Q_{n}^2 \psi_X'(-S_X)}{2 n}
-\frac{\beta_{H} Q_{n}^2 \psi_X'(-S_X)}{2 n}
+O(\frac{Q_{n}}{n})
+O(\frac{Q_{n}^3}{n^2})
=
Q_{n}
+O(\frac{Q_{n}}{n})
+O(\frac{Q_{n}^3}{n^2}),\Label{AQAx}
\end{align}
where 
$(a)$, $(b)$, and $(c)$
follow from \eqref{Anobu}, \eqref{A2keyofkey'}, and \eqref{A6-5-11}, respectively.
\end{proofof}


\section{Relative entropy between two distributions with subtle difference}
\Label{As14}
\subsection{Preparation}
To show Lemma \ref{Al5}, we need to asymptotically expand the 
relative entropy $D_X^n(m_n)$ between the two distributions 
$P^{n,\downarrow}_{\beta_{H}|h_{X}}$
and $\tilde{P}^{n,\downarrow}_{\beta_{H}|h_{X}}$,
which are very close to each other.
However, these two distributions have the same information spectrum
in the sense Han's book \cite{Han}
up to the second order.
That is, any real number $R$ satisfies that
\begin{align}
& \lim_{n \to \infty}
P^{n,\downarrow}_{\beta_{H}|h_{X}} 
\{j|
-\log P^{n,\downarrow}_{\beta_{H}|h_{X}} (j)\le 
n S_X
+\sqrt{n} R
\} \nonumber \\
=&
\lim_{n \to \infty}
\tilde{P}^{n,\downarrow}_{\beta_{H}|h_{X}} 
\{j|
-\log \tilde{P}^{n,\downarrow}_{\beta_{H}|h_{X}}(j) \le 
n S_X
+\sqrt{n} R
\} .
\end{align}
Hence, the conventional approaches for information theory and quantum information theory do not work well.
To resolve this problem, 
we employ strong large deviation by Bahadur-Rao and Blackwell-Hodges
\cite{LD1,BH1} that
brings us a more detailed evaluation for the tail probability.
In particular, 
since the RHSs of \eqref{Akeyofkey'} and \eqref{Akeyofkey2'}
are related to higher order moments of the sample mean,
we prepare several formulas for them.
After these preparations, we show Lemma \ref{Al5}.

We focus on the logarithmic likelihood of $P^{n\downarrow}_{\beta_{\tilde{A}}|h_{A}}$ ($\tilde{A}=H,L$ and $A=X,Y$),
which can be regarded as the 
the logarithmic likelihood ratio between 
the distribution $P^{n}_{\beta_{\tilde{A}}|h_{A}}$ and the counting measure $P_C$.
Now, we define the random variable 
$Z_A:= (\log P^{n,\downarrow}_{\beta_{\tilde{A}}|h_{A}}(\hat{j})+ n S_A)/ \sqrt{n}$,
where $\hat{j}$ is a random variable on the natural numbers $\{j\}^{d^n}_{j=1}$ that takes the value $j$ with the probability $P^{n\downarrow}_{\beta_{H}|h_{A}}(j)$.
Then, when 
$j= P_C \{ Z_A \ge a  \}$,
$j$ is the maximum integer satisfying 
$\log P^{n,\downarrow}_{\beta_{\tilde{A}}|h_{A}}(j) \ge - n S_A+ \sqrt{n} a$.
We have the relations
\begin{align}
\bbE [ Z_A] &=0 \Label{A6-5-6}\\
\bbE [ Z_A^2] &=\phi_A''(1) = \frac{1}{\psi_A'(-S_A)} \Label{A6-5-7}\\
\bbE [ Z_A^3] &= \frac{\phi_A'''(1)}{\sqrt{n}}
= 
-\frac{\psi_A''(-S_A)}{\psi_A'(-S_A)^3 \sqrt{n}}
,\Label{A6-5-8}\\
\bbE [ Z_A^4] &=3\phi_A''(1)^2+\frac{\phi_A''''(1)}{n}
=\frac{3}{\psi_A'(-S_A)^2}
+(- \frac{\psi_A'''(-S_A)}{\psi_A'(-S_A)^3}
+3\frac{\psi_A''(-S_A)^2}{\psi_A'(-S_A)^5})\frac{1}{n}
,\Label{A6-5-82}
\end{align}
where $\bbE$ is the expectation under the distribution $P^{n\downarrow}_{\beta_{\tilde{A}}|h_{A}}$.

Next, we focus on the Legendre transform of $\phi_A$, which is written as 
\begin{align}
\max_s s R - \phi_A(s)
=  \psi_A(R) R - \phi_A(\psi_A(R)).
\end{align}

Then, we have the following proposition,
which is the special case 
of strong large deviation by Bahadur-Rao and Blackwell-Hodges \cite{LD1,BH1}.
Indeed, the strong large deviation gives the limiting behavior of the tail probability for 
independent and identical distribution of 
an arbitrary probability distribution.
The following is the special case when 
the probability distribution is the counting measure.

\begin{propositiondash}
Assume that $\log P_{\beta_{\tilde{A}}|h_A} $ is a non-lattice function.
When we choose suitable smooth functions $f_{A,k} (R)$,
the following relations hold \cite{LD1,LD2}.
\begin{align}
&\log P_C \{ \log P^{n,\downarrow}_{\beta_{\tilde{A}}|h_{A}} (\hat{j}) \ge n R  \} \nonumber\\
=& 
-n (\psi_A(R) R - \phi_A(\psi_A(R)))
- \log (\sqrt{2 \pi n \phi_A''(\psi_A(R))} \psi_A(R)) \nonumber\\
&+\frac{1}{n\phi''(\psi_A(R))}
(-\frac{5 \phi'''(\psi_A(R))^2}{24\phi''(\psi_A(R))^2}
+\frac{\phi''''(\psi_A(R))}{8\phi''(\psi_A(R))}
-\frac{\phi'''(\psi_A(R))}{2 \psi_A(R)\phi''(\psi_A(R))}
-\frac{1}{\psi_A(R)^2}
)\nonumber\\
&+ \sum_{k=3}^{l} f_{A,k} (R) n^{1-k} +o(\frac{1}{n^{l-1}}) \nonumber\\
=& 
-n (\psi_A(R) R - \phi_A(\psi_A(R)))
- \log \sqrt{2 \pi n }
- \log \psi_A(R)
+ \frac{1}{2}\log {\psi_A}'(R)) \nonumber\\
&+\frac{1}{n}[
-\frac{1}{8}\frac{\psi_A'''(R)}{\psi_A'(R)^{2}}
+\frac{1}{6}\frac{\psi_A''(R)^2}{\psi_A'(R)^{3}}
+\frac{1}{2}\frac{\psi_A''(R)}{\psi_A(R)\psi_A'(R)}
-\frac{\psi_A'(R)}{\psi_A(R)^2}
]\nonumber\\
&+ \sum_{k=3}^{l} f_{A,k} (R) n^{1-k} +o(\frac{1}{n^{l-1}}) ,
\Label{A6-1-2}
\end{align}

The lattice case is given as follows.
Let $d$ be the lattice span of $\log P_{\beta_{\tilde{A}}|h_A} $.
When we choose suitable smooth functions $f_{A,k} (R)$,
the following relations hold\cite{BH1}\cite[Theorem 3.7.4]{Dembo98}.
\begin{align}
&\log P_C \{ \log P^{n,\downarrow}_{\beta_{\tilde{A}}|h_{A}} (\hat{j}) \ge n R  \} \nonumber\\
=& 
-n (\psi_A(R) R - \phi_A(\psi_A(R)))
- \log (
\frac{ \sqrt{2 \pi n \phi_A''(\psi_A(R))} 
(1-e^{-d\psi_A(R) })}{d}
) \nonumber\\
&+ \sum_{k=2}^{l} f_{A,k} (R) n^{1-k} +o(\frac{1}{n^{l-1}}) 
\Label{A6-1-2x}.
\end{align}
\end{propositiondash}

In the following, for a unified treatment, 
in the non-lattice case,
the functions $- (\psi_A(R) R - \phi_A(\psi_A(R)))$,
$- \log \sqrt{2 \pi }
- \log \psi_A(R)
+ \frac{1}{2}\log {\psi_A}'(R)) $,
and
$
-\frac{1}{8}\frac{\psi_A'''(R)}{\psi_A'(R)^{2}}
+\frac{1}{6}\frac{\psi_A''(R)^2}{\psi_A'(R)^{3}}
+\frac{1}{2}\frac{\psi_A''(R)}{\psi_A(R)\psi_A'(R)}
-\frac{\psi_A'(R)}{\psi_A(R)^2}$
 are written as
$f_{A,0} (R)$, $f_{A,1} (R)$, and $f_{A,2} (R)$.
So, (\ref{A6-1-2}) is simplified as
$\log P_C \{ \log P^{n,\downarrow}_{\beta_{H}|h_{X}} (\hat{j}) \ge n R  \} 
= \sum_{k=0}^{l} f_{A,k} (R) n^{1-k} -\frac{1}{2} \log n+o(\frac{1}{n^{l-1}}) $. 
In the lattice case,
$f_{A,0} (R)$ is defined as the same way, and
$f_{A,1} (R)$ is defined to be
$- \log (
\frac{ \sqrt{2 \pi n \phi_A''(\psi_A(R))} 
(1-e^{-d\psi_A(R)})}{d}
)$.

\subsection{First step}
Now, we start to prove Lemma \ref{Al5} based on the above preparation.
Although 
we show (\ref{A2keyofkey'}) and (\ref{A2keyofkey2'})
with the general case and 
(\ref{Akeyofkey'}) and (\ref{Akeyofkey2'}) with the non-lattice case,
the first step for our proof works commonly for both cases.

From the above proposition,
we have
\begin{align}
&
\log F_A(Z_A)
:=\log P_C \{ \log P^{n,\downarrow}_{\beta_{\tilde{A}}|h_{A}} (\hat{j}) \ge - n S_A+ \sqrt{n} Z_A \}
\nonumber \\
=& \sum_{k=0}^{l} \sum_{t=0}^{2(l-k)}
\frac{f_{A,k}^{(t)} (-S_A) n^{1-k-\frac{t}{2}}  }{t !} Z_A^t
-\frac{1}{2} \log n+o(\frac{1}{n^{l-1}}) . 
\end{align}
Now, we introduce the random variables
$\Delta_X Z_X$ and $\Delta_Y Z_Y$ as
\begin{align}
F_X(Z_X+{\Delta}_X Z_X )d^{m_n} &= F_X(Z_X),\Label{A6-4-1} \\
F_Y(Z_Y )d^{m_n} &= F_Y(Z_Y-\Delta_Y (Z_Y)).\Label{A6-5-1} 
\end{align}
These two conditions are equivalent with 
\begin{align}
m_n \log d &= \log F_X(Z_X)-\log F_X(Z_X+\Delta_X Z_X) \Label{A6-6-1}\\
m_n \log d &= \log F_Y(Z_Y-\Delta_Y Z_Y) - \log F_Y(Z_Y).\Label{A6-6-2}
\end{align}
Hence, 
${\Delta}_X Z_X$ 
and
${\Delta}_Y Z_Y$ 
satisfy the equations
\begin{align}
m_n \log d &=
\sum_{i=1}^{2l} 
\alpha_{X,i}(Z_X) \frac{(\Delta_X Z_X)^i}{i !}+o(n^{1-l})
\Label{A6-5-2} \\
m_n \log d &=
-\sum_{i=1}^{2l} 
\alpha_{Y,i}(Z_Y) \frac{(-\Delta_Y Z_Y)^i}{i !}+o(n^{1-l}),
\Label{A6-5-2b} 
\end{align}
where
$\alpha_{A,i}(Z_A):=-
\sum_{k=0}^{l-\frac{i}{2}}
\sum_{j=0}^{2(l-k)-i}
n^{1-k-\frac{i+j}{2}} \frac{Z_A^j}{j !} f_{A,k}^{(i+j)}(-S_A) $,
where $f_{A,k}^{(i)} $ is the $i$-th derivative of $f_{A,k}$ for
$A=X,Y$.
Due to the definition (\ref{A6-4-1}), we have
\begin{align}
D_{X}^n(m_n)
=&
\sum_{j=1}^{d^{m_n}} P^{n,\downarrow}_{\beta_{H}|h_{X}} ({j})
(m_n \log d )+ \log P^{n,\downarrow}_{\beta_{H}|h_{X}} ({j})
- \log P^{n,\downarrow}_{\beta_{H}|h_{X}} (\lceil d^{-m_n} j\rceil)\nonumber\\
=&
\bbE 
[
(m_n \log d )+ \log P^{n,\downarrow}_{\beta_{H}|h_{X}} (\hat{j})
- \log P^{n,\downarrow}_{\beta_{H}|h_{X}} (\lceil d^{-m_n} \hat{j}\rceil)]
\nonumber\\
=&
\bbE [(m_n \log d )+(-n S+\sqrt{n} Z_X)
- (-n S_X +\sqrt{n} (Z_X +\Delta_X Z_X))] 
=
\bbE [(m_n \log d )-\sqrt{n}\Delta_X Z_X] .\Label{A6-5-5}
\end{align}
Hence, it is needed to solve the equation \eqref{A6-5-2} with respect to 
$\Delta_X Z_X$.
Notice that $\alpha_{X,i}(Z_X)=O(n^{1-\frac{i}{2}})$.
We apply Lemma \ref{Al6-6} to the equation \eqref{A6-5-2} with
$x= \frac{\Delta_X Z_X}{\sqrt{n}}$,
$a_i= \alpha_{X,i}(Z_X) n^{\frac{i}{2}}$,
and
$\epsilon= \frac{m_n \log d }{n}$.
Then, we obtain
\begin{align}
\frac{\Delta_X Z_X}{\sqrt{n}} = 
\frac{m_n \log d}{\sqrt{n} \alpha_{X,1}(Z_X)}
-\frac{\alpha_{X,2} (Z_X) (m_n \log d)^2}{\sqrt{n}\alpha_{X,1}^3(Z_X)}
+\frac{2 \alpha_{X,2} (Z_X)^2 (m_n \log d)^3}{\sqrt{n}\alpha_{X,1}^5(Z_X)}
-\frac{\alpha_{X,3} (Z_X) (m_n \log d)^3}{\sqrt{n}\alpha_{X,1}^4(Z_X)}
+O(m_n^4 n^{-4}).
\end{align}
That is, we obtain
\begin{align}
\Delta_X Z_X = 
\frac{m_n \log d}{\alpha_{X,1}(Z_X)}
-\frac{\alpha_{X,2} (Z_X) (m_n \log d)^2}{\alpha_{X,1}^3(Z_X)}
+\frac{2 \alpha_{X,2} (Z_X)^2 (m_n \log d)^3}{\alpha_{X,1}^5(Z_X)}
-\frac{\alpha_{X,3} (Z_X) (m_n \log d)^3}{\alpha_{X,1}^4(Z_X)}
+O(m_n^4 n^{-7/2}).\Label{A4-6-1eq}
\end{align}

\subsection{Second step for general case}
From now, our discussion becomes specialized to the proofs of
\eqref{A2keyofkey'} and \eqref{A2keyofkey2'}.
That is, we discuss the general case.
After these proofs, 
we give a more detail discussion for the proofs of
\eqref{Akeyofkey'} and \eqref{Akeyofkey2'}.

Since $S_X= -\phi_X'(1)$,
using the relations \eqref{A6-6-14} and \eqref{A6-6-15},
we have
\begin{align}
f_{X,0}'(R) =& -\psi_X(R) , \quad
f_{X,0}''(R) = -\psi_X'(R) .
\end{align}
Thus,
\begin{align}
\alpha_{X,1}(Z_X)
=&
\sqrt{n}+\psi_X'(-S_X)Z_X + 
O(\frac{1}{\sqrt{n}} )
\\
\alpha_{X,2}(Z_X)
=&
\frac{\psi_X'(-S_X)}{2}
+\frac{\psi_X''(-S_X)}{2\sqrt{n}}Z_X 
+O(\frac{1}{n})
\\
\alpha_{X,3}(Z_X)
=&
 \frac{1}{\sqrt{n}} \frac{\psi_X''(-S_X)}{6} 
+O(n^{-1}).
\end{align}
We have
\begin{align}
\frac{m_n \log d}{\alpha_{X,1}(Z_X)}
=&
m_n \log d
\Bigl[n^{-1/2}-n^{-1}\psi_X'(-S_X)Z_X +O(n^{-3/2})\Bigr] 
\end{align}
and
\begin{align}
\frac{\alpha_{X,2} (Z_X) (m_n \log d)^2}{\alpha_{X,1}^3(Z_X)}
=&
(m_n \log d)^2
\Bigl[ n^{-3/2} \frac{\psi_X'(-S_X)}{2}
+ O(n^{-2})\Bigr] 
\nonumber\\
\frac{2 \alpha_{X,2} (Z_X)^2 (m_n \log d)^3}{\alpha_{X,1}^5(Z_X)}
=&
(m_n \log d)^3 n^{-5/2} \frac{\psi_X'(-S_X)^2}{2} 
+O(m_n^3 n^{-3})
\nonumber\\
\frac{\alpha_{X,3} (Z_X) (m_n \log d)^3}{\alpha_{X,1}^4(Z_X)}
=&
(m_n \log d)^3 n^{-5/2} \frac{\psi_X''(-S_X)}{6} 
+O(m_n^3 n^{-3}).
\end{align}
Therefore,
\begin{align}
&m_n \log d - \sqrt{n} \Delta Z_X\nonumber\\
=&
m_n \log d
\Bigl[n^{-1/2}\psi_X'(-S_X)Z_X
+O(n^{-1}) \Bigl] \nonumber\\
&+
(m_n \log d)^2
\Bigl[ n^{-1} \frac{\psi_X'(-S_X)}{2}
+ O(n^{-3/2})
\Bigr] 
+O((m_n \log d)^3 n^{-2}).
\end{align}
Now, we take the expectation of 
$m_n \log d - \sqrt{n} \Delta Z_X$
with use of (\ref{A6-5-6}), (\ref{A6-5-7}), (\ref{A6-5-8}), and (\ref{A6-5-82}).
We have
\begin{align}
\bbE[m_n \log d - \sqrt{n} \Delta Z_X]
=&
(m_n \log d)^2
 n^{-1} \frac{\psi_X'(-S_X)}{2}
+O(m_n^3 n^{-2} )
+O(m_n n^{-1} ) .
\end{align}
Hence, using \eqref{A6-5-5}, we obtain \eqref{A2keyofkey'}.

Similar to \eqref{A6-5-5}, due to the definition (\ref{A6-5-1}), we have
\begin{align}
D_{Y}^n(m_n)
=&
\bbE [-(m_n \log d )+\sqrt{n}\Delta_Y Z_Y] .
\end{align}
Following the same way as \eqref{A2keyofkey'},
we obtain \eqref{A2keyofkey2'} by solving \eqref{A6-5-2b}.

\subsection{Second step for non-lattice case}
Now, we proceed to the proofs of \eqref{Akeyofkey'} and \eqref{Akeyofkey2'} for the non-lattice case.
The following discussion continues \eqref{A4-6-1eq}.
Since $S_X= -\phi_X'(1)$,
using the relations \eqref{A6-6-14} and \eqref{A6-6-15},
we have
\begin{align}
f_{X,0}^{(j)}(R) =& -\psi_X^{(j-1)}(R) \\
f_{X,1}^{(1)}(R) =& 
-\frac{\psi_X'(R)}{\psi_X(R)}+\frac{\psi_X''(R)}{2\psi_X'(R)} \\
f_{X,1}^{(2)}(R) =& 
-\frac{\psi_X''(R)\psi_X(R)-\psi_X'(R)^2}{\psi_X(R)^2}
+\frac{\psi_X'''(R)\psi_X'(R)-\psi_X''(R)^2}{2\psi_X'(R)^2} \\
f_{X,1}^{(3)}(R) =& 
-\frac{\psi_X(R)^2\psi_X'''(R)-3 \psi_X(R)\psi_X'(R)\psi_X''(R) +2 \psi_X'(R)^3}{\psi_X(R)^3}
+\frac{\psi_X'(R)^2\psi_X''''(R)-3 \psi_X'(R)\psi_X''(R)\psi_X'''(R) +2 \psi_X''(R)^3}{2\psi_X'(R)^3}\\
f_{X,2}^{(1)}(R) =& 
-\frac{1}{8}\frac{\psi_X'(R)\psi_X''''(R)-2 \psi_X'''(R)\psi_X''(R)}{\psi_X'(R)^{3}}
+\frac{1}{6}\frac{2\psi_X'(R)\psi_X''(R)\psi_X'''(R)-3\psi_X''(R)^3}{\psi_X'(R)^{4}}
\nonumber \\
&+\frac{1}{2}\frac{\psi_X(R)\psi_X'(R)\psi_X'''(R)-\psi_X'(R)^2\psi_X''(R)
-\psi_X(R)\psi_X''(R)^2}{\psi_X(R)^2\psi_X'(R)^2}
-\frac{\psi_X(R)\psi_X''(R)-2 \psi_X'(R)^2}{\psi_X(R)^3}.
\end{align}
Thus,
\begin{align}
\alpha_{X,1}(Z_X)
=&
\sqrt{n}+\psi_X'(-S_X)Z_X + 
\frac{1}{\sqrt{n}} 
(\psi_X'(-S_X)+\frac{\psi_X''(-S_X)}{2}( Z_X^2- \frac{1}{\psi_X'(-S)}))
+\frac{1}{n}
(\frac{\psi_X'''(-S_X)}{6}Z_X^3 -D_{X,1}  Z_X) \nonumber \\
& +\frac{1}{n^{3/2}}
(\frac{\psi_X''''(-S_X)}{24}Z_X^4 
-\frac{D_{X,2}}{2} Z_X^2
-D_{X,3} ) 
+O(n^{-2})
\\
\alpha_{X,2}(Z_X)
=&
\frac{\psi_X'(-S_X)}{2}
+\frac{\psi_X''(-S_X)}{2\sqrt{n}}Z_X 
+\frac{1}{n} (\frac{\psi_X'''(-S_X)}{4} Z_X^2 -\frac{D_{X,1}}{2} )
+O(n^{-3/2})
\\
\alpha_{X,3}(Z_X)
=&
 \frac{1}{\sqrt{n}} \frac{\psi_X''(-S)}{6} 
+O(n^{-1}), 
\end{align}
where 
\begin{align}
D_{X,1}:=&
-\psi_X''(-S_X)+ \psi_X'(-S_X)^2
+ \frac{1}{2}
\frac{\psi_X'''(-S_X)\psi_X'(-S_X)-\psi_X''(-S_X)^2}{\psi_X'(-S_X)^2}\\
D_{X,2}:=&
-\psi'''(-S_X)+3 \psi'(-S_X)\psi''(-S_X) -2 \psi'(-S_X)^3\nonumber\\
&+\frac{\psi'(-S_X)^2\psi''''(-S_X)-3 \psi'(-S_X)\psi''(-S_X)\psi'''(-S_X) +2 \psi''(-S_X)^3}{2\psi'(-S_X)^3} \\
D_{X,3}:=& 
-\frac{1}{8}\frac{\psi'(-S_X)\psi''''(-S_X)
-2 \psi'''(-S_X)\psi''(-S_X)}{\psi'(-S_X)^{3}}
+\frac{1}{6}\frac{2\psi'(-S_X)\psi''(-S_X)\psi'''(-S_X)
-3\psi''(-S_X)^3}{\psi'(-S_X)^{4}}
\nonumber \\
&+\frac{1}{2}\frac{\psi'(-S_X)\psi'''(-S_X)-\psi'(-S_X)^2\psi''(-S_X)
-\psi''(-S_X)^2}{\psi'(-S_X)^2}
-\psi''(-S_X)+2 \psi'(-S_X)^2.
\end{align}
We have
\begin{align}
\frac{m_n \log d}{\alpha_{X,1}(Z_X)}
=&
m_n \log d
\Bigl[n^{-1/2}
-n^{-1}\psi_X'(-S_X)Z_X
+n^{-3/2}(-\frac{\psi_X''(-S_X)}{2}+\psi_X'(-S_X)^2)(Z_X^2 
- \frac{1}{\psi_X'(-S_X)}) \nonumber\\
&+n^{-2}
(-\frac{\psi_X'''(-S_X)}{6} -\psi_X'(-S_X)^3+ \psi_X''(-S_X)\psi_X'(-S_X))Z_X^3 \nonumber\\
&+n^{-2}(D_{X,1} +2 \psi_X'(-S_X)^2- \psi_X''(-S_X) ) Z_X \nonumber\\
&+n^{-5/2} 
\Bigl(
( \psi_X'(-S_X)^4 -\frac{3}{2} \psi_X'(-S_X)^2 \psi_X''(-S_X)
+\frac{1}{4} \psi_X''(-S_X)^2 
+\frac{1}{3} \psi_X'(-S_X)\psi_X'''(-S_X) 
-\frac{1}{24} \psi_X''''(-S_X) 
) Z_X^4 \nonumber\\ 
&+
( -3 \psi_X'(-S_X)^3
+\frac{5}{2} \psi_X'(-S_X) \psi_X''(-S_X)
-\frac{1}{2 \psi_X'(-S_X)} \psi_X''(-S_X)^2 
-2 \psi_X'(-S_X) D_{X,1}+ \frac{D_{X,2}}{2}
) Z_X^2 \nonumber\\
&+
( \psi_X'(-S_X) -\frac{\psi_X''(-S_X)}{2 \psi_X'(-S_X)} )^2
+D_{X,3}
\Bigr)
\Bigl]
+O(m_n n^{-3})
\end{align}
and
\begin{align}
\frac{\alpha_{X,2} (Z_X) (m_n \log d)^2}{\alpha_{X,1}^3(Z_X)}
=&
(m_n \log d)^2
\Bigl[ n^{-3/2} \frac{\psi_X'(-S_X)}{2}
+ n^{-2} (-\frac{3\psi_X'(-S_X)^2}{2} +\frac{\psi_X''(-S_X)}{2}) Z_X \nonumber\\
&+ n^{-5/2} 
(\frac{\psi_X'''(-S_X)}{4}-\frac{9}{4}\psi_X'(-S_X)\psi_X''(-S_X)+ 3 \psi_X'(-S_X)^3)Z_X^2
\nonumber \\
&-\frac{D_{X,1}}{2}-\frac{3}{2}\psi_X'(-S_X)^2+\frac{3}{4}\psi_X''(-S_X)
\Bigl] +O(m_n^2 n^{-3})
\nonumber\\
\frac{2 \alpha_{X,2} (Z_X)^2 (m_n \log d)^3}{\alpha_{X,1}^5(Z_X)}
=&
(m_n \log d)^3 n^{-5/2} \frac{\psi_X'(-S_X)^2}{2} 
+O(m_n^3 n^{-3})
\nonumber\\
\frac{\alpha_{X,3} (Z_X) (m_n \log d)^3}{\alpha_{X,1}^4(Z_X)}
=&
(m_n \log d)^3 n^{-5/2} \frac{\psi_X''(-S_X)}{6} 
+O(m_n^3 n^{-3}).
\end{align}
Therefore,
\begin{align}
&m_n \log d - \sqrt{n} \Delta Z_X\nonumber\\
&=
m_n \log d
\Bigl[n^{-1/2}\psi_X'(-S_X)Z_X
-n^{-1}(-\frac{\psi_X''(-S_X)}{2}+\psi_X'(-S_X)^2)(Z_X^2 - \frac{1}{\psi_X'(-S_X)}) \nonumber\\
&-n^{-3/2}
(-\frac{\psi_X'''(-S_X)}{6} -\psi_X'(-S_X)^3+ \psi_X''(-S_X)\psi_X'(-S_X))Z_X^3 \nonumber\\
&-n^{-3/2}( D_{X,1}+2 \psi_X'(-S_X)^2- \psi_X''(-S_X) ) Z_X \nonumber\\
&-n^{-2} 
\Bigl(
( \psi_X'(-S_X)^4 -\frac{3}{2} \psi_X'(-S_X)^2 \psi_X''(-S_X)
+\frac{1}{4} \psi_X''(-S_X)^2 
+\frac{1}{3} \psi_X'(-S_X)\psi_X'''(-S_X) 
-\frac{1}{24} \psi_X''''(-S_X) 
) Z_X^4 \nonumber\\ 
&+
( -3 \psi_X'(-S_X)^3
+\frac{5}{2} \psi_X'(-S_X) \psi_X''(-S_X)
-\frac{1}{2 \psi_X'(-S_X)} \psi_X''(-S_X)^2 
-2 \psi_X'(-S_X) D_{X,1}+ \frac{D_{X,2}}{2}
) Z_X^2 \nonumber\\
&+
( \psi_X'(-S_X) -\frac{\psi_X''(-S_X)}{2 \psi_X'(-S_X)} )^2
+D_{X,3}
\Bigr)
\Bigl] \nonumber\\
&+
(m_n \log d)^2
\Bigl[ n^{-1} \frac{\psi_X'(-S_X)}{2}
+ n^{-3/2} (-\frac{3\psi_X'(-S_X)^2}{2} +\frac{\psi_X''(-S_X)}{2}) Z_X \nonumber\\
&+ n^{-2} 
(\frac{\psi_X'''(-S_X)}{4}-\frac{9}{4}\psi_X'(-S_X)\psi_X''(-S_X)+ 3 \psi_X'(-S_X)^3)Z_X^2
-\frac{D_{X,1}}{2}-\frac{3}{2}\psi_X'(-S_X)^2+\frac{3}{4}\psi_X''(-S_X)
\Bigl] \nonumber\\
&+(m_n \log d)^3 n^{-2} (\frac{\psi_X''(-S_X)}{6} -\frac{\psi_X'(-S_X)^2}{2} )
+O(m_n^3 n^{-5/2})
+O(m_n^4 n^{-4}).
\end{align}
Now, we take the expectation of 
$m_n \log d - \sqrt{n} \Delta Z_X$
with use of (\ref{A6-5-6}), (\ref{A6-5-7}), (\ref{A6-5-8}), and (\ref{A6-5-82}).
Then, the coefficient of the term $(m_n \log d) n^{-2} $ equals 
\begin{align}
&(-\frac{\psi_X'''(-S_X)}{6} -\psi_X'(-S_X)^3+ \psi_X''(-S_X)\psi_X'(-S_X)) 
\frac{\psi_X''(-S_X)}{\psi_X'(-S_X)^3} \nonumber\\
& -3 \frac{1}{\psi_X'(-S_X)^2}
( \psi_X'(-S_X)^4 -\frac{3}{2} \psi_X'(-S_X)^2 \psi_X''(-S_X)
+\frac{1}{4} \psi_X''(-S_X)^2 
+\frac{1}{3} \psi_X'(-S_X)\psi_X'''(-S_X) 
-\frac{1}{24} \psi_X''''(-S_X) 
)  \nonumber\\
&- \frac{1}{\psi_X'(-S_X)}
( -3 \psi_X'(-S_X)^3
+\frac{5}{2} \psi_X'(-S_X) \psi_X''(-S_X)
-\frac{1}{2 \psi_X'(-S_X)} \psi_X''(-S_X)^2 
-2 \psi_X'(-S_X) D_{X,1}+ \frac{D_{X,2}}{2}
)  \nonumber\\
&-
( \psi_X'(-S_X) -\frac{\psi_X''(-S_X)}{2 \psi_X'(-S_X)} )^2
- D_{X,3} \nonumber\\
=&0.
\end{align}
Therefore, 
we have
\begin{align}
\bbE[m_n \log d - \sqrt{n} \Delta Z_X]
=&
(m_n \log d)^2
 n^{-1} \frac{\psi_X'(-S_X)}{2}
+(m_n \log d)^3 n^{-2} (\frac{\psi_X''(-S_X)}{6} -\frac{\psi_X'(-S_X)^2}{2} ) \nonumber\\
&+(m_n \log d)^2
 n^{-2} 
\Bigl[\frac{\psi_X'''(-S_X)}{4\psi_X'(-S_X)}-\frac{9}{4}\psi_X''(-S_X)+ 3 \psi_X'(-S_X)^2
-\frac{D_{X,1}}{2}-\frac{3}{2}\psi_X'(-S_X)^2+\frac{3}{4}\psi_X''(-S_X)
\Bigl] \nonumber\\
&
+O(m_n^3 n^{-5/2}) 
+O(m_n^4 n^{-4})
\nonumber\\
=&
(m_n \log d)^2
 n^{-1} \frac{\psi_X'(-S_X)}{2}
+(m_n \log d)^3 n^{-2} (\frac{\psi_X''(-S_X)}{6} -\frac{\psi_X'(-S_X)^2}{2} )\nonumber\\
&+(m_n \log d)^2
 n^{-2} 
(\frac{\psi_X''(-S_X)}{2\psi_X'(-S_X)}-\psi_X'(-S_X))^2
\nonumber \\
&
+O(m_n^3 n^{-5/2}) 
+O(m_n^4 n^{-4}).
\end{align}
Hence, we obtain \eqref{Akeyofkey'}.

Similar to \eqref{A6-5-5}, due to the definition (\ref{A6-5-1}), we have
\begin{align}
D_{Y}^n(m_n)
=&
\bbE [-(m_n \log d )+\sqrt{n}\Delta_Y Z_Y] .
\end{align}
Following the same way as \eqref{Akeyofkey'},
we obtain \eqref{Akeyofkey2'} by solving \eqref{A6-5-2b}.

\section{Evaluation of the entropy differences of the heat baths}\Label{AApH}
In the present section, we prove Proposition \ref{AEntP} in the main text.
Proposition \ref{AEntP} is given as a corollary of Lemma \ref{AL-H}.
Let us write down Proposition \ref{AEntP} again;
\begin{proposition}\Label{AEntP}
The following equalities hold with a proper positive number $\gamma$;
\begin{align}
S(\sigma_{H})-S(\rho_{\beta_{H}|\hat{H}^{(n)}_{H}})&=-m_{n}\log d +O(e^{-n\gamma})\Label{AentH}\\
S(\sigma_{L})-S(\rho_{\beta_{L}|\hat{H}^{(n)}_{L}})&=m_{n}\log d +O(e^{-n\gamma})\Label{AentL}
\end{align}
\end{proposition}

\begin{proofof}{Proposition \ref{AEntP}}
Because of 
$\Tr_{E_{X}}[U_{f_{n}}\rho_{\beta_{H}\beta_{L}}\otimes |e\rangle \langle e|U^{\dagger}_{f_{n}}]=\sum_{x,y}f_{n*}(P^{(n)}_{\beta_{H},\beta_{L}})(x,y)|x,y\rangle \langle x,y|$,
the following equalities clearly hold;
\begin{align}
S(\sigma_{H})=S(\sum_{y}f_{n*}(P^{(n)}_{\beta_{H},\beta_{L}})),\enskip S(\sigma_{L})=S(\sum_{x}f_{n*}(P^{(n)}_{\beta_{H},\beta_{L}})).
\end{align}
Moreover, $g_{1nx*}$ and $g_{1nY*}$, which are defined in \eqref{Ag1ndef}, do not change entropies of $X$ and $Y$, respectively.
Therefore, we only have to prove that the following equalities hold for a proper positive number $\gamma$; 
\begin{align}
S(\sum_{y}g_{2n*}(P^{n\downarrow}_{\beta_{H}|h_{X}}P^{n\downarrow}_{\beta_{H}|h_{Y}}))-S(P_{\beta_{H}|h^{(n)}_{X}})&=-m_{n}\log d+O(e^{-\gamma}),\Label{AH2}
\\
S(\sum_{x}g_{2n*}(P^{n\downarrow}_{\beta_{H}|h_{X}}P^{n\downarrow}_{\beta_{H}|h_{Y}}))-S(P_{\beta_{L}|h^{(n)}_{X}})&=m_{n}\log d+O(e^{-\gamma})\Label{AH3}
\end{align}
Because of Lemma \ref{AL-H} and Fannes's theorem, the following equalities hold for a positive number $\alpha_{1}$;
\begin{align}
S(\sum_{y}g_{2n*}(P^{n\downarrow}_{\beta_{H}|h_{X}}P^{n\downarrow}_{\beta_{H}|h_{Y}}))-S(\tilde{P}^{n\downarrow}_{\beta_{H}|h_{X}})&=O(ne^{-n\alpha_{1}}),
\\
S(\sum_{x}g_{2n*}(P^{n\downarrow}_{\beta_{H}|h_{X}}P^{n\downarrow}_{\beta_{H}|h_{Y}}))-S(\tilde{P}^{n\downarrow}_{\beta_{L}|h_{Y}})&=O(ne^{-n\alpha_{1}}),
\end{align}
where $\tilde{P}^{n\downarrow}_{\beta_{H}|h_{X}}$ and $\tilde{P}^{n\downarrow}_{\beta_{L}|h_{Y}}$ are defined in \eqref{AD11} and \eqref{AD12}.
Because of \eqref{AD11} and \eqref{AD12},
\begin{align}
S(\tilde{P}^{n\downarrow}_{\beta_{H}|h_{X}})-S(P_{\beta_{H}|h^{(n)}_{X}})&=-m_{n}\log d-\sum^{d^{n}}_{x=d^{n-m_{n}}+1}P^{n\downarrow}_{\beta_{H}|h_{X}}(x)\log P^{n\downarrow}_{\beta_{H}|h_{X}}(x),\Label{AH6}\\
S(\tilde{P}^{n\downarrow}_{\beta_{H}|h_{X}})-S(P_{\beta_{H}|h^{(n)}_{X}})&=m_{n}\log d.
\end{align}
Because $m_{n}=o(n)$, $\sum^{d^{n}}_{j=d^{n-m_{n}}}P(j)$ is exponentially small. 
Therefore, the second term of \eqref{AH6} satisfies $\sum^{d^{n}}_{x=d^{n-m_{n}}+1}P^{n\downarrow}_{\beta_{H}|h_{X}}(x)\log P^{n\downarrow}_{\beta_{H}|h_{X}}(x)=O(ne^{-n\alpha_{2}})$ with a proper positive number $\alpha_{2}$.
Thus, 
\begin{align}
S(\sum_{y}g_{2n*}(P^{n\downarrow}_{\beta_{H}|h_{X}}P^{n\downarrow}_{\beta_{H}|h_{Y}}))-S(P_{\beta_{H}|h^{(n)}_{X}})&=-m_{n}\log d+O(e^{-n\alpha_{1}})+O(ne^{-n\alpha_{2}}),
\\
S(\sum_{x}g_{2n*}(P^{n\downarrow}_{\beta_{H}|h_{X}}P^{n\downarrow}_{\beta_{H}|h_{Y}}))-S(P_{\beta_{L}|h^{(n)}_{X}})&=m_{n}\log d+O(e^{-n\alpha_{1}})
\end{align}
hold, and thus, \eqref{AH2} and \eqref{AH3} holds for the positive number $\gamma=\min\{\alpha_{1},\alpha_{2}\}/2$.
\end{proofof}

\section{Perturbation for higher order equation}\Label{AappB}

\begin{lemma}\Label{Al6-6}
Consider the equation
\begin{align}
\epsilon =\sum_{i=1}^l a_i x^i.\Label{A6-6-11}
\end{align}
When $\epsilon$ is sufficiently small, 
the solution $x$ is approximated as
\begin{align}
x= \sum_{i=1}^l \epsilon^i x_i + O(\epsilon^{l+1}).\Label{A6-6-10}
\end{align}
where
$x_1$ is given as $\frac{1}{a_1} $
and 
$x_l$ with $l \ge 2$ is inductively given as
$- \frac{1}{a_1}
\sum_{i_1,i_2,\ldots, i_{l-1}:\sum_{k=1}^{l-1}k i_k=l}
a_{\sum_{k=1}^{l-1}i_k} 
\frac{(\sum_{k=1}^{l-1}i_k)!}{\prod_{k=1}^{l-1} i_k! }
\prod_{k=1}^{l-1}x_k^{i_k}$.
Specially, $x_2$ and $x_3$ are given as
\begin{align}
x_2 &= -\frac{1}{a_1} x_1^2 =-\frac{1}{a_1^3}\Label{A6-6-10.5}\\
x_3 &
= -\frac{1}{a_1} (x_1^3+ 2 x_1 x_2)
= -\frac{1}{a_1^{4}}+\frac{2}{a_1^5}.
\end{align}
\end{lemma}
This lemma can be shown as follows.
First, we substitute \eqref{A6-6-10} into \eqref{A6-6-11}.
Then, compare the coefficients with the order $\epsilon^i$.
Hence, we obtain $x_l=
- \frac{1}{a_1}
\sum_{i_1,i_2,\ldots, i_{l-1}:\sum_{k=1}^{l-1}k i_k=l}
a_{\sum_{k=1}^{l-1}i_k} 
\frac{(\sum_{k=1}^{l-1}i_k)!}{\prod_{k=1}^{l-1} i_k! }
\prod_{k=1}^{l-1}x_k^{i_k}$.


\begin{thebibliography}{00}
\bibitem{Carnot}S. Carnot, \textit{Reflections on the Motive Power of Fire and on Machines Fitted to Develop that Power}, (1824).
\bibitem{Fermi}E. Fermi, {\it{Thermodynamics}} (Dover Books on Physics, 1956).
\bibitem{Bardeen}J. M. Bardeen, B. Carter, S. W. Hawking, Comm. Math. Phys. \textbf{31}, 161 (1973).
\bibitem{Maxwell}Maxwell, J. C.  Illustrations of the dynamical theory of gases.-Part 1. \textit{Philosophical Magazine} \textbf{19}, 19 (1860).
\bibitem{Maxwell'}Maxwell, J. C. Illustrations of the dynamical theory of gases.-Part 2. \textit{Philosophical Magazine} \textbf{20}, 21 (1860).
\bibitem{Boltzmann}Boltzmann, L. E.  Weitere Studien \"{u}ber das W\"{a}rmegleichgewicht unter Gasmolek\"{u}len. \textit{Wien Ber}. \textbf{66}, 275 (1872).
\bibitem{Boltzmann'}Boltzmann, L. E. \"{U}ber die Beziehung zwischen dem zweiten Hauptsatz der mechanischen W\"{a}rmetheorie und der Wahrscheinlichkeitsrechnung respektive den S\"{a}tzen \"{u}ber des W\"{a}rmegleichgewicht. \textit{Wien Ber}. \textbf{76}, 373 (1877).
\bibitem{ouyoutoukei111}Landau, L. D. and Lifshitz, E. M. {\it{Statistical Physics. Vol. 5 of the Course of Theoretical Physics}}, 3rd edition (1976).
\bibitem{Lieb1996}E. H. Lieb and J. Yngvason Phys. Pep. \textbf{310}, 1, (1996).
\bibitem{Sekimoto}K. Sekimoto, \textit{Stochastic Energetics (Lecture Notes in Physics)}, Springer, (2010).

\bibitem{Ehrenfest} P. Ehrenfest and T. Ehrenfest, \textit{The Conceptual Foundations of the Statistical Approach in Mechanics (Dover Books on Physics)},(2015).




\bibitem{Horodecki}M. Horodecki and J. Oppenheim, Nat. Commun. \textbf{4}, 2059 (2013).
\bibitem{oneshot1}O. C. O. Dahlsten, R. Renner, E. Rieper, and V. Vedral, New. J. Phys.\textbf{13}, 053015, (2011).  
\bibitem{oneshot3}J. Aberg, Nat. Commun. \textbf{4}, 1925 (2013).
\bibitem{Egloff}D. Egloff, O. C. O. Dahlsten, R. Renner and V. Vedral, arXiv:1207.0434, (2012).
\bibitem{Brandao}F. G. S. L. Brandao, M. Horodeck, N. H. Y. Ng, J. Oppenheim, and S. Wehner, PNAS, 112,3215(2015).
\bibitem{Car2}S. Popescu, arXiv:1009.2536.(2010).
\bibitem{Popescu2014}P. Skrzypczyk, A. J. Short and S. Popescu, Nature Communications \textbf{5}, 4185, (2014).
\bibitem{Popescu2015}Y. Guryanova,	S. Popescu,	A. J. Short,	R. Silva and P. Skrzypczyk, Nat. Comm. \textbf{7}, 12049, (2016).
\bibitem{oneshot2}L. Rio, J. Aberg, R. Renner, O. Dahlsten, and V. Vedral, \textit{Nature} \textbf{474}, 61, (2011). 

\bibitem{Jarzynski}C. Jarzynski, Phys. Rev. Lett. \textbf{78}, 2690, (1997). 
\bibitem{tasaki}H. Tasaki, arXiv:cond-mat/0009244 (2000).
\bibitem{Kurchan}J. Kurchan, arXiv:cond-mat/0007360(2000). 
\bibitem{Car1}S. DeLiberato and M. Ueda, Phys. Rev. E \textbf{84}, 051122 (2011).
\bibitem{Xiao}G. Xiao and J. Gong, arXiv:1503.00784, (2015).
\bibitem{Lenard}A. Lenard, J. Stat. Phys., \textbf{19}, 6, 575(1978).
\bibitem{TLH}
P. Talkner, E. Lutz, and P. H\"{a}nggi, {\em Phys. Rev. E} {\bf 75}, 050102 (R) (2007).
\bibitem{BBM1}
C. M. Bender, D. C. Brody, and B. K. Meister,
``Entropy and temperature of a quantum Carnot engine,''
{\em Proc. R. Soc. A} {\bf 458}, 1519-1526 (2002).
\bibitem{oldresult5}M. A. Nielsen, C. M. Caves, B. Schumacher, and H. Barnum, Proc. R. Soc. A \textbf{454}, 277 (1998).
\bibitem{Croocks}G. E. Crooks, Phys. Rev. E \textbf{60}, 2721 (1999).
\bibitem{Sagawa2010} T. Sagawa and M. Ueda, Phys. Rev. Lett. \textbf{104}, 090602 (2010).	
\bibitem{Ponmurugan2010}M. Ponmurugan, Phys. Rev. E \textbf{82}, 031129 (2010).
\bibitem{Horowitz2010}J. M. Horowitz and S. Vaikuntanathan, Phys. Rev. E \textbf{82}, 061120 (2010).
\bibitem{Horowitz2011}J. M. Horowitz and J. M. R. Parrondo, Europhys. Lett. \textbf{95}, 10005 (2011).
\bibitem{Sagawa2012}T. Sagawa and M. Ueda, Phys. Rev. Lett. \textbf{109}, 180602 (2012).
\bibitem{Ito2013}S. Ito and T. Sagawa, Phys. Rev. Lett. \textbf{111}, 180603 (2013).
\bibitem{sagawa1}T. Sagawa and M. Ueda, Phys. Rev. Lett, \textbf{100} 080403 (2008).
\bibitem{jacobs}K. Jacobs, Phys. Rev. A \textbf{80}, 012322 (2009). 
\bibitem{sagawa2}T. Sagawa, M. Ueda, Phys. Rev. Lett. \textbf{102} 250602 (2009).
\bibitem{Funo}K. Funo, Y. Watanabe and M. Ueda,Phys. Rev. A 88, 052319 (2013).
\bibitem{Morikuni}Y. Morikuni and H. Tasaki, J Stat Phys \textbf{143}, 1,(2011). 
\bibitem{Parrondo}J. M. R. Parrondo, J. M. Horowit and T. Sagawa, Nat. Phys. \textbf{11}, 131 (2015).
\bibitem{Izumida} Y. Izumida and K. Okuda, Phys. Rev. Lett. 112, 180603 (2014).
\bibitem{IE1}H. Tajima, Phys. Rev. E \textbf{88}, 042143 (2013).
\bibitem{IE2}H. Tajima, arXiv:1311.1285, (2013).
\bibitem{IE1.5}H. Tajima, JPS Conference Proceedings, \textbf{1}, 012129 (2014).


\bibitem{Reeb}D. Reeb and M. M. Wolf, New J. Phys. \textbf{16}, 103011, (2014).
\bibitem{Max}M. F. Frenzel, D. Jennings, and T. Rudolph, arXiv:1406.3937,(2014).
\bibitem{Malabarba}A. S. L. Malabarba, A. J. Short, P. Kammerlander, New. J. Phys. \textbf{17}, 045027 (2015).
\bibitem{openquantum}H. P. Breuer and F. Petruccione, {\it{The Theory of Open Quantum Systems}} (Oxford University Press, USA, 2007).
\bibitem{LD1}R. Bahadur and R. R. Rao, Ann. Math. Stat., \textbf{31}, 1015 (1960).
\bibitem{LD2} C. Joutard, Math. Methods Statist., \textbf{22}, 2, 155-164 (2013),
\bibitem{1ponme}
M. Hayashi and H. Tajima, arXiv:1504.06150, (2015).

\bibitem{X1}Raam Uzdin, Amikam Levy, and Ronnie Kosloff, Phys. Rev. X 5, 031044 (2015).
\bibitem{X2}Matteo Lostaglio, Kamil Korzekwa, David Jennings, and Terry Rudolph, Phys. Rev. X 5, 021001 (2015).



\bibitem{Tasakitext}H. Tasaki, \textit{Thermodynamics; A modern point of view}, (Baifukan (in Japanese), ISBN:4563024325, (2000).)

\bibitem{Shimizu}A. Shimizu, {\it{Basis of thermodynamics}}, (University of Tokyo Press (in Japanese), ISBN: 4130626094, (2007).)

\bibitem{AN}S. Amari and H. Nagaoka, {\it{Methods of Information Geometry}}. (Oxford University Press, 2000).

\bibitem{Fannes}M. A. Nielsen and I. L. Chuang, {\it{Quantum Computation and Quantum Information}} (Cambridge University Press, Cambridge, 2000).

\bibitem{Han}
T. S.\ Han:
{\it Information-Spectrum Methods in Information Theory},
(Springer, Berlin Heidelberg New York, 2002)
(originally appeared in Japanese in 1998).

\bibitem{BH1}
D. Blackwell and J. L. Hodges, 
{\em Ann.Math. Stat.}, \textbf{30}, 1113, (1959).

\bibitem{Dembo98}
A. Dembo and O. Zeitouni, 
{\em Large Deviations Techniques and Applications}, 
Stochastic Modelling and Applied Probability. Springer, 2 edition, (1998).

\bibitem{BBM2}
C. M. Bender, D. C. Brody, and B. K. Meister,
Proc. R. Soc. A London {\bf 461}, 733 (2005).

\bibitem{Baxter}
R.J. Baxter, {\em Exactly solved models in statistical mechanics}, 
London, Academic Press, 1982 

\bibitem{Ellis}
R. S. Ellis, {\em Entropy, Large Deviations and Statistical Mechanics}, 
Springer, 1985


\bibitem{catalyst}J. \r{A}berg, Phys. Rev. Lett. \textbf{113}, 150402 (2014).

\bibitem{Duelle}
D. Duelle, {\em Statistical Mechanics: Rigorous Results}, 
World Scientific \& Imperial College Press, 1999

\bibitem{pre}
M. Hayashi and H. Tajima, in preparation.

\bibitem{pre2}
H. Tajima and M. Hayashi, in preparation.

\end{thebibliography}
\end{document}